\documentclass[preprint, 12pt]{elsarticle}

\biboptions{sort&compress}
\usepackage{amssymb}
\usepackage{latexsym}
\usepackage{acronym}

\usepackage{url}
\usepackage{xcolor}
\definecolor{newcolor}{rgb}{.8,.349,.1}

\usepackage[T1]{fontenc}
\usepackage[utf8]{inputenc}
\usepackage{textcomp}
\usepackage{amsmath,amsfonts,braket,mathtools,graphicx,amsthm}
\usepackage{caption, subcaption}
\usepackage[font=footnotesize]{caption}
\usepackage{bm} %greek letters in textbf
\usepackage{blindtext}
\usepackage{hyperref}
\usepackage{colortbl} %cellcolor
\definecolor{cellbox}{RGB}{200, 230, 245} %cellbox color
\definecolor{cellbox_bestsampler}{RGB}{200, 230, 245} %cellbox color for best sampler
\definecolor{cellbox_worstsampler}{RGB}{230, 230, 230} %cellbox color for worst sampler

\definecolor{DarkViolet}{rgb}{0.58, 0.0, 0.83}
\definecolor{mydarkgreen}{RGB}{0, 165, 0}

\usepackage{multirow} %for multirow tables

\usepackage{autonum}

\usepackage{tikz}
\usetikzlibrary{shapes.geometric, arrows, arrows.meta, fadings}

\tikzstyle{arrow} = [thick,->,>=stealth]
\usepackage{varwidth}
\usepackage{colortbl}

\usetikzlibrary{calc}
\pgfdeclareverticalshading{arrowshading}{100bp}{
    color(0bp)=(blue);
    color(50bp)=(blue);
    color(50bp)=(green);
    color(100bp)=(green)
}
\usepackage{multicol} % for multicolumn support

\usetikzlibrary{shapes.arrows}
\makeatletter
\def\createshadingfromlist#1#2#3{%
  \pgfutil@tempcnta=0\relax
  \pgfutil@for\pgf@tmp:={#3}\do{\advance\pgfutil@tempcnta by1}%
  \ifnum\pgfutil@tempcnta=1\relax%
    \edef\pgf@spec{color(0)=(#3);color(100)=(#3)}%
  \else%
    \pgfmathparse{50/(\pgfutil@tempcnta-1)}\let\pgf@step=\pgfmathresult%
    \pgfutil@tempcntb=1\relax%
    \pgfutil@for\pgf@tmp:={#3}\do{%
      \ifnum\pgfutil@tempcntb=1\relax%
        \edef\pgf@spec{color(0)=(\pgf@tmp);color(25)=(\pgf@tmp)}%
      \else%
        \ifnum\pgfutil@tempcntb<\pgfutil@tempcnta\relax%
          \pgfmathparse{25+\pgf@step/4+(\pgfutil@tempcntb-1)*\pgf@step}%
          \edef\pgf@spec{\pgf@spec;color(\pgfmathresult)=(\pgf@tmp)}%
        \else%
          \edef\pgf@spec{\pgf@spec;color(75)=(\pgf@tmp);color(100)=(\pgf@tmp)}%
        \fi%
      \fi%
      \advance\pgfutil@tempcntb by1\relax%
    }%
  \fi%
  \csname pgfdeclare#2shading\endcsname{#1}{100}\pgf@spec%
}

\createshadingfromlist{shading1}{vertical}{red,yellow,green,cyan,blue}
\createshadingfromlist{shading2}{vertical}{blue!50,green!50}
\createshadingfromlist{shading3}{vertical}{black,blue,cyan,white}

\tikzset{
    *|/.style={
        to path={
            (perpendicular cs: horizontal line through={(\tikztostart)},
                                 vertical line through={(\tikztotarget)})
            -- (\tikztotarget) \tikztonodes
        }
    }
}

\newcommand{\bmp}{\boldsymbol{p}}

\newcommand{\btheta}{\boldsymbol{\theta}}
\newcommand{\bmtheta}{\boldsymbol{\theta}}
\newcommand{\by}{\boldsymbol{y}}

\acrodef{AT-HMC}[AT-HMC]{}
\acrodef{AT-GHMC}[AT-GHMC]{}

\usepackage{xcolor}

\date{\today}

\begin{document}

\begin{frontmatter}

\title{Adaptive tuning of Hamiltonian Monte Carlo methods}

\author[1,2]{Elena Akhmatskaya} \corref{cor1}
\ead{eakhmatskaya@bcamath.org}
\author[1,3]{Lorenzo Nagar} \corref{cor2}
\ead{lnagar@bcamath.org}
\author[4]{Jose Antonio Carrillo}
\author[1,3]{Leonardo Gavira Balmacz}
\author[5]{Hristo Inouzhe}
\author[6]{Martín Parga Pazos}
\author[7]{María Xosé Rodríguez Álvarez}
\cortext[cor1, cor2]{Corresponding authors:}
%Add other authors

\address[1]{BCAM -- Basque Center for Applied Mathematics, Bilbao, Spain}
\address[2]{Ikerbasque -- Basque Foundation for Science, Bilbao, Spain}
\address[3]{UPV/EHU -- Universidad del País Vasco / Euskal Herriko Unibertsitatea, Basque Country, Spain}
\address[4]{Mathematical Institute, University of Oxford, Oxford, United Kingdom}
\address[5]{Departamento de Matemáticas, Universidad Autónoma de Madrid, Madrid, Spain}
\address[6]{CIC bioGUNE -- Center for Cooperative Research in Biosciences, Derio, Spain}
\address[7]{Departamento de Estatística e Investigación Operativa, Universidade de Vigo, Vigo, Spain}
%Add other affiliations

\begin{abstract}
With the recently increased interest in probabilistic models, the efficiency of an underlying sampler becomes a crucial consideration. Hamiltonian Monte Carlo is one popular option for models of this kind.
Performance of the method, however, strongly relies on a choice of parameters associated with an integration for Hamiltonian equations. Up to date, such a choice remains mainly heuristic or introduces time complexity. We propose a novel computationally inexpensive and flexible approach (we call it Adaptive Tuning or ATune) that, by combining a theoretical analysis of the multivariate Gaussian model with simulation data generated during a burn-in stage of a Hamiltonian Monte Carlo simulation, detects a system specific splitting integrator with a set of reliable sampler's hyperparameters, including their credible randomization intervals, to be readily used in a production simulation. The method automatically eliminates those values of simulation parameters which could cause undesired extreme scenarios, such as resonance artifacts, low accuracy or poor sampling. The new approach is implemented in the in-house software package \textsf{HaiCS}, with no computational overheads introduced in a production simulation, and can be easily incorporated in any package for Bayesian inference with Hamiltonian Monte Carlo. The tests on popular statistical models reveal the superiority of adaptively tuned standard and generalized Hamiltonian Monte Carlo methods in terms of stability, performance and accuracy over conventional Hamiltonian Monte Carlo tuned heuristically and coupled with the well-established integrators. We also claim that the generalized Hamiltonian Monte Carlo is preferable for achieving high sampling performance. The efficiency of the new methodology is assessed in comparison with state-of-the-art samplers, e.g. the No-U-Turn-Sampler, in real-world applications, such as endocrine therapy resistance in cancer, modeling of cell-cell adhesion dynamics and influenza A epidemic outbreak.
\end{abstract}

\begin{keyword}
Hamiltonian Monte Carlo, Generalized Hamiltonian Monte Carlo, adaptive hyperparameters tuning, adaptive multi-stage splitting integration, randomization intervals, biomedical applications  
%{\color{blue} To be filled.}
\end{keyword}
\end{frontmatter}

\section{Introduction}

Hamiltonian Monte Carlo (HMC) \cite{duane1987hmc} has emerged as a powerful and widely adopted methodology for sampling from high-dimensional complex probability distributions, making it invaluable in Bayesian inference applications.
By leveraging the Hamiltonian equations of motion for generating Monte Carlo proposals, HMC exploits gradient information on posterior distributions to enhance the convergence of the Markov chain and enables efficient exploration of the parameter space. 

Further enhancing the sampling performance of HMC has attracted growing attention within the computational statistics community in recent years \cite{Sountsov2021, Apers2022, Chen2023, hmc_machine_learning, BouRabee2024GIST, BouRabee2024Adapt, Modi2024, BironLattes2024}.

The %introduction of a 
Markov kernel of HMC based on Hamiltonian dynamics induces a set of hyperparameters and tuning blocks that critically impact sampling performance and include a numerical integrator for solving Hamiltonian equations, 
an integration step size, 
%$\Delta t$,
a number of integrations per iteration, 
%$L$,
a mass matrix. %
%$M$. %, etc. 
For instance, while a 
smaller step size
%$\Delta t$ 
ensures high acceptance rates and reduces computational waste, it also shortens proposal moves, increasing correlation between samples. Conversely, a larger 
number of integration steps per iteration 
%$L$ 
enables longer trajectories and more independent samples but comes at the cost of increased computational overhead due to additional gradient evaluations per iteration. Consequently, significant research efforts have been devoted to identifying optimal parameter settings for HMC to balance computational efficiency and sampling performance. 

Early strategies for tuning the 
integration step size
%, $\Delta t$, 
focused on targeting an optimal acceptance rate for the Metropolis test \cite{beskos_optimal_tuning, Betancourt2014} or employing on-the-fly adaptation using primal-dual averaging algorithms \cite{hoffman_gelman2014}. 
Some sophisticated integration schemes have also been proposed, replacing the standard Verlet algorithm \cite{verlet1967} %\cite{verlet1967, swope1982} 
with more accurate multi-stage splitting integrators to enhance sampling performance and efficiency  
\cite{takaishi_deforcrand_2006, bcss_paper, AIApaper2016, sAIA_paper, Tamborrino2024, nagar_thesis}.
%\cite{mclachlan1995, takaishi_deforcrand_2006, bcss_paper, AIApaper2016, akhmatskaya_etal_2017, radivojevic_et_al_2018, sAIA_paper, Tamborrino2024}. 
Comprehensive reviews on numerical integration in HMC are available in \cite{%campos_sanz-serna2017, 
bou-rabee_sanz-serna_2018, 
%calvo2021hmc, 
Blanes2024}. 

Regarding the 
number of integration steps per iteration,
%, $L$, 
a common recommendation is randomization 
\cite{random_hmc2017},
%\cite{Mackenzie1989, random_hmc2017}, 
though more advanced algorithms have been proposed, such as automated trajectory length selection based on the double-backing criterion \cite{hoffman_gelman2014}. 

Additional contributions to HMC tuning include position-dependent mass matrix adaptation \cite{banana_girolami} and methods for determining the proper number of Monte Carlo iterations to ensure convergence \cite{Margossian2023}. 

More recent advancements in optimization algorithms for HMC hyperparameters address both integration step size \cite{Apers2022, BouRabee2024GIST, BouRabee2024Adapt, Tamborrino2024, Modi2024, BironLattes2024} and number of integration steps per iteration \cite{Sountsov2021, Chen2023, BouRabee2024GIST, Modi2024, BironLattes2024}. 

In this study, we propose a computationally inexpensive adaptive tuning procedure that identifies an optimal numerical integrator along with a set of HMC hyperparameters tailored to a given simulation system. We call this procedure ATune and refer to an HMC method reinforced with such tuning as Adaptively Tuned HMC (\acs{AT-HMC}).

Furthermore, we extend our focus to Generalized Hamiltonian Monte Carlo (GHMC) 
\cite{kennedy_pendleton2001},
%\cite{horowitzGHMC, kennedy_pendleton2001}, 
a promising but relatively unexplored variant of standard HMC. 

The key distinction between HMC and GHMC lies in the update mechanism for the momenta. 
In HMC, momenta are completely discarded after the Metropolis test and resampled at the beginning of each iteration from a 
normal distribution with mass matrix covariance.
%Gaussian distribution $\mathcal{N} (0, M)$. 
On the contrary, GHMC performs a partial momentum update (PMU) \cite{horowitzGHMC}.
The PMU introduces an 
additional hyperparameter
%$\varphi$, 
which, if not carefully chosen, can degrade sampling performance. This likely explains why GHMC has been less commonly used than HMC. 

However, when properly tuned, GHMC can offer significant improvements in sampling accuracy and efficiency. There are two primary reasons for that. 

First, in contrast to HMC, GHMC produces irreversible chains by construction 
 \cite{ottobreIrrevers2016}.
%\cite{Diaconis2000, neal2011mcmc, ottobreIrrevers2016}. 
This usually implies a faster convergence to equilibrium and a reduction in asymptotic variance  \cite{Neal2004, Sun2010, Song2022}.
%\cite{Neal2004, Sun2010, duncanIrrevers2016, ottobreIrrevers2016, Song2022}. 
Second, the partial momentum update in GHMC with a small momentum mixing angle makes a choice of the number of steps per integration leg less critical than in the case of HMC, enabling efficient sampling with shorter Hamiltonian trajectories \cite{skeel2014}, thereby enhancing both the speed and efficiency of the sampling process.

Although theoretical studies recognize potential of GHMC, little effort has been devoted to developing effective tuning strategies. Only recently, GHMC has begun to attract growing interest within the scientific community \cite{Gouraud2022, rioudurand2022, Hoffman2022, rioudurand2023tuningMALT}. 

Building on these insights, we extend our Adaptive Tuning to GHMC to derive \acs{AT-GHMC}. 
Apart from the previously discussed in the literature choices of HMC and GHMC simulation parameters, 
we add to our optimal settings list randomization intervals for simulation parameters. Indeed, randomization of simulation parameters is part of any HMC-based algorithm \cite{neal2011mcmc}, but until now its choice remains purely heuristic.  

In summary, we aim to find optimal, or close to optimal choices for:
\begin{itemize}
\item an integration scheme for the Hamiltonian equations of motion 
\item a dimensional integration step size
%$\Delta t$ 
and its randomization interval
\item a number of integration steps
% $L$ 
per Monte Carlo iteration and its randomization interval
\item a noise control parameter
%$\varphi$ 
(along with the appropriate randomization interval) in the PMU.
\end{itemize}
Our approach relies on the adaptive s-AIA integration method \cite{sAIA_paper}, which 
%The adaptive s-AIA integration method \cite{sAIA_paper} 
provides useful guidance on choices of integrators and optimal step size locations. %for both a choice of a numerical integrator and a location of an optimal step size. 
According to \cite{sAIA_paper}, s-AIA integrators outperform the standard Velocity Verlet and other fixed-parameter multi-stage integrators in terms of both integration accuracy and their impact on sampling efficiency of HMC, for any simulation step size within a system-specific stability interval.
Additionally, %as previously suggested in \cite{bcss_paper, mazur}, 
a key observation in \cite{sAIA_paper} is that the highest performance of HMC is achieved near the center of the stability interval, provided a stability limit is accurately estimated. This also was suggested in \cite{bcss_paper, mazur}. 
Using these findings and the analysis underlying the s-AIA approach, we identify optimal integration schemes and step size randomization intervals for both \acs{AT-HMC} and \acs{AT-GHMC}, as well as an optimal randomization scheme for
a number of integration steps
%$L$ 
in \acs{AT-GHMC}.  

Moreover, we revisit the optimization strategies previously proposed for Modified HMC in \cite{akhmatskaya_etal_2017}, and adapt them to GHMC, refining the momentum refreshment procedure to derive an optimal, system-dependent noise parameter 
in the PMU.
%$\varphi$. 
Using the proposed settings, GHMC achieves efficient phase-space exploration even with a small 
number of integration steps
%$L$ 
\cite{skeel2014}. 

The new tuning methodology does not incur additional computational overhead, as the optimal values are either system-specific (and can be determined a priori) or computed during a burn-in stage.
Numerical experiments on standard benchmarks 
confirm the positive effect of proposed optimal settings on HMC and GHMC performance. 

Additionally, we assess the efficiency of \acs{AT-HMC} and \acs{AT-GHMC} on three real-world applications, each relying on a different probabilistic model, 
and compare the results with 
those obtained using the state-of-the-art samplers commonly employed in Bayesian inference on such models. The applications include:

\begin{itemize}
\item Patient resistance to endocrine therapy in breast cancer -- a Bayesian Logistic Regression (BLR) model
\item Cell-cell adhesion dynamics -- a Partial Differential Equation (PDE) model
\item Influenza A (H1N1) epidemics outbreak -- an Ordinary Differential Equation (ODE) model.
\end{itemize}
The paper is organized as follows. Section~\ref{sec:HMC} presents the essential background, covering HMC, GHMC and the numerical integration of Hamiltonian dynamics. Section~\ref{sec:TopPerformanceAroundHSL} revisits the numerical experiments in \cite{sAIA_paper}, showing that optimal performance is consistently achieved near the center of a stability interval. In Section~\ref{sec:DeltatOptimalRandomization}, we present the refined randomization interval for 
the optimal simulation step size
%$\Delta t$ 
and support our choice by numerical experiments.

Sections~\ref{sec:TuningPhi} and \ref{sec:TuningL} introduce %and validate on the set of benchmarks (see Table \ref{tab:Benchmarks})
the optimization procedures for 
the random noise parameter in the PMU
%$\varphi$ 
and the number of integration steps per iteration.
%$L$. 
The Adaptive Tuning algorithm ATune is outlined in Section~\ref{sec:Algorithm}, validated on the set of benchmarks in Section~\ref{sec:NumericalExperiments}, and systematically tested in the three real-world application studies in Section~\ref{sec:Applications}. Section~\ref{sec:Conclusion} summarizes our conclusions.
Additional technical details and simulations are provided in the appendices.

\section{Hamiltonian Monte Carlo}\label{sec:HMC}

\subsection{Standard formulation}
Hamiltonian Monte Carlo \cite{duane1987hmc} is a Markov Chain Monte Carlo (MCMC) method for obtaining $N$ correlated samples $\{\bm{\theta}_n\}_{n = 1, ..., N}$ from a target probability distribution $\pi (\bmtheta)$ in $\mathbb{R}^D$, $D$ being the dimension of $\bmtheta$. HMC generates a Markov chain in the joint phase space $\Omega \subseteq \mathbb{R}^{2D}$ with invariant distribution
\begin{equation}\label{eq:InvariantJointDistribution}
\pi(\bm{\theta}, \bm{p}) = \pi(\bm{\theta}) n(\bm{p}) \propto \exp (- H(\bm{\theta}, \bm{p})),
\end{equation}
and recovers $\pi (\bmtheta)$ by marginalizing out the auxiliary momentum variable $\bm{p}$ (distributed according to $n(\bm{p})$). 

In \eqref{eq:InvariantJointDistribution}, $H(\bm{\theta}, \bm{p})$ is the Hamiltonian
\begin{equation}\label{eq:HamiltonianSeparable}
H(\bmtheta, \bmp) = \frac{1}{2} \bmp^T M^{-1} \bmp + U(\bmtheta),
\end{equation}
where $M$ is the symmetric positive definite mass matrix, and $U(\bmtheta)$ is the potential energy, which contains information about the target distribution $\pi(\bm{\theta})$:
\begin{equation}%\label{eq:PotentialFunctionHMC}
U(\bmtheta) = - \log \pi(\bmtheta) + \text{const} .
\end{equation}
HMC generates new proposals $(\bmtheta', \bmp')$ by alternating momentum update steps, where $\bmp$ is drawn from a Gaussian $\mathcal{N} (0, M)$, with the steps that update position $\bmtheta$ and momenta $\bmp$ through the numerical integration of the Hamiltonian equations 
\begin{equation}\label{eq:HamiltonianSystemSeparable}
\frac{d \bm{\theta}}{dt} = M^{-1} \bm{p}, \qquad \frac{d \bm{p}}{dt} = - \nabla_{\theta} U(\bm{\theta}),
\end{equation}
%The latter is 
performed $L$ times using an explicit symplectic and reversible integrator $\Psi_h$ \cite{numerical_hamiltonian_problems} ($h$ being an integration step size). Thus, 
\begin{equation}\label{eq:Integration}
(\bmtheta', \bmp') = \underbrace{\Psi_h \circ ... \circ \Psi_h}_\text{$L$ times} (\bmtheta, \bmp).
\end{equation}
Symplecticity and reversibility of $\Psi_h$ ensure that the generated Markov chain is ergodic, thus that $\pi(\bmtheta, \bmp)$ is an invariant measure for the generated Markov chain.  

Due to numerical integration errors, the Hamiltonian energy and hence the target density \eqref{eq:InvariantJointDistribution} are not exactly preserved. To address this issue, the invariance of the target density is secured through a Metropolis test with the acceptance probability
\begin{equation}\label{eq:MetropolisTestAR}
\alpha = \min \{1, \exp (-\Delta H) \},
\end{equation}
where
\begin{equation}\label{eq:EnergyError}
\Delta H = H (\bmtheta', \bmp') - H (\bmtheta, \bmp)
\end{equation}
is the energy error resulting from the numerical integration \eqref{eq:Integration}. In case of acceptance, $\bmtheta'$ becomes the starting point for the subsequent iteration. Conversely, if the proposal is rejected, the previous state $\bmtheta$ is retained for the next iteration. Regardless of acceptance or rejection, the momentum variable is discarded, and a new momentum $\bmp$ is drawn from $\mathcal{N} (0, M)$. 

\subsection{Generalized Hamiltonian Monte Carlo}
Generalized Hamiltonian Monte Carlo \cite{kennedy_pendleton2001} incorporates a partial momentum update (PMU) \cite{horowitzGHMC} in standard Hamiltonian Monte Carlo to replace the full refreshment of auxiliary momenta. The PMU reduces random-walk behavior by partially updating the auxiliary momenta $\boldsymbol{p}$ through mixing with an independent and identically distributed (i.i.d) Gaussian noise $\boldsymbol{u} \sim \mathcal{N} (0, M)$:

\begin{equation}\label{eq:PMUIntro}
%\boldsymbol{p}^{\text{new}} = \sqrt{1 - \varphi} \, \boldsymbol{p} + \sqrt{\varphi} \, \boldsymbol{u}.
\bm{p} \leftarrow \sqrt{1 - \varphi} \, \bm{p} + \sqrt{\varphi} \, \bm{u}.
\end{equation}
Here, $\varphi \in (0, 1]$ is a random noise parameter which determines the extent to which the momenta are refreshed. When $\varphi = 1$, the update \eqref{eq:PMUIntro} reduces to the standard HMC momentum resampling scheme.

Notice that the PMU produces configurations that preserve the joint target distribution, as the orthogonal transformation in \eqref{eq:PMUIntro} maintains the distributions of $\boldsymbol{p} \sim \mathcal{N} (0, M)$.
Since the momenta are not fully discarded, GHMC performs a momentum flip after each rejected proposal.
As shown in \cite{neal2011mcmc} and formally proved in \cite{skeel2014}, the three core steps of GHMC (Hamiltonian dynamics combined with the Metropolis test, the PMU, and momentum flip) each individually satisfy detailed balance and maintain reversibility, while their combination does not. This means that the resulting Markov chains are not reversible. Nevertheless, \cite{skeel2014} demonstrated that GHMC fulfils the weaker modified detailed balance condition, ensuring the stationarity of the generated Markov chains.

\subsection{Numerical integration of Hamiltonian dynamics}
The most commonly used numerical scheme for solving the Hamiltonian equations \eqref{eq:HamiltonianSystemSeparable} is the \emph{Velocity Verlet} integrator \cite{verlet1967}. This scheme is widely favored due to its simplicity, good stability, and energy conservation properties. The algorithm reads
\begin{align}
\bmp & \leftarrow \bmp - \frac{h}{2} \nabla_{\theta} U(\bmtheta), \nonumber \\
\bmtheta & \leftarrow \bmtheta + h M^{-1} \bmp, \nonumber \\
\bmp & \leftarrow \bmp - \frac{h}{2} \nabla_{\theta} U(\bmtheta). \label{eq:1sVV}
\end{align}
Here, $h$ represents the
dimensionless
integration step size. 
We will use a notation $\Delta t$ for its dimensional counterpart.
The updates of momenta and position in \eqref{eq:1sVV} are often called momentum \emph{kick} and position \emph{drift}, respectively.

More sophisticated integration schemes have been proposed in literature, including \emph{multi-stage palindromic splitting integrators}
\cite{Blanes2024}.
% \cite{blanes_casas_murua2008, bcss_paper, campos_sanz-serna2017}. 
These methods perform multiple splits of the separable Hamiltonian system \eqref{eq:HamiltonianSystemSeparable}, applying a sequence of kicks and drifts \eqref{eq:1sVV} with arbitrary chosen step sizes, in contrast to the half-kick, drift, half-kick structure of the standard Verlet method. By denoting a position drift and a momentum kick, respectively, with
\begin{equation}\label{eq:SeparateHamiltonianFlows}
\varphi^A_h (\bmtheta, \bmp) = (\bmtheta + h M^{-1} \bmp, \bmp), \qquad \varphi^B_h (\bmtheta, \bmp) = (\bmtheta, \bmp - h \nabla_{\theta} U(\bmtheta)),
\end{equation}
the family of $k$-stage splitting integrators is defined as \cite{bou-rabee_sanz-serna_2018}
\begin{equation}\label{eq:kStageSchemeEven}
\Psi_h^k = \varphi^B_{b_1 h} \circ \varphi^A_{a_1 h} \circ \dots \circ \varphi^A_{a_{k'} h} \circ \varphi^B_{b_{k'+1} h} \circ \varphi^A_{a_{k'} h} \circ \dots \circ \varphi^A_{a_1 h} \circ \varphi^B_{b_1 h}, \quad b_i, a_j \in \mathbb{R}^+,
\end{equation}
if $k = 2 k'$, and
\begin{equation}\label{eq:kStageSchemeOdd}
\Psi_h^k = \varphi^B_{b_1 h} \circ \varphi^A_{a_1 h} \circ \dots \circ \varphi^B_{b_{k'} h} \circ \varphi^A_{a_{k'} h} \circ \varphi^B_{b_{k'} h} \circ \dots \varphi^A_{a_1 h} \circ \varphi^B_{b_1 h}, \quad b_i, a_j \in \mathbb{R}^+,
\end{equation}
if $k = 2k'-1$. The coefficients $b_i$, $a_j$ in \eqref{eq:kStageSchemeEven}--\eqref{eq:kStageSchemeOdd} have to satisfy the conditions $2 \sum_{i=1}^{k'} b_i + b_{k'+1} = 2 \sum_{j=1}^{k'} a_j = 1$, and $2 \sum_{i=1}^{k'} b_i = 2 \sum_{j=1}^{k'-1} a_j + a_{k'}= 1$, respectively, to ensure an integration step of length $h$. A term \emph{stage} ($k$) refers to the number of gradient evaluations per integration. It is straightforward to verify that the only 1-stage splitting integrator of the form \eqref{eq:kStageSchemeOdd} is the Velocity Verlet (VV) scheme  \eqref{eq:1sVV}, which can be expressed as
\begin{equation}\label{eq:1sVVmap}
\Psi_h^{\text{VV}} = \varphi^B_{\frac{h}{2}} \circ \varphi^A_h \circ \varphi^B_{\frac{h}{2}}.
\end{equation}
Various choices of integration coefficients $b_i$, $a_j$ in \eqref{eq:kStageSchemeEven}--\eqref{eq:kStageSchemeOdd} yield integrators that reach their pick performance at different specific values of a simulation step size $h$. 
To address the step size dependency, adaptive integration approaches AIA \cite{AIApaper2016} and s-AIA \cite{sAIA_paper} were developed for molecular simulations and Bayesian inference applications, respectively. For a given system and simulation step size $h$, these methods select the most suitable 2- (AIA, s-AIA2) and 3- (s-AIA3) stage splitting integrator ensuring optimal energy conservation for harmonic forces. 
Examples of 2- and 3-stage integrators, including those used in this study, are %summarized in Table~\ref{tab:IntegratorsTableGHMC} and
reviewed in \cite{bcss_paper, campos_sanz-serna2017, sAIA_paper}.

\begin{table}[t]
\centering
\resizebox{\textwidth}{!}{
\begin{tabular}{l r l r r}
Integrator & N. of stages ($k$) & Feature & References \\
\hline
VV & $1$ & longest stability interval & \cite{verlet1967} \\
BCSS$k$ & 2, 3 & minimizes $\mathbb{E} [\Delta H]$ in $(0, h_\text{CoLSI})$ & \cite{bcss_paper} \\
ME$k$ & 2, 3 & minimizes truncation error for $h \to 0$  & \cite{mclachlan1995, predescu2012} \\
sAIA$k$ & 2, 3 & minimizes $\mathbb{E} [\Delta H]$ $\forall h \in (0, 2 k)$ & \cite{sAIA_paper} \\
\hline
\end{tabular}
}
\caption{\label{tab:IntegratorsTableGHMC} Multi-stage splitting integrators considered for this study. $\mathbb{E} [\Delta H]$ is the expected energy error resulting from the numerical integration of Hamiltonian dynamics, $h_\text{CoLSI} = k$ is the centre of the longest stability interval.}
\end{table}

\section{Tracking the location of optimal step size}\label{sec:TopPerformanceAroundHSL}
At first, we aim to demonstrate that top performance of HMC is typically achieved with a step size  being near the center of the longest stability interval (CoLSI) for an integrator in use, as previously suggested in \cite{mazur} and lately confirmed in \cite{bcss_paper, sAIA_paper}. We will use notations $h_{\text{CoLSI}}$ and $\Delta t_{\text{CoLSI}}$ for such dimensionless and dimensional step sizes, respectively. We recall that for a $k$-stage integrator, the length of the longest dimensionless stability interval equals 2$k$, hence, $h_\text{CoLSI}$ = $k$ \cite{bcss_paper}. %\cite{bcss_paper, sAIA_paper}.

We consider the numerical experiments presented in \cite{sAIA_paper} for the three benchmarks -- 1000-dimensional multivariate Gaussian model, G1000,
% Gaussian 1 (for simplicity, from now on we call it G1000) 
and two BLR models, German and Musk -- performed with the numerical integrators summarized in Table \ref{tab:IntegratorsTableGHMC}. 
For further details on benchmarks, see Table~\ref{tab:Benchmarks} and \cite{sAIA_paper}.

For each benchmark, we identify the integrator and integration step size which lead to the lowest ratio between the number of gradient evaluations, grad (which represents the bulk of the computational cost in an HMC simulation), and the minimum effective sample size \cite{geyer1992} across variates, minESS, computed for HMC production chains of the length of $N_\text{conv} + 1000$ iterations. Here $N_\text{conv}$ is the number of samples in the production chains, required for reaching convergence \cite{vethari_etal_2021}.
%
% \cite{gelman_rubin1992, brooks_gelman1998, vethari_etal_2021}.
%, and thus the smaller $N_{1.01}$ corresponds to faster convergence. 
%normalized to the computational time minimum ESS across variates in HMC simulations
%%%%%%%%%%%%%%%%%%%%%%%%%%%%%%%%%%%%%%%%%%%%%%
We choose minESS (calculated through the \texttt{effectiveSize} function of the \texttt{coda} package of \texttt{R} \cite{coda}) as the most demanding one among the metrics considered in \cite{sAIA_paper}. 

\begin{sloppypar}
We refer to the performance metric introduced in such a way, i.e. $\text{grad/minESS}$, as $\text{gESS}$. In Section \ref{sec:NumericalExperiments}, we provide a detailed discussion of this metric and explore more performance metrics for validation of the proposed methodology. We emphasize that lower values of grad and higher minESS suggest faster convergence and better sampling efficiency, respectively, meaning that lower $\text{gESS}$ indicates the improved overall performance.
\end{sloppypar}

\begin{figure}[t]
    \centering
    \includegraphics[width=
    %0.48
    \textwidth]{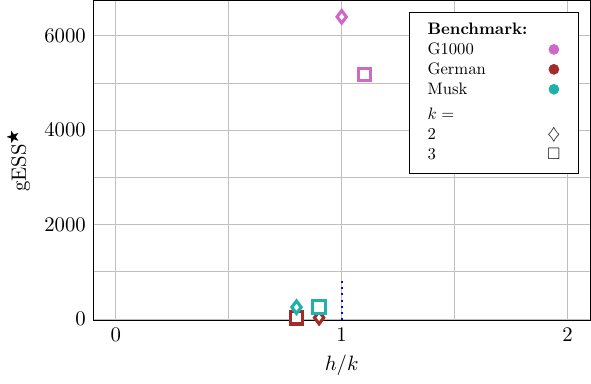} \hfill
        \caption{Best grad/minESS performance ($\text{gESS}^{\bigstar}$) obtained for the tested benchmarks -- G1000 (mauve), German (dark red) and Musk (turquoise) (Table \ref{tab:MetricsAroundHSL}). $k$ is the number of stages of an integrator. For any benchmark, top performance is achieved around the CoLSI, i.e. $h = k$.} %\textcolor{red}{EA: Figure has to be updated.}}
    \label{fig:TopPerformanceAroundHSL}
\end{figure}

Table \ref{tab:MetricsAroundHSL} displays the best grad/minESS performance, i.e. the lowest $\text{gESS} \coloneqq \text{gESS}^{\bigstar}$  achieved for each benchmark with $k$-stage integrators, $k = 2, 3$, and specifies the integrators along with the corresponding step sizes (normalized to the number of stages for clarity, i.e. normalized $h_\text{CoLSI}$ is equal to 1 for all $k$-stage integrators) responsible for such performance. The results are visualized in Figure \ref{fig:TopPerformanceAroundHSL} using normalized step sizes.
\begin{table}[t]
\centering
\begin{tabular}{l r r r l}
Benchmark & $k$ & $\text{gESS}^{\bigstar}$ & $h_{\text{gESS}^{\bigstar}}/k$ & Integrator \\ 
  \hline
\multirow{2}{*}{G1000} & 2 & 6402.3 & 1.0 & s-AIA2 \\
 & 3 & 5177.1 & 1.1 & BCSS3 \\
\hline
\multirow{2}{*}{German} & 2 & 28.8 & 0.9 & AIA2 \\
 & 3 & 26.4 & 0.8 & ME3 \\
\hline
\multirow{2}{*}{Musk} & 2 & 249.3 & 0.8 & BCSS2 \\
 & 3 & 252.8 & 0.9 & s-AIA3 \\
\hline
\end{tabular}
\caption{\label{tab:MetricsAroundHSL} Best grad/minESS ($\text{gESS}^{\bigstar}$)
%/minMCSE$^{-1}$
performance, detected in the HMC simulations in \cite{sAIA_paper} for the three benchmarks, using best performed integrator at $h_{\text{gESS}^{\bigstar}}$ integration step size (normalized to the number of stages $k$).} %\textcolor{red}{EA: Table has to be updated.} }
\end{table}

\begin{sloppypar}
From Table \ref{tab:MetricsAroundHSL}, one can observe that top performance is consistently achieved around the CoLSI, $h_\text{CoLSI} = 1$, and thus $h_\text{CoLSI}$ is an obvious candidate for an optimal step size. The deviation of $h_{\text{gESS}^{\bigstar}}$ from $h_\text{CoLSI}$ is not critical at this stage as an optimal step size is to be randomized yet, and only performance at a randomized step matters. 
\end{sloppypar}

Furthermore, 3-stage integrators always outperform 2-stage ones, as already noticed in \cite{radivojevic_et_al_2018} and \cite{sAIA_paper}. Hence, it is natural to search for an optimal integrator within the class of 3-stage integrators. Unfortunately, Table \ref{tab:MetricsAroundHSL} does not suggest a unique candidate for the best performed integrator in this class. 

We remark that according to \cite{sAIA_paper}, BCSS3 and s-AIA3 around CoLSI give rise to very similar accuracy and performance of HMC for all considered benchmarks. For the German benchmark, the efficiencies of all the three 3-stage integrators, i.e. BCSS3, ME3 and s-AIA3, at CoLSI are almost indistinguishable. Thus, it is not surprising that all those integrators appear in Table \ref{tab:MetricsAroundHSL}. However, only s-AIA3 adapts its performance to a choice of a step size, and this property becomes important when randomization of a step size is applied. In this context, s-AIA3 looks like the best pick, but this has to be confirmed yet. 
In the next Sections, we propose a randomization interval for $h_\text{CoLSI}$ and use the 3-stage integrators (BCSS3, ME3, s-AIA3) and the standard 1-stage Velocity Verlet for its validation.  

\section{Randomizing optimal step size}\label{sec:DeltatOptimalRandomization}
Further, we propose an approach for finding an optimal randomization interval for an integration step size $h_\text{CoLSI}$ (and for its dimensional counterpart $\Delta {t_\text{CoLSI}}$) and compare the resulted HMC performance with the best grad/minESS performance in Table \ref{tab:MetricsAroundHSL}. 

\subsection{Randomization interval}\label{sec:rho_maxima}
\begin{figure}[t]
    \centering
    \includegraphics[width=
    %0.48
    \textwidth]{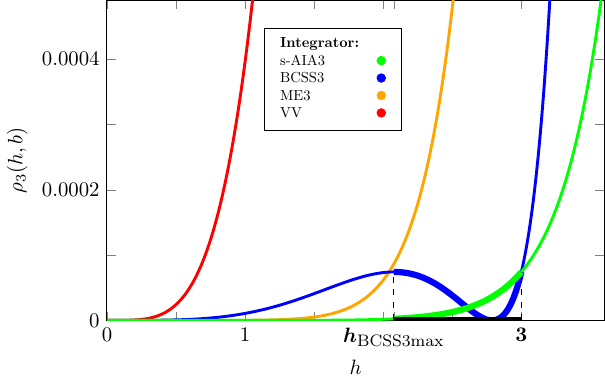} \hfill
        \caption{The upper bound $\rho_{\text{3-stage}} (h, b)$ of the expected energy error for 3-stage integrators.}
    \label{fig:Rho3}
\end{figure}

To identify endpoints of a randomization interval for an optimal step size we refer to the upper bound of the expected energy error for 3-stage integrators $\rho_3 (h, b)$ derived in \cite{sAIA_paper} and plotted for the range of integrators considered in this study in Figure \ref{fig:Rho3}.

When searching for an optimal randomization interval for $h_\text{CoLSI}$ the following considerations have to be taken into account. First, the interval should include the step sizes that guarantee the high accuracy of the numerical integration, i.e. $\rho_3 (h, b) \ll 1$. Then, one should keep interval narrow enough to stay close to an estimated optimal value of a step size and to avoid including too small step sizes, which usually satisfy the aforementioned inequality but lead to correlated samples in HMC. Finally, the approximate analysis proposed in \cite{sAIA_paper} may likely result in the overestimated values of the stability limit and thus $\Delta {t_\text{CoLSI}}$ (see \ref{app:3stageCorrection}). 
This implies a reasonable choice for the right endpoint to be $h_\text{CoLSI}$.

On the other hand, we recall that BCSS3 was derived to minimize the maximum of $\rho_3 (h, b)$ in $0 < h < h_\text{CoLSI}$ and thus to achieve the best performance near $h_\text{CoLSI}$. The inspection of Figure \ref{fig:Rho3} reveals that $\rho_3 (h, b)$ with $b = b_\text{BCSS3}$ exhibits a local maximum $h_\text{BCSS3max}$ on the left side of $h_\text{CoLSI}$. Within the interval ($h_\text{BCSS3max}$, $h_\text{CoLSI}$) both s-AIA3 and BCSS3 demonstrate very high accuracy, which deteriorates after $h$ passes the center of the stability interval. Thus, such an interval obeys all requirements on an optimal randomization interval summarized above. 

The local maxima of $\rho_{\text{BCSS3}} (h)$, i.e. $h_\text{BCSS3max}$, can be obtained by setting $\frac{\partial \rho_3 (h, b)}{\partial h} = 0$ and $b = b_{\text{BCSS3}}$ \cite{bcss_paper} 
 %in \eqref{eq:rhoBCSS3deriv} 
 (see \ref{app:rho_maxima}) 
 %the local maxima of $\rho_{\text{BCSS3}} (h)$ 
 %$h / k \approx 0.7$, i.e.
%\begin{equation*}
%h_\text{BCSSmax} \approx 0.69241,
%\end{equation*}
%for which $3 h_\text{BCSS3max} 
which yields $h_{\text{lower}}\equiv h_\text{BCSS3max}\approx 2.0772$. 

Therefore, the proposed randomization interval for the nondimensional optimal integration step size is
\begin{equation}\label{eq:h_BCSS3maxtoHSL}
h \sim \mathcal{U} (h_\text{lower}, 3) \approx \mathcal{U} (2.0772, 3),
\end{equation}
where $\mathcal{U}$ denotes the uniform distribution, and the corresponding interval for the dimensional step size is 
\begin{equation}\label{eq:t_BCSS3maxtoHSL}
\Delta {t} \sim \mathcal{U} (\Delta {t_\text{lower}}, \Delta {t_\text{CoLSI}}),
\end{equation}
where $\Delta {t_\text{lower}}$ and $\Delta {t_\text{CoLSI}}$ are calculated as proposed in \cite{sAIA_paper} and briefly described in \ref{app:NondimensionalizationAppendix}.

\begin{table}[t!]
\centering
\resizebox{0.90\width}{!}{\begin{tabular}{l l l r r r}
Benchmark & Sampler & Integrator & $L$ & $\Delta t$ & $\varphi$ \\
\hline
\multirow{8}{*}{G1000} & \multirow{4}{*}{HMC} & s-AIA3 & \multirow{3}{*}{$\mathcal{U} \{1, ..., 2666\}$} 
& \multirow{3}{*}{$\mathcal{U} (0.036, 0.052)$} & \multirow{4}{*}{ 1.0} \\
 & & BCSS3 &  
 & \\
 & & ME3 &  
 & \\ 
 & & VV & $\mathcal{U} \{1, ..., 7998\}$ 
 & $\mathcal{U} (0.012, 0.017)$ \\ \cline{2-6}
 & \multirow{4}{*}{GHMC} & s-AIA3 & \multirow{3}{*}{$\mathcal{U} \{1, ..., 1332\}$} 
& \multirow{3}{*}{$\mathcal{U} (0.036, 0.052)$} & \multirow{4}{*}{$\mathcal{U} (0.0, 0.5)$} \\
 & & BCSS3 &  
 & & \\
 & & ME3 &
 & & \\
 & & VV & $\mathcal{U} \{1, ..., 3999\}$
 & $\mathcal{U} (0.012, 0.017)$ & \\
\hline
\multirow{8}{*}{German} &  \multirow{4}{*}{HMC} & s-AIA3 & \multirow{3}{*}{$\mathcal{U} \{1, ..., 16\}$} 
& \multirow{3}{*}{$\mathcal{U} (0.111, 0.160)$} & \multirow{4}{*}{1.0} \\
 & & BCSS3 &  
 & \\
 & & ME3 &  
 & \\
 & & VV & $\mathcal{U} \{1, ..., 49\}$
 & $\mathcal{U} (0.037, 0.053)$ \\ \cline{2-6}
 & \multirow{4}{*}{GHMC} & s-AIA3 & \multirow{3}{*}{$\mathcal{U} \{1, ..., 7\}$}
 & \multirow{3}{*}{$\mathcal{U} (0.111, 0.160)$} & \multirow{4}{*}{$\mathcal{U} (0.0, 0.5)$} \\
 & & BCSS3 &  
 & & \\
 & & ME3 &
 & & \\
 & & VV & $\mathcal{U} \{1, ..., 24\}$
 & $\mathcal{U} (0.037, 0.053)$ & \\
\hline
\multirow{8}{*}{Musk} &  \multirow{4}{*}{HMC} & s-AIA3 & \multirow{3}{*}{$\mathcal{U} \{1, ..., 110\}$} 
 & \multirow{3}{*}{$\mathcal{U} (0.083, 0.110)$} & \multirow{4}{*}{1.0} \\
 & & BCSS3 &  
 &  \\
 & & ME3 &  
 &  \\ 
 & & VV & $\mathcal{U} \{1, ..., 333\}$
 & $\mathcal{U} (0.028, 0.040)$ \\ \cline{2-6}
 & \multirow{4}{*}{GHMC} & s-AIA3 & \multirow{3}{*}{$\mathcal{U} \{1, ..., 55\}$}
 & \multirow{3}{*}{$\mathcal{U} (0.084, 0.121)$} & \multirow{4}{*}{$\mathcal{U} (0.0, 0.5)$} \\
 & & BCSS3 &  
 & & \\
 & & ME3 &  
 & & \\
 & & VV & $\mathcal{U} \{1, ..., 166\}$
  & $\mathcal{U} (0.028, 0.040)$ & \\
\hline
\end{tabular}}
\caption{\label{tab:HMCExperimentsRandomIntervalInput} HMC and GHMC parameter settings for the numerical experiments presented in Section \ref{sec:DeltaTRandomizationNumericalExperiments}. Integration step sizes for each randomization interval are provided in dimensional units.}
\end{table}
  
 \subsection{Validation}
 \label{sec:DeltaTRandomizationNumericalExperiments}

To validate the proposed optimal randomized step size and to choose an optimal integrator among the 3-stage integrators, we run HMC simulations with s-AIA3, BCSS3, ME3 and VV at $\Delta {t}$
%_\text{CoLSI}}$
in \eqref{eq:t_BCSS3maxtoHSL} (in 1-stage units for VV) for all benchmarks introduced in Section \ref{sec:TopPerformanceAroundHSL} and listed in Table~\ref{tab:MetricsAroundHSL}. 

%{\color{red} To evaluate $\Delta t$, for all numerical experiments, we use the more accurate $S_\omega$ fitting factor approach, which requires the computation of the Hessian matrix and its eigenvalues (see Appendix~C in \cite{AtuneArxiv} and \cite{sAIA_paper} for further details).
%In these and subsequent numerical experiments, eigenvalues are computed via a full dense eigendecomposition using the \texttt{dsyev} routine from the LAPACK package \cite{lapack_netlib}. 
%For the benchmarks considered in this Section, a computational overhead introduced by Hessian evaluations are insignificant compared with the overall cost of HMC/GHMC simulations.}

The remaining HMC parameters, namely, a number of integration steps per Monte Carlo iteration, $L$, and numbers of burn-in and production iterations, are kept the same as in the numerical experiments in \cite{sAIA_paper}.
% and Table \ref{tab:MetricsAroundHSL}. 
In addition, we conduct GHMC simulations to evaluate the efficacy of the proposed randomization interval with this sampler in comparison with HMC. 
 For the PMU in Eq. \eqref{eq:PMUIntro}, we set $\varphi$ to be uniformly selected from the interval $(0.0, 0.5)$ (a standard recommendation) and use twice shorter trajectories than for HMC.
We remark that this choice of GHMC parameters is based on heuristic intuition. The optimal selection of these parameters are investigated in the following sections.

The simulation parameters for HMC and GHMC used in this section are summarized in Table \ref{tab:HMCExperimentsRandomIntervalInput}. 

We emphasize that the comparison between the adaptive s-AIA3 integrator and the other fixed-parameter integrators is fair in terms of computational cost and does not require further normalization with respect to computational time. Indeed, the optimal interval \eqref{eq:t_BCSS3maxtoHSL} is determined once and shared across all integrators; the average number of gradient evaluations per iteration $\overline{L}k$ ($\overline{L}$ being the mean of the number of integration steps per Monte Carlo iteration) is kept the same for all of them (see Table~\ref{tab:HMCExperimentsRandomIntervalInput}), and the adaptive integration coefficients in s-AIA3 are identified via a pretabulated map \cite{sAIA_tables} (see also \cite{sAIA_paper} for further details), which incurs no additional computational overhead during the simulation.

%\ref{app:TablesNumericalExperimentsDeltat}. %Table \ref{tab:HMCExperimentsRandomIntervalInput}. 
%-\ref{tab:GHMCExperimentsRandomIntervalInput}, respectively. 

\begin{figure}[t!]
    \centering
    \includegraphics[width=\textwidth]{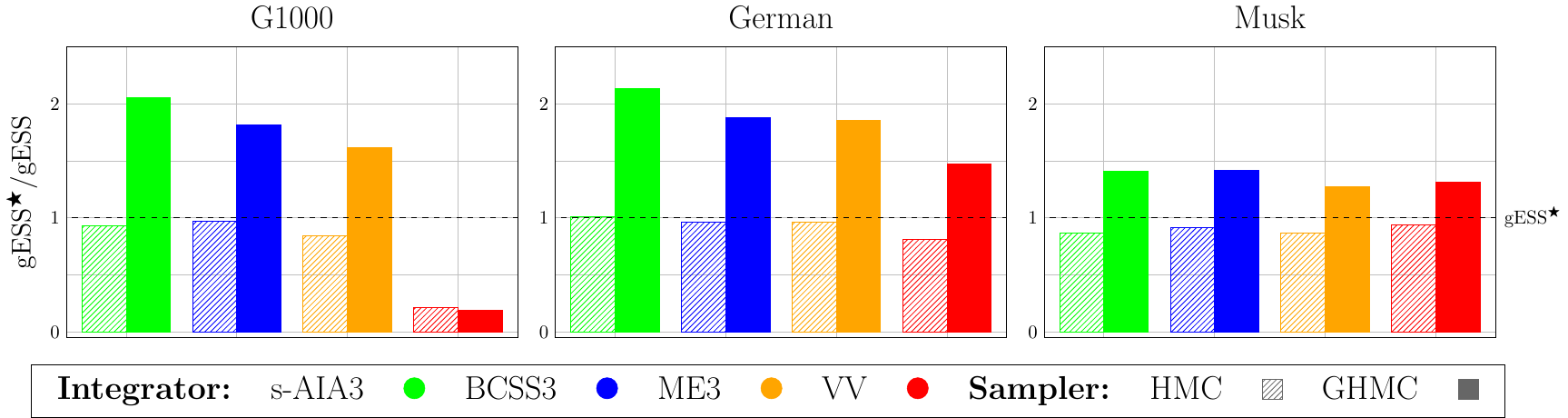}
        \caption{Comparison of $\text{gESS}^{\bigstar}$ performance (Table \ref{tab:MetricsAroundHSL}) with $\text{gESS}$ observed in HMC (dashed bars) and GHMC (filled bars)  with $\Delta {t} \sim \mathcal{U} (\Delta {t_\text{lower}}, \Delta {t_\text{ColSI}})$ for G1000 (left), German (center) and Musk (right) benchmarks using s-AIA3 (green), BCSS3 (blue), ME3 (orange), VV (red) integrators. GHMC combined with s-AIA3 (green filled bars) always outperforms $\text{gESS}^{\bigstar}$ in Table \ref{tab:MetricsAroundHSL} and demonstrates best performance in all three benchmarks. HMC (green dashed bars) also shows the best performance with s-AIA3, which is close to $\text{gESS}^{\bigstar}$ in Table \ref{tab:MetricsAroundHSL} but not necessarily better.} %\textcolor{red}{Figure has to be updated.}}
    \label{fig:ESShBCSS3max_plots}
\end{figure}

Figure \ref{fig:ESShBCSS3max_plots} compares performance (in terms of $\text{gESS}$) of HMC and GHMC using $\Delta {t_\text{CoLSI}}$ and a randomization interval \eqref{eq:t_BCSS3maxtoHSL} with $\text{gESS}^{\bigstar}$ in Table \ref{tab:MetricsAroundHSL} through $\text{gESS}^{\bigstar}$/$\text{gESS}$ relations. 
%The values of minESS are provided in \ref{app:TablesNumericalExperimentsDeltat}. 
%%Table \ref{tab:HMCExperimentsRandomIntervalResults}.

First, we recognize that, as expected, for all benchmarks and for both HMC and GHMC, the adaptive s-AIA3 integrator yields either better or similar sampling performance (in terms of $\text{gESS}$) compared with other tested integrators. Moreover, for GHMC, this performance is visibly higher (up to 2x) than $\text{gESS}$ observed in HMC with the newly proposed settings (see Table \ref{tab:HMCExperimentsRandomIntervalInput}) 
%Table \ref{tab:HMCExperimentsRandomIntervalResults}) 
and with the previously reported setup (Table \ref{tab:MetricsAroundHSL}). We notice that the new settings for HMC with s-AIA3 do not necessarily improve the results in Table \ref{tab:MetricsAroundHSL} for each benchmark,  but they always stay close to them. We recall that the $\text{gESS}^{\bigstar}$ values in Table \ref{tab:MetricsAroundHSL} for each benchmark were achieved with various integrators and at different step sizes by trial and error, whereas our new settings were generated by the procedure universal for all simulated systems. 

Additionally, we perform a sensitivity analysis on the lower endpoint of the randomization interval \eqref{eq:h_BCSS3maxtoHSL} to validate our choice of $h_\text{lower}$. 
We show that variations of $\pm 5\%$ around $h_\text{lower}$ produce similar values for $\text{gESS}$ and the additional performance metrics proposed in Section~\ref{sec:PerformanceMetrics}, and thus confirm the robustness of such a choice, despite its numerical nature. 
The results are summarized in \ref{app:RandomizationSensitivityCheck}.

In summary, the numerical experiments not only justify the proposed optimal choices for a numerical integrator (s-AIA3), a step size ($h_\text{CoLSI}$ / $\Delta {t_\text{CoLSI}}$), and a randomization interval (\eqref{eq:h_BCSS3maxtoHSL} / \eqref{eq:t_BCSS3maxtoHSL}), which we will consider for the rest of this study, but also reveal the performance superiority of GHMC over HMC. 

We remark that, in contrast to the HMC experiments which relied on the recommended choices of $L$, the results for GHMC in Figure \ref{fig:ESShBCSS3max_plots} 
%\ref{tab:HMCExperimentsRandomIntervalResults} 
were obtained without a proper tuning of $L$ and its additional parameter $\varphi$ for the PMU. Obviously, refining these two parameters can further improve performance of GHMC. 

Our next objective is to search for optimal, or close to optimal choices of the remaining GHMC parameters.

\section{Randomization interval for $\varphi$}\label{sec:TuningPhi}
We note that GHMC is not the sole HMC-based method which partially refreshes the momenta. The Modified Hamiltonian Monte Carlo, or MHMC, methods for sampling with modified Hamiltonians also rely on the PMU procedure, though the PMU is Metropolized to secure sampling in a modified ensemble \cite{akhmatskaya_etal_2017}. Both GHMC and MHMC use Hamiltonian dynamics for generating proposals but they are accepted more frequently in MHMC due to the better conservation of modified Hamiltonians by symplectic integrators. On the contrary, partially refreshed momentum is always accepted in GHMC, whereas for MHMC it is accepted according to the modified Metropolis test. Clearly, the MHMC method with very high acceptance rates in the PMU and the GHMC method with very high acceptance rate in the Metropolis test will behave almost indistinguishably.   

Here, we aim to derive an optimal 
%interval $\mathcal{U} (\varphi_{\text{lower}}, \varphi_{\text{upper}})$
$\varphi$ parameter for a GHMC method with its optimal settings, i.e. when using the s-AIA3 integrator with \eqref{eq:h_BCSS3maxtoHSL} or \eqref{eq:t_BCSS3maxtoHSL}. %\textcolor{red}{Thus, in our case, the high acceptance rates in the Metropolis test of GHMC are ensured (see, e.g. Figure \ref{fig:gradESS_random_interval_comparison}).} 
With this in mind, we refer to the procedure proposed in \cite{akhmatskaya_etal_2017} which links $\varphi$ in MHMC with the target acceptance rate in the modified Metropolis test in the PMU (see Eq.~(22) in the aforementioned study). Following this idea, we express $\varphi$ in terms of a step size, parameters of the integrator in use and the target acceptance rate to obtain

\begin{equation}\label{eq:OptimalPhi}
\varphi = - \ln \tilde{\alpha} \frac{1 + 2 {h}^2 \lambda}{2 D {h}^4 \lambda^2}.
\end{equation}
Here, $\tilde{\alpha}$ is the expected acceptance rate for updated momentum, $D$ is the dimension of the system, $h$ is the dimensionless counterpart of the integration step size $\Delta t$ in use, $\lambda = \lambda_k(\boldsymbol{z})$ is the coefficient depending on $k$ -- a number of stages of a numerical integrator -- and $\boldsymbol{z}$ -- the integrator coefficients. In particular \cite{radivojevic_et_al_2018},
\begin{equation}\label{eq:Lambdak}
\lambda_2 (b) = \frac{6 b - 1}{24}, \qquad \lambda_3 (b, a) = \frac{1 - 6 a (1 - a) (1 - 2 b)}{12}.
\end{equation}
We remark that in the notations of this study $\varphi$ corresponds to %${\sin\varphi}^2$ in \cite{akhmatskaya_etal_2017}.
$\sin^2 \phi$ in \cite{akhmatskaya_etal_2017}.
Next, we adapt \eqref{eq:OptimalPhi} to the conditions of this work. More precisely, we are interested in optimizing performance of GHMC with s-AIA3 (i.e. $k = 3$)  in $(h_\text{lower}, h_\text{CoLSI})$. We recall that for s-AIA3, the integrator parameters $a$ and $b$ are step size dependent and so is $\lambda{_3}$.

\begin{figure}[t]
    \centering
    \includegraphics[width=\textwidth]{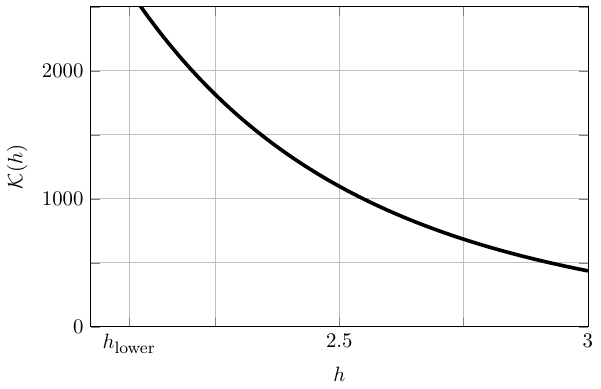}
        \caption{Plot of $\mathcal{K}(h)$ (Eq. \ref{eq:Kh}) for $h$ $\in$ $(h_\text{lower}, h_\text{CoLSI})$ \eqref{eq:h_BCSS3maxtoHSL}.}
    \label{fig:K_vs_h}
\end{figure}

Given that for GHMC the expected acceptance rates for momentum update are equal to 1, in order to mimic the behaviour of $\varphi$ \eqref{eq:OptimalPhi} in GHMC we set $\tilde{\alpha} = 0.999$. Therefore, we obtain
\begin{equation}\label{eq:PhiVsK}
\varphi(h)  = - \ln {0.999} \frac{\mathcal{K}(h)}{D},
\end{equation}
where
\begin{equation}\label{eq:Kh}
\mathcal{K}(h) = \frac{1 + 2 {h}^2 \lambda_3(h)}{2{h}^4 {\lambda_3(h)}^2}.
\end{equation} 
Moreover, from \eqref{eq:PMUIntro}, $\varphi (h) \in (0, 1]$, which yields
\begin{equation}\label{eq:OptimalPhiAsMin}
\varphi_\text{opt} (h) = \min \left\{ 1, - \ln {0.999} \frac{\mathcal{K}(h)}{D} \right\}. 
\end{equation}
We notice that $\mathcal{K}(h)$ is a monotonically decreasing function of $h$ in the interval $(h_\text{lower}, h_\text{CoLSI})$ (Figure \ref{fig:K_vs_h}), and does not depend on a simulated model. Hence, we can choose the bounds of the randomization interval for optimal $\varphi$ as 
\begin{equation}\label{eq:PhiLowerUpper}
\varphi_\text{lower} = \varphi_\text{opt} (h_\text{CoLSI}), \qquad \varphi_\text{upper} = \varphi_\text{opt} (h_\text{lower}).
\end{equation}
Thus, an optimal randomization interval for the random noise $\varphi$ in the PMU \eqref{eq:PMUIntro} reads as
\begin{equation}\label{eq:PhiOptimalIntervalFinal}
\varphi \sim \mathcal{U} \left(\varphi_\text{lower}, \varphi_{\text{upper}} \right).
\end{equation}
Optionally, one can adapt $\varphi_{\text{opt}}$ \eqref{eq:OptimalPhiAsMin} in each Monte Carlo iteration for $h$ within the randomization interval \eqref{eq:h_BCSS3maxtoHSL} and its corresponding parameters of s-AIA3. In this case, no additional randomization of $\varphi$ is required.   

In what follows, we first describe a procedure for an optimal choice of a number of integration steps per Monte Carlo iteration, $L$, and then compare our proposed scheme \eqref{eq:PhiOptimalIntervalFinal} combined with the various choices of $L$ with the commonly recommended $\varphi \sim \mathcal{U} (0, 0.5)$ and two additional randomization intervals, $\varphi \sim \mathcal{U} (0, 0.1)$ (close to Metropolized Molecular Dynamics: $\varphi = 0$) and $\varphi \sim \mathcal{U} (0, 0.9)$ (close to HMC: $\varphi = 1$).

\section{Optimising Hamiltonian trajectory lengths}\label{sec:TuningL}

A PMU \eqref{eq:PMUIntro} allows for maintaining a favourable direction of sampling, provided (i) the acceptance rate for proposed Hamiltonian trajectories is high and (ii) momentum mixing angle $\varphi$ at the PMU is small. Under these conditions, choosing the smallest possible $L$, i.e. $L = 1$ helps to explore in full the area containing a direction once accepted by the Metropolis test, thus mimicking a long but flexible trajectory. So, in contrast to HMC, GHMC may benefit from short Hamiltonian trajectories. 

The optimal settings proposed in this study tend to satisfy both conditions, (i) and (ii). Indeed, the randomization interval for optimal step size $h_\text{CoLSI}$ and the integrator s-AIA3 are chosen to guarantee small expected energy errors (see Figure \ref{fig:Rho3}) and hence, high acceptance rates. %and Figure \ref{fig:gradESS_random_interval_comparison} (right)). 
Furthermore, as follows from \eqref{eq:OptimalPhiAsMin}, $\varphi \sim 1/D$, where $D$ is the dimension of a simulated system. Thus, the reasonable suggestion for an optimal choice of $L$ might be $L_\text{opt} = 1$, provided that the underlying analysis for finding an optimal integrator and a step size is accurate. 

However, as was indicated in \cite{sAIA_paper}, the accuracy of the approach for estimating $h_\text{CoLSI}$ %, first proposed in \cite{sAIA_paper} and followed in this work, 
may suffer when it is applied to systems with clear anharmonic behaviour. Such behaviour can be caught by inspecting a system-specific value of a \emph{fitting factor} $S_f$, computed using the simulation data collected at a burn-in stage of a GHMC simulation as
%(for details, see Eq.... and \cite{sAIA_paper}). 
\begin{equation}\label{eq:SfCases}
S_f =
\begin{cases}
S_\omega, \quad &\text{if the full set of $\omega_i, i = 1, ..., D$ is available}, \\
S, \quad &\text{otherwise},
\end{cases}
\end{equation}
where
\begin{equation}
S_{\omega} = \max \left(1, \frac{1}{\Delta t_{\text{VV}}} \sqrt[6]{\frac{32 \mathbb{E}_{\text{VV}} [\Delta H]}{\sum_{j=1}^D \omega_j^6}} \right), \, \, S = \max \left(1, \frac{1}{\tilde{\omega} \Delta t_{\text{VV}}} \sqrt[6]{\frac{32 \mathbb{E}_{\text{VV}} [\Delta H]}{D}} \right).
\end{equation}
%
%\begin{equation}\label{eq:FittingFactorS}
%S_f = S = \max \left(1, \frac{1}{\tilde{\omega} \Delta t_{\text{VV}}} \sqrt[6]{\frac{32 \mathbb{E}_{\text{VV}} [\Delta H]}{D}} \right),
%\end{equation}
Here, $\omega_i$, $i = 1, ..., D$ are the angular frequencies of the system ($\tilde{\omega}$ is the highest), $\Delta t_{\text{VV}}$ is the step size used during the burn-in stage performed with the VV integrator, and $\mathbb{E}_{\text{VV}} [\Delta H]$ is the expectation of the energy error due to the integration with VV. %, and $D$ is the dimension of the system.
Further details on fitting factors are provided in \cite{sAIA_paper} and in \ref{app:NondimensionalizationAppendix}.

Its clear deviation from 1 implies the deviation from harmonic behaviour. As a result, for such systems, the expected energy error at the 
estimated $h_\text{CoLSI}$ may differ from the one predicted by the analysis of a harmonic oscillator. Moreover, using the expected energy error for 1-stage integrator at $L = 1$ for the analysis of 3-stage integrators in \cite{sAIA_paper} leads to an overestimation of $h_\text{CoLSI}$, as discussed in \ref{app:3stageCorrection}. 
Even though, in the settings proposed in this study, it is not only possible to quantify the error introduced by the suggested analysis but also to compensate for it by choosing parameter $L$ in the appropriate way.  

In particular, let $\mathbb{E}^{\text{$L=1$}}_{\text{$S_{f}=1$}} [\Delta H]$ be the expected energy error for 1-stage Velocity Verlet derived at $L=1$ for systems with harmonic behaviour ($S_f = 1$) and $\mathbb{E}^{\text{$L$}}_{\text{$S_{f}>1$}} [\Delta H]$ be the expected energy error for 1-stage Velocity Verlet at an arbitrary $L \ge 1$ for anharmonic systems ($S_f > 1$). As shown in \cite{sAIA_paper},  

\begin{equation}\label{eq:Rho_VV_1}
\mathbb{E}^{\text{$L=1$}}_{\text{$S_{f}=1$}} [\Delta H] = \frac{h^6}{32},  \quad  h = S_f \tilde{\omega} \Delta t.
\end{equation}
We denote   
\begin{equation}\label{eq:FractionX}
X = \frac{\mathbb{E}^{\text{$L$}}_{\text{$S_{f}>1$}} [\Delta H]}{\mathbb{E}^{\text{$L=1$}}_{\text{$S_{f}=1$}} [\Delta H]}.
\end{equation}
By setting $X = 1$ and solving the equation for $L \ge 1$, one can identify values of $L$ that level the difference between the expected errors in a GHMC simulation for harmonic and anharmonic systems. In \ref{app:LoptCalculation}, 
we follow this idea and obtain such values of $L$ as a function of $S_f$:
%Moreover, taking the suggestion in 
%\cite{random_hmc2017}, we randomize $L > 1$ %as 
%so that $L_\text{opt}$ is the mean of the randomization interval, i.e. 
%\begin{equation}\label{eq:LRandomization}
%L \sim \mathcal{U} \left\{ 2, ..., L_\text{upper} \right\}, \qquad L_\text{upper} = 2 L_\text{opt} - 2,
%\end{equation}
%
%where (see \ref{app:LoptCalculation} 
%for details):
%
\begin{equation}\label{eq:Lopt}
L_\text{opt} =
\begin{cases}
1, \quad &\text{for} \quad 1.0\le S_f < 1.5,\\
2, 5, 7, \quad  &\text{otherwise.}
\end{cases}
\end{equation}
%
%The resulting randomization scheme for $L$ will be the following:
Finally, taking the suggestion in \cite{random_hmc2017}, we randomize $L$ by selecting the optimal values in \eqref{eq:Lopt} with equal probability. The resulting randomization scheme is as follows:
\begin{equation}\label{eq:LoptRandScheme}
L =
\begin{cases}
1, \quad &\text{for} \quad 1.0\le S_f < 1.5,\\
\mathcal{U} \{2, 5, 7\}, \quad  &\text{otherwise.}
\end{cases}
\end{equation}

\section{Adaptive Tuning algorithm}\label{sec:Algorithm}
Here, we summarize the  Adaptive Tuning algorithm (ATune)
%procedure 
that, given a simulation system, generates a set of optimal parameters for a GHMC/HMC simulation to give rise to Adaptively Tuned GHMC/HMC (AT-GHMC/HMC). %(\acs{AT-GHMC}/\acs{AT-HMC}). 

Similarly to the s-AIA algorithm presented in \cite{sAIA_paper}, ATune relies on the simulation data collected during the GHMC/HMC burn-in stage. The latter is run with 1-stage Velocity Verlet using $L = 1$ and a simulation step size $\Delta t_{\text{VV}}$ tuned on the fly, in order to reach a target acceptance rate (AR) for the burn-in (more details in Appendix~B in \cite{sAIA_paper}). Finally, the random noise $\varphi$ for the PMU \eqref{eq:PMUIntro} to run the GHMC burn-in stage is selected according to \eqref{eq:PhiLowerUpper}--\eqref{eq:PhiOptimalIntervalFinal}.

On the completion of the GHMC/HMC burn-in stage, the data needed for calculation of the fitting factor $S_f$ \eqref{eq:SfCases} of the simulated system are computed and stored:
\begin{itemize}
\item AR
\item maximum frequency $\tilde{\omega}$
\item (optional) frequencies, $\omega_i$, $i = 1, ..., D$ and their standard deviation $\sigma$.
\end{itemize}
$S_f$ is critical for the nondimensionalization procedure as well as for finding optimal integration coefficients for s-AIA3 integrators and optimal parameter settings for a GHMC/HMC production simulation, as described in \cite{sAIA_paper} and summarized in \ref{app:NondimensionalizationAppendix}. 
If the full set of frequencies $\omega_i$, $i = 1, ..., D$, $\tilde{\omega}$ and $\sigma$ are available during the burn-in stage (which is optional and requires an extra computational effort due to the calculation of Hessians, see \cite{sAIA_paper}), then $S_f = S_\omega$ (see \eqref{eq:SfCases}). Otherwise, a cheaper option of $S_f$, i.e. $S_f = S$ \eqref{eq:SfCases}, which only needs $\tilde{\omega}$, is chosen. 
Then, the parameters of a s-AIA3 integrator are computed according to the procedure introduced
in \cite{sAIA_paper} with the use of the pretabulated map $h \to b_\text{opt}$ available in \cite{sAIA_tables}. 

%Therefore, t
The GHMC/HMC optimal parameters for the production stage are:
\begin{itemize}
\item \textbf{Step size $\Delta t$} \\
$\Delta t$ is picked uniformly randomly from $(\Delta t_\text{lower}, \Delta t_\text{CoLSI})$, where
\begin{equation}
\Delta t_\text{lower} = \frac{h_\text{lower}}{\text{CF}}, \qquad \Delta t_\text{CoLSI} = \frac{3}{\text{CF}},
\end{equation}
%and CF, 
$h_\text{lower}$ is defined in \eqref{eq:h_BCSS3maxtoHSL} and
\begin{equation}\label{eq:CFcases}
\text{CF} =
\begin{cases}
S_f(\tilde{\omega} - \sigma), & \text{if} \, \sigma > 1, \\
S_f \tilde{\omega}, & \text{otherwise.}
\end{cases}
\end{equation}
See \ref{app:NondimensionalizationAppendix} for details.
%are defined in {\color{mydarkgreen} \cite{AtuneArxiv}, Eqs.(C.3)--(C.4)}, %\eqref{eq:CF}--\eqref{eq:CFnoSigmaCorrection} 
%and \eqref{eq:h_BCSS3maxtoHSL}, respectively.
\item \textbf{PMU random noise $\varphi$} (for GHMC only) \\
As at the burn-in stage, $\varphi$ is randomized following Eqs. \eqref{eq:PhiLowerUpper}--\eqref{eq:PhiOptimalIntervalFinal}.
\item \textbf{Number of integration steps per Monte Carlo iteration $L$} (recommended for GHMC but can be used with HMC) \\
$L$ is chosen following \eqref{eq:LoptRandScheme}. 
\end{itemize}
Finally, the production stage is run using s-AIA3 integrator \cite{sAIA_paper} and the set of optimal parameters for a chosen number of iterations $N_\text{prod}$.

\begin{figure}[!]
\centering
{\resizebox{\textwidth}{!}{\begin{tikzpicture}[node distance=2cm]
\node (Burnin) [rectangle, draw=DarkViolet, align=left, fill=blue!10, rounded corners, font=\usefont{T1}{pag}{m}{n}, minimum width=21cm, minimum height=10.5cm, text width=21cm] {\large
    \begin{varwidth}{\linewidth}
    \centering \Huge{\textbf{INPUT}}\\[-0.5em]
    \Large
    \begin{minipage}[t]{0.49\textwidth}
        	%\Large
            \begin{itemize}
            \item \Large \usefont{T1}{qhv}{m}{n}{\textbf{Number of iterations:}}
            
            \usefont{T1}{pcr}{m}{n}{$N_{\usefont{T1}{cmr}{m}{n}{\text{burn-in}}}$ (burn-in)
            
            $N_{\usefont{T1}{cmr}{m}{n}{\text{prod}}}$ (production)}
            \item \usefont{T1}{qhv}{m}{n}{\textbf{Calculation of frequencies:}}
            
            \usefont{T1}{pcr}{m}{n}{$I_{\omega} = 0 \, \text{(no)} / \, 1 \, \text{(yes)}$}
            \item \usefont{T1}{qhv}{m}{n}{\textbf{Integrator:}}
            
            \usefont{T1}{pcr}{m}{n}{1-stage Velocity Verlet}
            \item \usefont{T1}{qhv}{m}{n}{{\color{orange}\textbf{Trajectory length:}}}
            
            \usefont{T1}{pcr}{m}{n}{$L = 1$}
            \end{itemize}
     \end{minipage}
     \hfill
	 \begin{minipage}[t]{0.49\textwidth}
	 		\Large
            \begin{itemize}
            \item[•] \usefont{T1}{qhv}{m}{n}{\textbf{Dimensional step size:}}
            
            \usefont{T1}{pcr}{m}{n}{$\Delta t = \Delta t_{\usefont{T1}{cmr}{m}{n}{\text{VV}}}$} tuned on the fly
            \item[•] \usefont{T1}{qhv}{m}{n}{\textbf{{\color{blue}{PMU random noise:}}}}
            
            \color{magenta} \usefont{T1}{pcr}{m}{n}{$\varphi = \varphi_i \sim \mathcal{U} (\varphi_{\usefont{T1}{cmr}{m}{n}{\text{lower}}}, \varphi_{\usefont{T1}{cmr}{m}{n}{\text{upper}}})$}
            
            \color{black} $i = 1, ..., N_{\usefont{T1}{cmr}{m}{n}{\text{burn-in}}}$
            
            $\varphi_{\usefont{T1}{cmr}{m}{n}{\text{lower}}}, \varphi_{\usefont{T1}{cmr}{m}{n}{\text{upper}}}$ \usefont{T1}{pcr}{m}{n}{in \eqref{eq:PhiLowerUpper}}
            \item[•] \usefont{T1}{qhv}{m}{n}{\textbf{Integrator parameters:}}
            
            %\usefont{T1}{pcr}{m}{n}{Map} $h \to b_\text{opt}$ \usefont{T1}{cmr}{m}{n}{{\color{mydarkgreen} (\cite{AtuneArxiv}, App. C,} \cite{sAIA_tables})}
            \usefont{T1}{pcr}{m}{n}{Map $h \to b_\text{opt}$} \usefont{T1}{cmr}{m}{n}{\cite{sAIA_tables}}
            \end{itemize}
    \end{minipage}
    \end{varwidth}
};
\node (BurninText) [text width=2cm, align=left, left of=Burnin, xshift = -11.5cm, yshift = 4cm, font=\Huge\bfseries\usefont{T1}{ugq}{m}{n}, text width=10cm] {\fontsize{35}{40}\selectfont \textbf{GHMC/HMC} \\ \textbf{(Burn-in)}};
\node (ComputeBurnin) [rectangle, draw=DarkViolet, rounded corners, fill=blue!10, below of=Burnin, yshift=-7.275cm, xshift=-7cm, font=\usefont{T1}{pag}{m}{n}, minimum height=6cm, minimum width=13cm, text height = 0.975cm] {\Large
    \begin{varwidth}{\linewidth}
    \centering
    \Huge{\textbf{OUTPUT}}
    \Large
    \begin{itemize}
        \item[•] \usefont{T1}{qhv}{m}{n}{\textbf{Acceptance rates:}} \usefont{T1}{cmr}{m}{n}{AR}
        \item[•] \usefont{T1}{pcr}{m}{n}{\textbf{if} $I_\omega = 1  \implies \omega_i, \tilde{\omega}, \sigma$
        
        \textbf{else} $\tilde{\omega}$, set $\sigma = 0$}
    \end{itemize}
    \end{varwidth}
};
\node (AnalyseFittingFactor) [rectangle, draw=DarkViolet, rounded corners, fill=blue!10, below of=Burnin, yshift=-8.9cm, xshift=7cm, font=\usefont{T1}{pcr}{m}{n}, minimum height=9.25cm, minimum width=13cm, align = left] {\large
    \begin{varwidth}{\linewidth}
    \centering
    \usefont{T1}{pag}{m}{n}{\Huge{\textbf{ANALYSIS: s-AIA3}}} 
    \Large
    \begin{itemize}
        \item \usefont{T1}{qhv}{m}{n}{\textbf{Nondimensionalization:}}
        \begin{itemize}
        	\item \usefont{T1}{pcr}{m}{n}{\textbf{if}} $I_\omega = 0 \implies S_f = S$ {\usefont{T1}{cmr}{m}{n}{\eqref{eq:FittingFactorsBothAsMaxAppendix}}} %\eqref{eq:FittingFactorsBothAsMaxAppendix}
        
        \textbf{else} $S_f = S_{\omega}$ {\usefont{T1}{cmr}{m}{n}{\eqref{eq:FittingFactorsBothAsMaxAppendix}}} %\eqref{eq:FittingFactorsBothAsMaxAppendix}
        %\item \usefont{T1}{pcr}{m}{n}{\textbf{if}} $\sigma < 1 \implies \usefont{T1}{cmr}{m}{n}{\text{CF}}$ {\color{mydarkgreen} \usefont{T1}{cmr}{m}{n}{\cite{AtuneArxiv}, Eq.~(C.4)}} %\eqref{eq:CFnoSigmaCorrection}
        \item \usefont{T1}{pcr}{m}{n}{\textbf{if}} $\sigma < 1 \implies \usefont{T1}{cmr}{m}{n}{\text{CF}}$ {\usefont{T1}{cmr}{m}{n}{\eqref{eq:CFcases}}} %\eqref{eq:CFnoSigmaCorrection}
        
        %\textbf{else} $\usefont{T1}{cmr}{m}{n}{\text{CF}}$ {\color{mydarkgreen} \usefont{T1}{cmr}{m}{n}{\cite{AtuneArxiv}, Eq.~(C.3)}} %\eqref{eq:CF} %= S_f \left( \tilde{\omega} - \sigma \right)$
        \textbf{else} $\usefont{T1}{cmr}{m}{n}{\text{CF}}$ {\usefont{T1}{cmr}{m}{n}{\eqref{eq:CFcases}}} %\eqref{eq:CF} %= S_f \left( \tilde{\omega} - \sigma \right)$
        \end{itemize}
        \Huge
        \item \usefont{T1}{qhv}{m}{n}{\Large \textbf{Integrator parameters:}} 
        \Large
        \begin{itemize}
        	\item $a_{\usefont{T1}{cmr}{m}{n}{\text{opt}}}, b_{\usefont{T1}{cmr}{m}{n}{\text{opt}}}$ {\usefont{T1}{cmr}{m}{n}{\cite{sAIA_tables}}} 
        \end{itemize}
    \end{itemize}
    \end{varwidth}
};
\node (Phi) [rectangle, draw=DarkViolet, rounded corners, align=center, fill=blue!10, below of=Burnin,  yshift = -18cm, font=\usefont{T1}{pag}{m}{n}, minimum height=6.75cm, minimum width = 11cm, text width=10cm, text height = -0.4cm] {\Large
	\centering
	\Huge{\textbf{{\color{blue}{PMU RANDOM NOISE $\varphi$}}}}
    \Large
	\begin{itemize}
	\centering
			\item[•] {\color{magenta} $\varphi_i \sim \mathcal{U} (\varphi_{\usefont{T1}{cmr}{m}{n}{\text{lower}}}, \varphi_{\usefont{T1}{cmr}{m}{n}{\text{upper}}})$} \eqref{eq:PhiOptimalIntervalFinal}
			
			$i = 1, ..., N_{\usefont{T1}{cmr}{m}{n}{\text{prod}}}$
	\end{itemize}
	
	$\varphi_{\usefont{T1}{cmr}{m}{n}{\text{lower}}}, \varphi_{\usefont{T1}{cmr}{m}{n}{\text{upper}}}$ \usefont{T1}{pcr}{m}{n}{in \eqref{eq:PhiLowerUpper}}
};

\node (TimeStep) [rectangle, draw=DarkViolet, rounded corners, align=center, fill=blue!10, left of=Phi, xshift = -10cm, font=\usefont{T1}{pag}{m}{n}, minimum height=6.75cm, minimum width=12cm, text width = 11cm, text height = 0.5cm] {\Large
    \centering
    \Huge{\textbf{STEP SIZE $\Delta t$}}
    \Large
    \begin{itemize}
    \centering
    		\item[•] {\color{magenta} $\Delta t_i \sim \mathcal{U} \left( \Delta t_{\usefont{T1}{cmr}{m}{n}{\text{lower}}}, \Delta t_{\usefont{T1}{cmr}{m}{n}{\text{upper}}} \right)$} \eqref{eq:t_BCSS3maxtoHSL}
    		
    		$i = 1, ..., N_{\usefont{T1}{cmr}{m}{n}{\text{prod}}}$ 
    \end{itemize}
    
    $\Delta t_{\usefont{T1}{cmr}{m}{n}{\text{lower}}} = \frac{h_{\usefont{T1}{cmr}{m}{n}{\text{lower}}}}{\usefont{T1}{cmr}{m}{n}{\text{CF}}}$ \eqref{eq:h_BCSS3maxtoHSL}
    
    $\Delta t_{\usefont{T1}{cmr}{m}{n}{\text{upper}}}\!=\!\frac{3}{\usefont{T1}{cmr}{m}{n}{\text{CF}}}$
};
\node (L) [rectangle, draw=DarkViolet, rounded corners, align=center, fill=blue!10, right of=Phi,  xshift = 9.65cm, font=\usefont{T1}{pag}{m}{n}, minimum height=6.75cm, minimum width=11cm, text width = 11cm, text height = -0.5cm] {\Large
	\centering
	\Huge{\textbf{{\color{orange}{NUMBER OF INTEGRATION STEPS PER ITERATION $L$}}}}
	\begin{itemize}
    \Large
	\centering
			\item[•] {\color{magenta} $L_i \sim $ \usefont{T1}{cmr}{m}{n}{\eqref{eq:LoptRandScheme}}}
			
			$i = 1, ..., N_{\usefont{T1}{cmr}{m}{n}{\text{prod}}}$
	\end{itemize}
	%\Large
	%$L_{\usefont{T1}{cmr}{m}{n}{\text{opt}}}$ \usefont{T1}{pcr}{m}{n}{in \eqref{eq:Lopt}}%$
};
\node (ProductionSettingsText) [text width=2cm, align=left, above of=TimeStep, xshift = -5.25cm, yshift = 2.9cm, font=\Huge\bfseries\usefont{T1}{ugq}{m}{n}] {\fontsize{35}{40}\selectfont 
	\begin{varwidth}{\linewidth}
	\textbf{Production} \\ \textbf{settings}
    \end{varwidth}
};
\node (Production) [rectangle, draw = mydarkgreen, rounded corners, fill=green!10, below of=Phi, yshift=-5.75cm, font=\usefont{T1}{pag}{m}{n}, minimum width=15cm, text width=16cm, minimum height=6.5cm] {\large
    \begin{varwidth}{\linewidth}
    \centering \Huge{\textbf{INPUT}}
    
    \begin{minipage}[t]{0.49\textwidth}
    \centering
        	\Large
            \begin{itemize}
            \item[•] \usefont{T1}{qhv}{m}{n}{\textbf{Integrator:}}
            
            \usefont{T1}{pcr}{m}{n}{\color{magenta}{s-AIA3}}
            \item[•] \usefont{T1}{qhv}{m}{n}{\textbf{Dimensional step size:}}
            
            \usefont{T1}{pcr}{m}{n}{$\Delta t = {\color{magenta} \Delta t_i}$, $i = 1, ..., N_{\usefont{T1}{cmr}{m}{n}{\text{prod}}}$}
            \end{itemize}
     \end{minipage}
     \hfill
	 \begin{minipage}[t]{0.49\textwidth}
	 \centering
	 		\Large
            \begin{itemize}
            \item[•] \usefont{T1}{qhv}{m}{n}{\textbf{{\color{orange}{Trajectory length:}}}}
            
            \usefont{T1}{pcr}{m}{n}{$L = {\color{magenta} L_i}$, $i = 1, ..., N_{\usefont{T1}{cmr}{m}{n}{\text{prod}}}$}
            \item[•] \usefont{T1}{qhv}{m}{n}{\textbf{{\color{blue}{PMU random noise:}}}}
            
            \usefont{T1}{pcr}{m}{n}{$\varphi = {\color{magenta} \varphi_i}$, $i = 1, ..., N_{\usefont{T1}{cmr}{m}{n}{\text{prod}}}$}
            \end{itemize}
    \end{minipage}
    \end{varwidth}
};
\node (ProductionText) [text width=2cm, align=left, left of=Production, xshift = -15.25cm, yshift = 2cm, font=\Huge\bfseries\usefont{T1}{ugq}{m}{n}] {\fontsize{35}{40}\selectfont \textbf{GHMC/HMC} \\ \textbf{(Production)}};
\node (Averages) [rectangle, draw = mydarkgreen, rounded corners, fill=green!10, align = center, below of=Production, yshift=-4.25cm, font=\usefont{T1}{pag}{m}{n}, minimum height=4cm, minimum width=8cm] {\Large
    \begin{varwidth}{\linewidth}
    \centering \Huge{\textbf{OUTPUT}}\\
    \Large    
    \begin{itemize}
        \item[•] \usefont{T1}{qhv}{m}{n}{\textbf{GHMC/HMC trajectories}}
    \end{itemize}
    \end{varwidth}
};
\draw [arrow, *|, DarkViolet, line width = 2pt] (Burnin.south) to (ComputeBurnin.north);
\draw [arrow, DarkViolet, line width = 2pt] (ComputeBurnin.east) -- ([yshift=1.62cm]AnalyseFittingFactor.west);
\draw [arrow, DarkViolet, line width = 2pt] (AnalyseFittingFactor) -- ([xshift=3cm]TimeStep.north);
\draw [arrow, *|, DarkViolet, line width = 2pt] (AnalyseFittingFactor.south) to ([xshift=3cm]Phi.north);
\draw [arrow, *|, DarkViolet, line width = 2pt] (AnalyseFittingFactor.south) to (L);
\draw [arrow, DarkViolet, line width = 2pt, anchor = south] (TimeStep) -- (Production);
\draw [arrow, DarkViolet, line width = 2pt] (Phi) -- (Production);
\draw [arrow, DarkViolet, line width = 2pt] (L) -- (Production);
\draw [arrow, mydarkgreen, line width = 2pt] (Production) -- (Averages);
\end{tikzpicture}}}
\caption{Schematic representation of the ATune algorithm for generating a set of optimal parameters for a GHMC/HMC simulation. The features specific to \acs{AT-GHMC} only are highlighted in blue, whereas the features optional for \acs{AT-HMC} are displayed in orange. Optimal parameters are shown in magenta.}\label{fig:GHMCoptsettingScheme}
\end{figure}
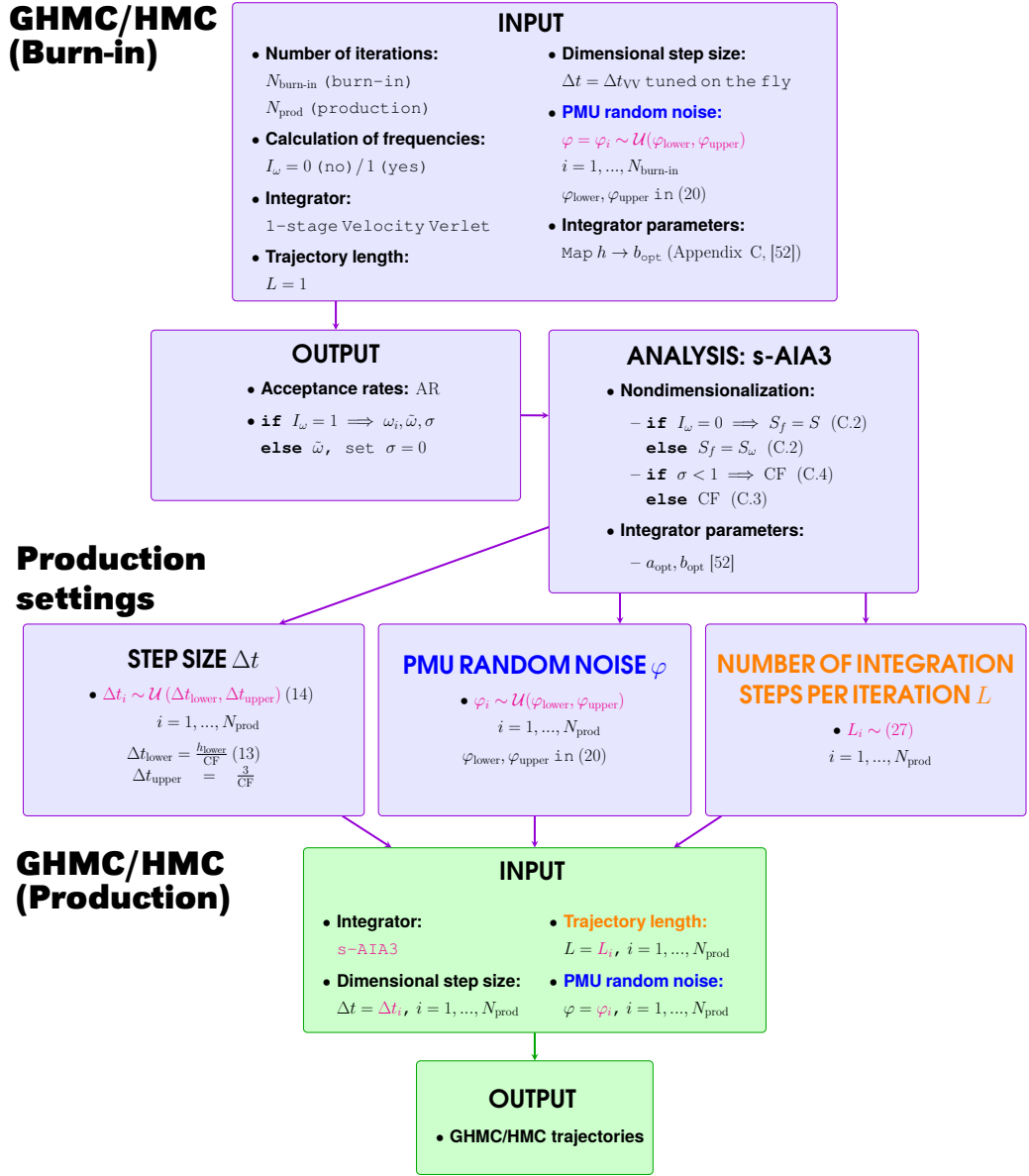

The ATune algorithm is outlined in Figure \ref{fig:GHMCoptsettingScheme}. The algorithm is implemented (with no computational overheads introduced in a production simulation) in the BCAM in-house software package \textsf{HaiCS} (Hamiltonians in Computational Statistics), designed for statistical sampling of high dimensional and complex distributions and parameter estimation in Bayesian  models using MCMC and HMC-based methods. 
The package provides analytical gradients for popular statistical models, including all models but one considered in this study. In addition, finite difference approximation of gradients is supported in \textsf{HaiCS}.

A comprehensive overview of the package can be found in \cite{tijana_thesis}, whereas applications of \textsf{HaiCS} software are presented in \cite{radivojevic_akhmatskaya_MHMC_2020, radivojevic_et_al_2018, inouzhe2023, sAIA_paper}.

%\newpage

\section{Numerical experiments}\label{sec:NumericalExperiments}
%\label{sec:OptimalPhiNumericalExperiments}
The next objective is to numerically justify the efficiency of our new ATune algorithm using a set of standard benchmarks summarized in Table \ref{tab:Benchmarks}. 
%both the optimal randomization interval for the PMU random noise $\varphi$ presented in Section \ref{sec:TuningPhi} and the randomization scheme for $L$ discussed in Section \ref{sec:TuningL}. 
%The optimal settings for the tested benchmarks (Table~\ref{tab:Benchmarks}) are summarized in Table~\ref{tab:OptimalLPhiInput}. 

\begin{table}[t]
\centering
\resizebox{\textwidth}{!}{\begin{tabular}{l r r r r}
Benchmark & Model & Data & $D$ (dimension) & $K$ (observations) \\ 
  \hline
G1000 & \multirow{3}{*}{$\mathcal{N}(0, \Sigma)$} & \multirow{3}{*}{$\Sigma^{-1} \sim \mathcal{W}(\mathbb{I}_D, D)$ \cite{hoffman_gelman2014}} & 1000 & - \\
G500 &  &  & 500 & - \\
G2000 &  &  & 2000 & - \\
\hline
German & \multirow{2}{*}{Bayesian Logistic Regression \cite{liu2001}} & \multirow{2}{*}{\cite{lichman2013uci}} & 25 & 1000 \\
Musk &  &  & 167 & 476 \\
\hline
Banana \cite{radivojevic_akhmatskaya_MHMC_2020} & $\mathcal{N}(0, \sigma_\theta^2)$ & $\sigma_\theta^2 = 1$, $\boldsymbol{y} \sim \mathcal{N}(1, 2)$ & 2 & 100 \\
  \hline
\end{tabular}}
\caption{\label{tab:Benchmarks} List of benchmarks used in this study. $\mathcal{W}$ is the Wishart distribution (more details in \cite{hoffman_gelman2014}), $\mathbb{I}_D$ is the identity matrix in $\mathbb{R}^{D \times D}$, $\boldsymbol{y} \in \mathbb{R}^K$ is the set of observations.}
\end{table}

\subsection{Performance metrics}\label{sec:PerformanceMetrics}
The irreversibility of GHMC may allow it to reach convergence faster, compared to reversible samplers, like HMC. To check this, we choose the performance metric that includes information about the speed of convergence of the Markov chains. The metric is briefly introduced in Section \ref{sec:TopPerformanceAroundHSL}. Here we provide more details on the metric and extend it to the alternative approaches for computing effective sample size (ESS).    

Following the recommendation in \cite{vethari_etal_2021}, we consider a chain to be converged if $\max \text{PSRF} < 1.01$, where PSRF is the Potential Scale Reduction Factor \cite{gelman_rubin1992}, calculated as described in \cite{brooks_gelman1998}.
%
%. The Potential Scale Reduction Factor (PSRF) \cite{gelman_rubin1992} can be calculated as described in \cite{brooks_gelman1998}. 

Starting from that, let us define the number of samples in the production chains, required for reaching $\max \text{PSRF} < 1.01$, as $N_{1.01}$,
%In this way, the smaller $N_{1.01}$ corresponds to faster convergence.
%Afterwards, we 
and evaluate the ratio grad/ESS between the number of gradient computations and ESS  for production chains of the length of $N_{1.01} + 1000$ iterations. Such a metric combines both the information about convergence and sampling efficiency.

The number of gradient evaluations reads as
\begin{equation}
\text{grad} = (N_{1.01} + 1000) \, \overline{L} \, k,
\end{equation}
where $\overline{L}$ is the mean of the number of integration steps per Monte Carlo iteration (it varies depending on a chosen randomization scheme for $L$) and $k$ is the number of stages of an integrator in use (in our case, $k = 3$). We recall that the number of stages $k$ indicates the number of times the algorithm performs an evaluation of gradients per step size. Therefore, $\overline{L} k$ represents the theoretical average number of gradient evaluations per Monte Carlo iteration. 

\begin{sloppypar}
%Regarding the evaluation of the ESS, so far 
In Section \ref{sec:TopPerformanceAroundHSL}, we calculate ESS using the \texttt{effectiveSize} function of the \texttt{coda} package of \texttt{R} \cite{coda} and consider the minimum ESS across variates, i.e. minESS. We want to include two additional ESS metrics in our list, which now  appears like: 
\begin{itemize}
\item minESS (\texttt{coda}) %-- the minimum ESS across variates
\item meanESS (\texttt{coda}) -- the average ESS over the sampled variables calculated through the \texttt{effectiveSize} function of the \texttt{coda} package of \texttt{R} \cite{coda}
\item multiESS -- the metric calculated in the \texttt{multiESS} function of the \texttt{mcmcse} package \cite{mcsmcse} using the approach proposed in \cite{vats_multiess}.
\end{itemize}
To summarize, we propose three metrics for assessing convergence and sampling performance, such as grad/minESS, grad/meanESS, and grad/multiESS. For each metric, a lower gradient value (grad) indicates better convergence, while a higher ESS suggests improved sampling efficiency. Consequently, the smallest grad/ESS ratio represents the best overall performance.
\end{sloppypar}

To enable comparison of efficiency between two samplers, a Relative Efficiency Factor (REF) is introduced as:
\begin{gather}
\text{REF} (\text{Sampler}_1, \text{Sampler}_2) = \frac{\mathcal{M}_2}{\mathcal{M}_1}, \label{eq:REF}\\ 
\mathcal{M} = \frac{\text{grad}}{\text{ESS}}, \qquad \text{ESS} \in \{\text{minESS}, \text{meanESS}, \text{multiESS}\}. \nonumber 
\end{gather}
The REF value quantifies the extent to which $\text{Sampler}_1$ outperforms $\text{Sampler}_2$ for a particular $\mathcal{M} = \text{grad}/\text{ESS}$ metric, where ESS is picked from the set \{minESS, meanESS, multiESS\}.

\subsection{Benchmark tests}\label{sec:BenchmarkTests}
We use the metrics described in \ref{sec:PerformanceMetrics} for the full set of simulation benchmarks (Table \ref{tab:Benchmarks}). For the biggest benchmark (G2000), we consider the ESS metrics evaluated over the first $N_{1.01} + 2000$ iterations, in order to take into account enough iterations ($>D = 2000$) for the multiESS calculation.
%\textcolor{red}{(EA: I don't understand the paragraph below. Did you mean maxPSRF was always bigger than 1.01 (you said minPSRF twice)? I corrected it according to my understanding but please check!)}
We also remark that for Gaussian benchmarks the grad/ESS metrics for the simulations with $L = 1$ were calculated over the first $\tilde{N}_{1.01} + 1000$ ($2000$ for G2000), where $\tilde{N}_{1.01}$ is the number of iterations required to achieve $\text{avgPSRF} < 1.01$. The relaxation of the PSRF threshold helped to avoid extremely long simulations ($N_\text{prod} \gg 10^6$) needed for reaching convergence in terms of $\max \text{PSRF} < 1.01$.
Numerical experiments for all benchmarks considered here, as well as the case studies in Section~\ref{sec:Applications}, were performed using 10 chains, except for Banana, which was run with 100 chains.
%We considered a relaxation of the PSRF threshold since we observed that even for very long production simulations ($N_\text{prod} = 10^6$) maxPSRF was always bigger than 1.01 (and also bigger than 1.1 -- the wider threshold proposed in \cite{gelman_rubin1992}).%minPSRF was always bigger than 1.01 (and, for all, bigger than the wider $\text{minPSRF} < 1.1$ threshold proposed in \cite{gelman_rubin1992}).

\begin{table}[t]
\centering
\begin{tabular}{l r r}
\hline
Benchmark & $L$ & $\varphi$ \\
\hline
G1000 & 1 & $\mathcal{U}(0.00044, 0.00264)$ \\
G500 & 1 & $\mathcal{U}(0.00088, 0.00527)$ \\
G2000 & 1 & $\mathcal{U}(0.00022, 0.00132)$ \\
German & 1 & $\mathcal{U}(0.01752, 0.10545)$ \\
%Musk & $\mathcal{U} \{2, 6 \}$ & $\mathcal{U}(0.00262, 0.01579)$ \\
Musk & $\mathcal{U} \{2, 5, 7\}$ & $\mathcal{U}(0.00262, 0.01579)$ \\
%Banana & $\mathcal{U} \{2, 6 \}$ & $\mathcal{U}(0.21904, 1)$ \\
Banana & $\mathcal{U} \{2, 5, 7\}$ & $\mathcal{U}(0.21904, 1.00000)$ \\
\hline
\end{tabular}
\caption{\label{tab:OptimalLPhiInput} Optimal randomization intervals for $L$ \eqref{eq:LoptRandScheme} and $\varphi$ \eqref{eq:PhiOptimalIntervalFinal} for each tested benchmark -- G1000, G500, G2000, German, Musk, Banana.}
\end{table}

\begin{table}[!]
\centering
%\resizebox{\textwidth}{!}{\begin{tabular}{c c c c c c c c c c}
\resizebox{\textwidth}{!}{\begin{tabular}{l r l r r r r r r r}
Benchmark & $\varphi$ & ESS & $L = 1$ & $L_\text{opt}$ & $\mathcal{U}\{1, \frac{D}{3}\}$ & $\mathcal{U}\{1, \frac{2}{3}D\}$ & $\mathcal{U}\{1, \frac{4}{3}D\}$ & $\mathcal{U}\{1, \frac{8}{3}D\}$ \\
\hline
\multirow{6}{*}{G1000} & \multirow{3}{*}{$\mathcal{U} (\varphi_\text{lower}, \varphi_\text{upper})$} & minESS & \multicolumn{2}{r}{\cellcolor{cellbox_bestsampler} 39373.1 $\quad$} & 2387.2 & \color{red} \textbf{1491.1} & 2509.4 & 5021.0 \\
% & & meanESS & \multicolumn{2}{c}{\cellcolor{cellbox_bestsampler} \color{red} \textbf{0.4596}} & 532.1 & 1071 & 2142 & 4285 \\
 & & meanESS & \multicolumn{2}{r}{\cellcolor{cellbox_bestsampler} \color{red} \textbf{0.5} $\quad$} & 532.1 & 1071.4 & 2142.4 & 4285.3 \\
% & & multiESS & \multicolumn{2}{c}{\cellcolor{cellbox_bestsampler} \color{red} \textbf{0.3787}} & 299.8 & 402.5 & 895.4 & 3069 \\\cline{2-9}
 & & multiESS & \multicolumn{2}{r}{\cellcolor{cellbox_bestsampler} \color{red} \textbf{0.4} $\quad$} & 299.8 & 402.5 & 895.4 & 3069.2 \\\cline{2-9}
% & \multirow{3}{*}{1 (HMC)} & minESS & \multicolumn{2}{c}{\cellcolor{cellbox_worstsampler} 132650} & 46712 & 21620 & 11740 & \textbf{5554} \\
 & \multirow{3}{*}{1 (HMC)} & minESS & \multicolumn{2}{r}{\cellcolor{cellbox_worstsampler} 132650.2 $\quad$} & 46712.1 & 21620.1 & 11740.0 & \textbf{5554.2} \\
% & & meanESS & \multicolumn{2}{c}{\cellcolor{cellbox_worstsampler} \textbf{2.530}} & 533.6 & 1054 & 2093 & 4198 \\
 & & meanESS & \multicolumn{2}{r}{\cellcolor{cellbox_worstsampler} \textbf{2.5} $\quad$} & 533.6 & 1054.3 & 2093.2 & 4198.1 \\
% & & multiESS & \multicolumn{2}{c}{\cellcolor{cellbox_worstsampler} \textbf{3.712}} & 301.3 & 1001 & 2000 & 4001 \\
 & & multiESS & \multicolumn{2}{r}{\cellcolor{cellbox_worstsampler} \textbf{3.7} $\quad$} & 301.3 & 1001.0 & 2000.0 & 4001.0 \\
\hline
\multirow{6}{*}{G500} & \multirow{3}{*}{$\mathcal{U} (\varphi_\text{lower}, \varphi_\text{upper})$} & minESS & \multicolumn{2}{r}{\cellcolor{cellbox_bestsampler} 31792.2 $\quad$} & 1553.4 & \color{red} \textbf{919.8} & 1090.2 & 2339.1 \\
% & & meanESS & \multicolumn{2}{c}{\cellcolor{cellbox_bestsampler} \color{red} \textbf{0.3429}} & 260.0 & 520.7 & 1043 & 2093 \\
 & & meanESS & \multicolumn{2}{r}{\cellcolor{cellbox_bestsampler} \color{red} \textbf{0.3} $\quad$} & 260.0 & 520.7 & 1043.2 & 2093.1 \\
% & & multiESS & \multicolumn{2}{c}{\cellcolor{cellbox_bestsampler} \color{red} \textbf{0.2553}} & 170.6 & 320.8 & 384.0 & 1422 \\\cline{2-9}
 & & multiESS & \multicolumn{2}{r}{\cellcolor{cellbox_bestsampler} \color{red} \textbf{0.3} $\quad$} & 170.6 & 320.8 & 384.0 & 1422.3 \\\cline{2-9}
% & \multirow{3}{*}{1 (HMC)} & minESS & \multicolumn{2}{c}{\cellcolor{cellbox_worstsampler} 116471} & 42111 & 21821 & 10094 & \textbf{4790} \\
 & \multirow{3}{*}{1 (HMC)} & minESS & \multicolumn{2}{r}{\cellcolor{cellbox_worstsampler} 116471.2 $\quad$} & 42111.1 & 21821.2 & 10094.4 & \textbf{4790.2} \\
% & & meanESS & \multicolumn{2}{c}{\cellcolor{cellbox_worstsampler} \textbf{2.363}} & 261.3 & 515.5 & 1035 & 2046 \\
 & & meanESS & \multicolumn{2}{r}{\cellcolor{cellbox_worstsampler} \textbf{2.4} $\quad$} & 261.3 & 515.5 & 1035.3 & 2046.0 \\
% & & multiESS & \multicolumn{2}{c}{\cellcolor{cellbox_worstsampler} \textbf{3.666}} & 107.8 & 321.6 & 563.3 & 1325 \\
 & & multiESS & \multicolumn{2}{r}{\cellcolor{cellbox_worstsampler} \textbf{3.7} $\quad$} & 107.8 & 321.6 & 563.3 & 1325.2 \\
\hline
\multirow{6}{*}{G2000} & \multirow{3}{*}{$\mathcal{U} (\varphi_\text{lower}, \varphi_\text{upper})$} & minESS & \multicolumn{2}{r}{\cellcolor{cellbox_bestsampler} 63451.2 $\quad$} & 3305.0 & \color{red} \textbf{2515.3} & 5036.4 & 9195.1 \\
% & & meanESS & \multicolumn{2}{c}{\cellcolor{cellbox_bestsampler} \color{red} \textbf{0.6518}} & 1104 & 2207 & 4417 & 8853 \\
 & & meanESS & \multicolumn{2}{r}{\cellcolor{cellbox_bestsampler} \color{red} \textbf{0.7} $\quad$} & 1104.1 & 2207.4 & 4417.2 & 8853.4 \\
% & & multiESS & \multicolumn{2}{c}{\cellcolor{cellbox_bestsampler} \color{red} \textbf{0.5029}} & 428.5 & 860.1 & 1723 & 7096 \\\cline{2-9}
 & & multiESS & \multicolumn{2}{r}{\cellcolor{cellbox_bestsampler} \color{red} \textbf{0.5} $\quad$} & 428.5 & 860.1 & 1723.0 & 7096.3 \\\cline{2-9}
% & \multirow{3}{*}{1 (HMC)} & minESS & \multicolumn{2}{c}{\cellcolor{cellbox_worstsampler} 174435} & 47859 & 22770 & 10870 & \textbf{9582} \\
 & \multirow{3}{*}{1 (HMC)} & minESS & \multicolumn{2}{r}{\cellcolor{cellbox_worstsampler} 174435.1 $\quad$} & 47859.0 & 22770.3 & 10870.1 & \textbf{9582.1} \\
% & & meanESS & \multicolumn{2}{c}{\cellcolor{cellbox_worstsampler} \textbf{2.791}} & 1099 & 2178 & 4348 & 8671 \\
 & & meanESS & \multicolumn{2}{r}{\cellcolor{cellbox_worstsampler} \textbf{2.8} $\quad$} & 1099.1 & 2178.4 & 4348.1 & 8671.2 \\
% & & multiESS & \multicolumn{2}{c}{\cellcolor{cellbox_worstsampler} \textbf{3.739}} & 677.3 & 2000 & 4001 & 8000 \\
 & & multiESS & \multicolumn{2}{r}{\cellcolor{cellbox_worstsampler} \textbf{3.7} $\quad$} & 677.3 & 2000.0 & 4001.0 & 8000.0 \\
\hline
%\multirow{6}{*}{German} & \multirow{3}{*}{$\mathcal{U} (\varphi_\text{lower}, \varphi_\text{upper})$} & minESS & \multicolumn{2}{c}{\cellcolor{cellbox_bestsampler} \color{red} \textbf{1.734}} & 13.94 & 25.93 & 53.38 & 113.6 \\
\multirow{6}{*}{German} & \multirow{3}{*}{$\mathcal{U} (\varphi_\text{lower}, \varphi_\text{upper})$} & minESS & \multicolumn{2}{r}{\cellcolor{cellbox_bestsampler} \color{red} \textbf{1.7} $\quad$} & 13.9 & 25.9 & 53.4 & 113.6 \\
% & & meanESS & \multicolumn{2}{c}{\cellcolor{cellbox_bestsampler} \color{red} \textbf{0.3463}} & 10.15 & 23.84 & 49.05 & 101.0 \\
 & & meanESS & \multicolumn{2}{r}{\cellcolor{cellbox_bestsampler} \color{red} \textbf{0.3} $\quad$} & 10.2 & 23.8 & 49.1 & 101.0 \\
% & & multiESS & \multicolumn{2}{c}{\cellcolor{cellbox_bestsampler} \color{red} \textbf{0.1218}} & 10.62 & 25.10 & 49.47 & 100.5 \\\cline{2-9}
 & & multiESS & \multicolumn{2}{r}{\cellcolor{cellbox_bestsampler} \color{red} \textbf{0.1} $\quad$} & 10.6 & 25.1 & 49.5 & 100.5 \\\cline{2-9}
% & \multirow{3}{*}{1 (HMC)} & minESS & \multicolumn{2}{c}{\cellcolor{cellbox_worstsampler} 24.92} & \textbf{13.25} & 26.18 & 53.09 & 112.6 \\ 
 & \multirow{3}{*}{1 (HMC)} & minESS & \multicolumn{2}{r}{\cellcolor{cellbox_worstsampler} 24.9 $\quad$} & \textbf{13.3} & 26.2 & 53.1 & 112.6 \\ 
% & & meanESS & \multicolumn{2}{c}{\cellcolor{cellbox_worstsampler} \textbf{5.860}} & 10.56 & 24.45 & 49.23 & 102.9 \\ 
 & & meanESS & \multicolumn{2}{r}{\cellcolor{cellbox_worstsampler} \textbf{5.9} $\quad$} & 10.6 & 24.5 & 49.2 & 102.9 \\ 
% & & multiESS & \multicolumn{2}{c}{\cellcolor{cellbox_worstsampler} \textbf{2.334}} & 10.88 & 25.44 & 49.44 & 100.8 \\
 & & multiESS & \multicolumn{2}{r}{\cellcolor{cellbox_worstsampler} \textbf{2.3} $\quad$} & 10.9 & 25.4 & 49.4 & 100.8 \\
\hline
\multirow{6}{*}{Musk} & \multirow{3}{*}{$\mathcal{U} (\varphi_\text{lower}, \varphi_\text{upper})$} & minESS & 216.8 & \cellcolor{cellbox_bestsampler} \color{red} \textbf{143.7} & 229.4 & 389.7 & 601.3 & 1458.3 \\
% & & meanESS & 145.4 & \cellcolor{cellbox_bestsampler} \color{red} \textbf{52.08} & 68.44 & 165.0 & 464.5 & 1031 \\
 & & meanESS & 145.4 & \cellcolor{cellbox_bestsampler} \color{red} \textbf{52.1} & 68.4 & 165.0 & 464.5 & 1031.2 \\
% & & multiESS & 19.79 & \cellcolor{cellbox_bestsampler} \color{red} \textbf{12.79} & 36.00 & 160.4 & 342.0 & 774.5 \\\cline{2-9}
 & & multiESS & 19.8 & \cellcolor{cellbox_bestsampler} \color{red} \textbf{12.8} & 36.0 & 160.4 & 342.0 & 774.5 \\\cline{2-9}
 & \multirow{3}{*}{1 (HMC)} & minESS & 6026.1 & 1156.4 & \cellcolor{cellbox_worstsampler} \textbf{187.7} & 290.3 & 525.7 & 1215.2 \\ 
 & & meanESS & 3068.0 & 705.6 & \cellcolor{cellbox_worstsampler} \textbf{143.5} & 176.3 & 454.6 & 1047.4 \\ 
% & & multiESS & 375.8 & 72.99 & \cellcolor{cellbox_worstsampler} \textbf{50.60} & 159.5 & 279.5 & 506.9  \\
 & & multiESS & 375.8 & 73.0 & \cellcolor{cellbox_worstsampler} \textbf{50.6} & 159.5 & 279.5 & 506.9  \\
\hline
 &  &  & \multicolumn{2}{c}{$L = 1$} & $\mathcal{U}\{1, D\}$ & $L_\text{opt}$ & $\mathcal{U}\{1, 5D\}$ & $\mathcal{U}\{1, 6D\}$ \\
\hline
\multirow{6}{*}{Banana} & \multirow{3}{*}{$\mathcal{U} (\varphi_\text{lower}, \varphi_\text{upper})$} & minESS & \multicolumn{2}{r}{283.9} & 245.2 & \cellcolor{cellbox_bestsampler} \color{red} \textbf{160.2} & 168.0 & 184.1 \\
 & & meanESS & \multicolumn{2}{r}{185.9} & 164.6 & \cellcolor{cellbox_bestsampler} \color{red} \textbf{149.0} & 167.3 & 176.9 \\
 & & multiESS & \multicolumn{2}{r}{192.2} & 176.3 & \cellcolor{cellbox_bestsampler} \color{red} \textbf{166.1} & 182.8 & 183.0 \\\cline{2-9}
 & \multirow{3}{*}{1 (HMC)} & minESS & \multicolumn{2}{r}{587.7} & 426.7 & 229.8 & \cellcolor{cellbox_worstsampler} \textbf{210.0} & 206.9 \\
 & & meanESS & \multicolumn{2}{r}{392.6} & 275.0 & 191.9 & \cellcolor{cellbox_worstsampler} \textbf{177.9} & 199.6 \\
 & & multiESS & \multicolumn{2}{r}{421.4} & 292.3 & 214.9 & \cellcolor{cellbox_worstsampler} \textbf{198.7} & 218.7 \\
\hline
\end{tabular}}
\caption{\label{tab:NumericalExperimentsGHMCvsHMC} Performance comparison between GHMC with optimal randomization scheme $\varphi \in \mathcal{U}(\varphi_\text{lower}, \varphi_\text{upper})$ \eqref{eq:PhiOptimalIntervalFinal} and standard HMC, in terms of grad/ESS metrics ($\text{ESS} \in \{\text{minESS}, \text{meanESS}, \text{multiESS}\})$ for the range of randomization schemes for $L$ in all the tested benchmarks. The overall top values for each metric in each benchmark are highlighted in red. The best values for the sampler, but not overall the best for each metric, are shown in bold. The results achieved with the most successful simulation settings for the most successful sampler are shaded in light blue, whereas the results obtained with the most successful simulation settings for less successful sampler are highlighted in light grey.}
\end{table}

In order to evaluate the optimal settings proposed in this study, we compare our recommended choice $\varphi \sim$ $\mathcal{U} (\varphi_\text{lower}, \varphi_\text{upper})$ \eqref{eq:PhiOptimalIntervalFinal} with the fixed choice of $\varphi = 1$ (which corresponds to standard HMC) and with three randomization intervals $\varphi \sim$ $\mathcal{U} (0, 0.1)$, $\mathcal{U} (0, 0.5)$, $\mathcal{U} (0, 0.9)$. For each choice of $\varphi$, in addition to the proposed optimal randomization interval for $L$ \eqref{eq:LoptRandScheme}, we consider five different $\mathcal{U} \{1, L_\text{upper}\}$ intervals (four for Banana), where the values of $L_\text{upper}$ are proportional to the dimension of the simulated system. %We provide the results  standard HMC.
The optimal randomization intervals for $\varphi$ and $L$ are summarized in Table~\ref{tab:OptimalLPhiInput}. %s \ref{tab:OptimalPhiInput} and \ref{tab:OptimalLInput} respectively.

As put forward at the end of Section \ref{sec:DeltaTRandomizationNumericalExperiments}, we use s-AIA3 as a numerical integrator for Hamiltonian dynamics and the step size randomization interval proposed in Section \ref{sec:rho_maxima} (Eq. \eqref{eq:t_BCSS3maxtoHSL}). 
With respect to the fitting factor approach \eqref{eq:SfCases}, all benchmarks tested here use the more accurate $S_\omega$ (see \ref{app:NondimensionalizationAppendix} and \cite{sAIA_paper} for further details). Finally, in these and subsequent numerical experiments, eigenvalues are computed via a full dense eigendecomposition using the \texttt{dsyev} routine from the LAPACK package \cite{lapack_netlib}.

\begin{figure}[!t]
    \centering
    \includegraphics[width=\textwidth]{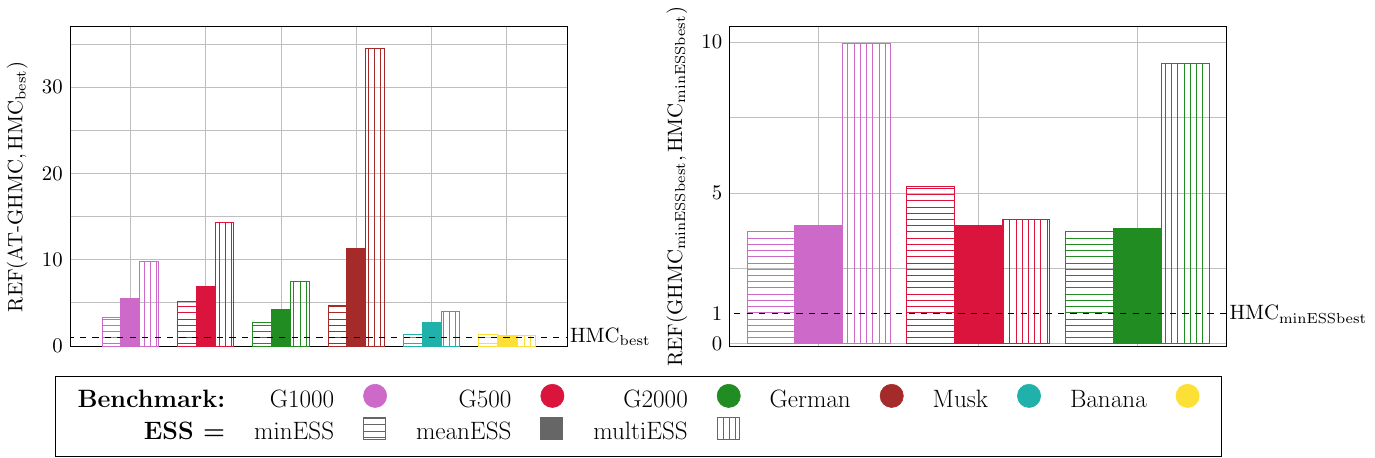}
        \caption{On the left: Relative efficiency (REF \eqref{eq:REF}) of GHMC with the optimal parameter settings, \acs{AT-GHMC} (light blue boxes in Table \ref{tab:NumericalExperimentsGHMCvsHMC}) with respect to HMC with its top parameter setting, HMC$_\text{best}$ (light grey boxes in Table \ref{tab:NumericalExperimentsGHMCvsHMC}) among all the tested benchmarks. All the benchmarks show the superiority (up to 34x) of \acs{AT-GHMC} over HMC$_\text{best}$ with each performance metric. 
        \\
        On the right: Relative efficiency (REF \eqref{eq:REF}) of GHMC with respect to HMC with their best settings for minESS (see Table \ref{tab:NumericalExperimentsGHMCvsHMC}) for the Gaussian benchmarks -- G1000, G500, G2000. All the benchmarks show the superiority (up to 10x) of GHMC$_\text{minESSbest}$ over HMC$_\text{minESSbest}$ with each performance metric.}
    \label{fig:phivsl_comparison_merged}
\end{figure}

Table \ref{tab:NumericalExperimentsGHMCvsHMC} presents the performance comparison between GHMC with  $\varphi \sim$ $\mathcal{U} (\varphi_\text{lower}, \varphi_\text{upper})$ and standard HMC ($\varphi = 1$) across the tested benchmarks, with varying values of $L_\text{upper}$. For any benchmark and for each metric, the best performance is consistently achieved by GHMC (shown in red). Combining GHMC with our proposed choice of $L_\text{opt}$ leads to the best GHMC results for the average metrics, i.e. grad/meanESS and grad/multiESS. Moreover, for non-Gaussian benchmarks (also, with smaller dimensions), grad/minESS shows the best results with such a choice of $L$. For Gaussian benchmarks, the best results for grad/minESS are reached for larger values of $L_\text{upper}$ -- they are lower for GHMC (around $D$) and higher for HMC ($> \text{2}D$).   

The results obtained with other tested randomization intervals for $\varphi$, $\mathcal{U} (0, 0.1)$, $\mathcal{U} (0, 0.5)$, $\mathcal{U} (0, 0.9)$, support our choice for the optimal randomization interval ($\varphi_\text{lower}$, $\varphi_\text{upper}$) and can be examined in \ref{app:PhivsLNumericalExperiments}.

Finally, in Figure \ref{fig:phivsl_comparison_merged}, we provide performance comparison in terms of the REF \eqref{eq:REF} between GHMC with our optimal settings, i.e. \acs{AT-GHMC}, and HMC with its best settings found heuristically (see Table \ref{tab:NumericalExperimentsGHMCvsHMC}) using all proposed ESS metrics and all tested benchmarks. For the Gaussian benchmarks, %in Figure \ref{fig:phivsl_comparison_topminESS}, 
we also present the comparison at the $L_\text{upper}$ values that optimize the minESS metric in HMC and GHMC (see Table \ref{tab:NumericalExperimentsGHMCvsHMC}). Both plots in Figure~\ref{fig:phivsl_comparison_merged} confirm the improvements (up to 34x and 10x, respectively) gained with our proposed GHMC settings across all metrics and benchmarks. They also demonstrate that greater improvements are achieved with the average metrics, i.e. meanESS and multiESS.

Last of all, we investigate the ability of AT-GHMC to explore the phase space in comparison with both tuned and standard HMC. 
We consider the Banana benchmark and compare AT-GHMC against the manually tuned HMC of \cite{Lan2015}, as well as standard HMC with $L = D$ and two choices of the step size: $\Delta t = 0.05$, comparable with the value identified by AT-GHMC, and $\Delta t = D^{-1}$. 
The parameter settings and AR are reported in Table~\ref{tab:ContourPlotsSettings}.

\begin{table}[t]
\centering
\renewcommand{\arraystretch}{1.2}
\begin{tabular}{l r r r r}
\hline
Sampler & $\Delta t$ & $L$ & $\varphi$ & AR \\
\hline
AT-GHMC & $\mathcal{U}(0.152, 0.219)$ & $\mathcal{U}(2, 8)$ & $\mathcal{U}(0.21904, 1.00000)$ & 0.92 \\
\hline
HMC tuned & $0.111$ & $7$ & 1.00000 & 0.69 \\
\hline
\multirow{2}{*}{HMC} & $0.050$ & \multirow{2}{*}{$2$} & \multirow{2}{*}{$1.00000$} & 0.91 \\
 & $0.500$ &  &  & 0.02 \\
\hline
\end{tabular}
\caption{\label{tab:ContourPlotsSettings} Parameter settings and AR for the contour plots in Figure~\ref{fig:ContourPlots}.}
\end{table}

\begin{figure}[!t]
    \centering
    \includegraphics[width=\textwidth]{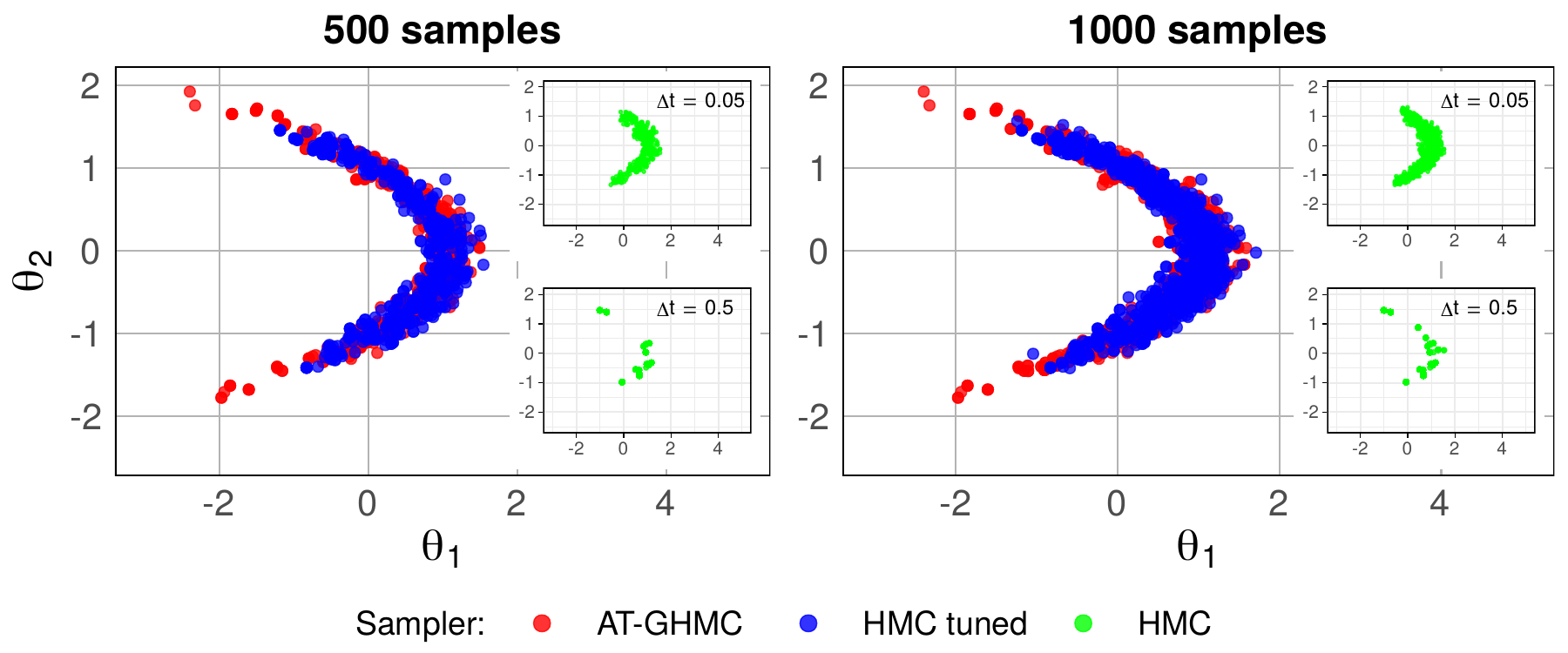}
        \caption{First 50 and 100 samples from 10 independent chains obtained with AT-GHMC (red), tuned HMC (blue), and standard HMC (green) with two choices of $\Delta t$. 
        %As the number of samples increases, AT-GHMC exhibits broader exploration of the phase space, while both AT-GHMC and tuned HMC consistently outperform standard HMC using untuned hyperparameters.
        }
    \label{fig:ContourPlots}
\end{figure}

Figure~\ref{fig:ContourPlots} illustrates the first 50 and 100 samples from 10 independent chains for each sampler. 
The results clearly show that AT-GHMC explores a broader region of the phase space than tuned HMC. Moreover, both AT-GHMC and tuned HMC substantially outperform untuned HMC, highlighting the critical role of hyperparameter selection in achieving efficient sampling. 
Finally, we emphasize that the tuning strategy employed in \cite{Lan2015} relies on a manual, trial-and-error procedure, whereas AT-GHMC leverages the fully automated ATune approach.

\section{Applications -- Case studies}\label{sec:Applications}
Next, we apply ATune 
%our optimal GHMC parameter setting 
to three case studies:
\begin{itemize}
\item Patient resistance to endocrine therapy in breast cancer %Breast cancer therapy 
\item Cell-cell adhesion dynamics
\item Influenza A (H1N1) epidemics outbreak.
\end{itemize}

\subsection{Patient resistance to endocrine therapy in breast cancer}\label{sec:BreastCancer}
Estrogen-positive (ER+) breast cancer is the most common subtype of breast cancer, accounting for over 70\% of all breast tumours diagnosed. Hormone therapies such as tamoxifen are the gold-standard treatment. However, a significant amount of patients develops resistance to the treatment, which constitutes a serious clinical problem. The development of this resistance is not yet well understood, and plenty of clinical and laboratory research aims to find which genes may be relevant to the problem.

For one such clinical study presented in \cite{piva2014sox2}, ER+ breast cancer patients treated with tamoxifen for 5 years were tested for the presence of a specific gene marker, SOX2. The  objective of the study was to analyse the connection between SOX2 and the emergence of resistance to tamoxifen treatment. An overexpression of this gene was previously identified in bioinformatic analyses of resistant breast cancer cells as a key contributor for resistance, but clinical evidence was presented in \cite{piva2014sox2} for the first time. 

The dataset used in our work stems from that clinical trial. It contains numerical data for SOX2 values and 7 other gene-related covariates taken from 75 tumours samples. The recorded response variable was the event of relapse after 5 years. Notably, the SOX2 variable perfectly separates the dataset for this response variable, as all cases with SOX2 values below 3 respond well to the treatment and all those above this value present a resistant response. The appearance of separation implies that maximum likelihood estimates do not exist for some binary classification models such as logistic regression \cite{mansournia2018separation} in the classical frequentist interpretation. The perfect separation makes the Bayesian framework natural to model such a dataset, as it can provide estimates for the parameters involved without the need to use regularization techniques.

The model of choice is a Bayesian Logistic Regression, where the dataset has 75 observations and dimension $D=8$ (including intercept parameter). We test the effect of three priors, informative $\mathcal{N}(0,1)$, weakly informative $\mathcal{N}(0,2.5)$ (according to \cite{gelman2008weakly}) and low informative $\mathcal{N}(0,5)$.

\begin{table}[t]
\centering
\resizebox{\textwidth}{!}{\begin{tabular}{l l r r r r}
Prior & Sampler & $N_{1.01}$ & $L$ & $\Delta t$ & $\varphi$ \\
\hline
\multirow{4}{*}{$\mathcal{N}(0, 5)$} & \acs{AT-GHMC} & 300 & 1 & $\mathcal{U} \{0.889, 1.284\}$ & $\mathcal{U} \{0.055,  0.330\}$ \\ %\cline{2-6} 
 & \acs{AT-HMC} & 200 & 1 & \multirow{2}{*}{$\mathcal{U} \{0.890, 1.285\}$} & - \\ %\cline{2-6} 
 & HMC$_{L_\text{NUTS}}$ & 100 & $\mathcal{U} \{1,  6 \}$ & & - \\ %\cline{2-6} 
% & NUTS & 100 & $\overline{L} = 14.8$ & $\overline{\Delta t} = 0.247$ & - \\
 & NUTS & 100 & $\overline{L} = 15$ & $\overline{\Delta t} = 0.247$ & - \\
 \hline
\multirow{4}{*}{$\mathcal{N}(0, 2.5)$} & \acs{AT-GHMC} & 100 & 1 & $\mathcal{U} \{0.725, 1.047\}$ &  $\mathcal{U} \{0.055,  0.330\}$ \\ %\cline{2-6} 
 & \acs{AT-HMC} & 100 & 1 & \multirow{2}{*}{$\mathcal{U} \{0.723, 1.045\}$} & - \\% \cline{2-6} 
 & HMC$_{L_\text{NUTS}}$ & 200 & $\mathcal{U} \{1, 8\}$ &  & - \\ %\cline{2-6} 
% & NUTS & 100 & $\overline{L} = 11.8$ & $\overline{\Delta t} = 0.354$ & - \\
 & NUTS & 100 & $\overline{L} = 12$ & $\overline{\Delta t} = 0.354$ & - \\
 \hline
\multirow{4}{*}{$\mathcal{N}(0, 1)$} & \acs{AT-GHMC} & 100 & 1 & $\mathcal{U} \{0.545, 0.787\}$ &  $\mathcal{U} \{0.055,  0.330\}$\\ %\cline{2-6} 
 & \acs{AT-HMC} & 200 & 1 & \multirow{2}{*}{$\mathcal{U} \{0.545,  0.788\}$} & - \\ %\cline{2-6} 
 & HMC$_{L_\text{NUTS}}$ & 200 & $\mathcal{U} \{1,  11 \}$ &  & - \\ %\cline{2-6} 
% & NUTS & 100 & $\overline{L} = 8.1$ & $\overline{\Delta t} = 0.474$ & - \\
 & NUTS & 100 & $\overline{L} = 8$ & $\overline{\Delta t} = 0.474$ & - \\
 \hline
\end{tabular}}
\caption{Simulation parameters for the numerical experiments on the breast cancer dataset with the BLR model combined with three different priors -- $\mathcal{N} (0, 5)$, $\mathcal{N} (0, 2.5)$, $\mathcal{N} (0, 1)$. The optimal settings for \acs{AT-GHMC} and \acs{AT-HMC}, as well as the optimal $\Delta t$ randomization interval for HMC$_{L_\text{NUTS}}$ are found using the procedure described in Section \ref{sec:Algorithm}. \acs{AT-GHMC}, \acs{AT-HMC}, and HMC$_{L_\text{NUTS}}$ employ s-AIA3 integrator while NUTS uses standard Verlet. The mean values of $L$ and $\Delta t$ used by NUTS optimal routine are denoted as $\overline{L}$ and $\overline{\Delta t}$, respectively.}
\label{tab:CancerSimulationParams}
\end{table}

%We compare the performance of \acs{AT-GHMC} proposed in this study against the No-U-Turn-Sampler (NUTS) \cite{hoffman_gelman2014}, the optimized HMC sampler implemented in the popular \textsf{R} package Stan (\texttt{rstan}) \cite{RStan}.
%\textcolor{magenta}{
We compare the performance of \acs{AT-GHMC} proposed in this study against the No-U-Turn-Sampler (NUTS) \cite{hoffman_gelman2014}, the optimized HMC sampler implemented in Stan \cite{stan2024} using its \texttt{R} interface \texttt{rstan} \cite{RStan}.
%}
%{\color{blue} In other cases, I cited \cite{STAN}, which one should I use?}
NUTS in Stan uses the leapfrog/Verlet integrator \cite{verlet1967} and the original optimization routine to determine both a step size, $\Delta t$, and a number of Hamiltonian dynamics integration steps per iterations, $L$. 
Additionally, we perform numerical experiments with HMC 
%as implemented in \textsf{HaiCS} 
using 
%two trajectory length choices, 
$L_\text{opt}$ from Eq. \eqref{eq:LoptRandScheme} (\acs{AT-HMC}) and setting $L = L_\text{NUTS}$ (HMC$_{L_\text{NUTS}}$). The latter is determined such that the integration leg $\Delta t L$ matches that found by NUTS.
AT-GHMC, AT-HMC and HMC$_{L_\text{NUTS}}$ use s-AIA3 and the more accurate $S_\omega$ fitting factor approach \eqref{eq:SfCases} (see \ref{app:NondimensionalizationAppendix} for further details).

%RStan uses default parameters and the NUTS sampler \cite{hoffman_gelman2014} and the leapfrog/Verlet integrator \cite{verlet1967}.
All samplers are run with the same number of burn-in iterations, $N_\text{burn-in}$, to ensure a fair comparison. We assess performance using grad/ESS metrics introduced in Section~\ref{sec:PerformanceMetrics}. 
%, $\text{ESS} \in \{\text{minESS}, \texŧ{meanESS}, \text{multiESS}\}$ (see Section~\ref{sec:PerformanceMetrics}), calculated over the first $N_{1.01} + 1000$ iterations after the burn-in. Here, $N_{1.01}$ is the minimum number of iterations required to achieve $\text{maxPSRF} < 1.01$, as outlined in Section \ref{sec:PerformanceMetrics}. 
Small values of $N_{1.01}$ in NUTS are expected, given its optimization routine aims to bring PSRF close to 1.
The simulation settings are summarized in Table \ref{tab:CancerSimulationParams}.

\begin{figure}[t]
	    \centering
        \includegraphics[width=\textwidth]{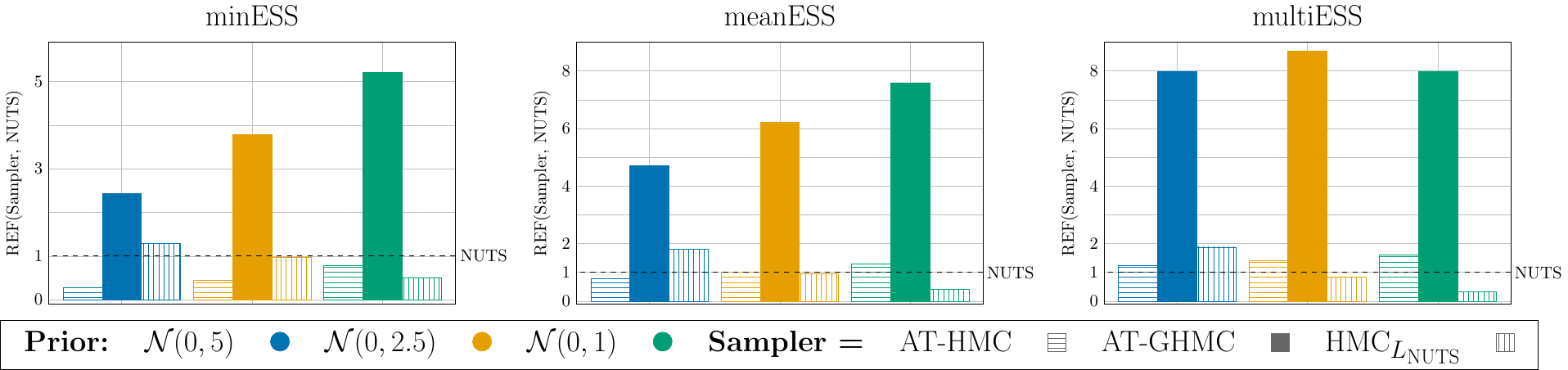}
	    \caption{Relative efficiency REF \eqref{eq:REF} for minESS (left), meanESS (center), and multiESS (right) of \acs{AT-GHMC}, \acs{AT-HMC}, and HMC$_{L_\text{NUTS}}$ with respect to Stan NUTS obtained in the numerical experiments run with the settings presented in Table \ref{tab:CancerSimulationParams}. The performance of \acs{AT-GHMC} (filled bars), \acs{AT-HMC} (horizontal bars), and
	    %with two lengths of $L$ --   
	    %optimal settings 
	    %$L_\text{opt}$ \eqref{eq:LoptRandScheme} (horizontal bars) and 
	    HMC$_{L_\text{NUTS}}$ (vertical bars) 
	    %and HMC with $L_\text{opt}$ \eqref{eq:Lopt} 
	    %the two values of $L_\text{upper}$ 
	    are compared with NUTS for the three examined priors: $\mathcal{N} (0, 5)$ (blue), $\mathcal{N} (0, 2.5)$ (yellow), $\mathcal{N} (0, 1)$ (green). \acs{AT-GHMC} significantly outperforms NUTS for any metric and prior. 
	    \acs{AT-HMC} demonstrates good results for average metrics (meanESS, multiESS) with all priors but it is not convincing for minESS. 
	    Using $L_\text{NUTS}$ instead of $L_\text{opt}$ \eqref{eq:LoptRandScheme} in HMC leads to performance degradation in HMC simulations with $\mathcal{N} (0, 1)$, whereas for $\mathcal{N} (0, 5)$ the results are visibly better than with NUTS HMC and with \acs{AT-HMC} for all metrics.}
	    % but grad/minESS (left) for $\mathcal{N} (0, 0.5)$ (blue). The biggest improvement is achieved by $\mathcal{N} (0, 0.25)$ (yellow) for grad/multiESS (right).}
     \label{fig:GHMC_STAN_Comparison_breast_cancer}
	\end{figure}

\begin{sloppypar}
Figure \ref{fig:GHMC_STAN_Comparison_breast_cancer} compares the sampling performance of \acs{AT-GHMC}, \acs{AT-HMC}, and HMC$_{L_\text{NUTS}}$ with NUTS using REF metrics \eqref{eq:REF} for $\text{ESS} \in \{\text{minESS}, \text{meanESS}, \text{multiESS}\}$. \acs{AT-GHMC} consistently outperforms NUTS across all metrics and priors, achieving more than 8 times improvement (e.g., multiESS for the $\mathcal{N} (0, 2.5)$ prior).
%up to 3.5 times improvement (e.g., multiESS for the $\mathcal{N} (0, 5)$ prior).
\end{sloppypar}

We emphasize that
%in this paper, no 
the ATune algorithm does not offer an optimal choice of $L$ for HMC. The values of $L_\text{opt}$ \eqref{eq:LoptRandScheme} used in \acs{AT-HMC} are the recommended values for GHMC. However, the strong meanESS and multiESS performance of \acs{AT-HMC} in all experiments confirms their applicability in HMC.
On the other hand, the relative efficiency of HMC$_{L_\text{NUTS}}$ with respect to NUTS shows its strong dependence on a chosen prior. 
Specifically, HMC$_{L_\text{NUTS}}$ demonstrates visible improvements over NUTS with the low informative prior $\mathcal{N}(0, 5)$; the comparable performance with NUTS if the chosen prior is $\mathcal{N} (0, 2.5)$, and the poor sampling efficiency with the highly informative prior $\mathcal{N} (0, 1)$. 
For $\mathcal{N} (0, 5)$, the improved performance can be attributed to the higher accuracy and longer optimal step size of the 3-stage integrator (s-AIA3) employed in HMC$_{L_\text{NUTS}}$ compared to those achievable with the standard 1-stage Verlet integrator used in NUTS. 
On the contrary, the poor performance of HMC$_{L_\text{NUTS}}$ with $\mathcal{N} (0, 1)$ is likely caused by the underestimation of $\Delta t_\text{CoLSI}$ hinted by the comparison of its value with $\overline{\Delta t}$ (see Table \ref{tab:CancerSimulationParams}). Indeed, the values are close, which is not 
%the case for $\mathcal{N}(0, 5)$ and $\mathcal{N} (0, 2.5)$, and it is not 
what one expects when comparing optimal step sizes for 1- and 3- stage integrators. 
It is also not the case for $\mathcal{N}(0,5)$ and $\mathcal{N}(0,2.5)$, where optimal step sizes are longer than the NUTS optimal step sizes (less pronounced for $\mathcal{N} (0, 2.5)$ though), and where \acs{AT-GHMC} and \acs{AT-HMC} 
%with $L_\text{NUTS}$ 
have never been outperformed by NUTS. 
These observations are encouraging. They mean that even in the worst scenario, when the estimated optimal parameters are less accurate, they still guarantee the good sampling performance (see \acs{AT-GHMC} and \acs{AT-HMC} for $\mathcal{N} (0,1))$ and are not inferior to the optimal parameters offered by the state-of-the-art NUTS approach. 
Moreover, the sampling advantages of GHMC over HMC, enhanced by the stochastic procedure for finding its optimal settings, disregard the inaccuracy in the estimation of an optimal step size and, as a result, \acs{AT-GHMC} demonstrates a significant improvement over NUTS, \acs{AT-HMC} and HMC$_{L_\text{NUTS}}$ as happened in the case of $\mathcal{N}(0,1)$. The fact that the performance of \acs{AT-HMC} in this case is comparable with, or better than the NUTS performance also confirms the robustness of the proposed methodology.

\subsection{Parameter inference on a cell-cell adhesion model}\label{sec:CellCellModel}

%Many phenomena on the cellular level such as tissue formation or cell sorting can be explained by introducing differential cell-cell adhesion \cite{Armstrong-Painter_2006, Gerisch-Chaplain_2008, MURAKAWA-Togashi,Carillo_Huang_2018, Murakawa-Carillo-Sato_2019, Trush_etal_2019}. 
%This phenomena has been used to explain pattern formation for zebrafish lateral lines \cite{Volkening2015,Martinson_Volkening_2023}, cancer invasion modeling \cite{Gerisch-Chaplain_2008}, and cell sorting in early fly brain \cite{Trush_etal_2019} and early mice brain development \cite{Mastusnafa_Murakawa_etal_2017}. 
%
Many phenomena on the cellular level can be explained by introducing differential cell-cell adhesion \cite{Armstrong-Painter_2006, MURAKAWA-Togashi, Carillo_Huang_2018,Murakawa-Carillo-Sato_2019}. 
This approach has been used to describe, e.g., pattern formation for zebrafish lateral lines, cancer invasion modeling and cell sorting in early fly brain and early mice brain development (see \cite{Carrillo2025} and references therein).
When modeling such biological systems using PDEs, the adhesion can be represented by means of a non-local interaction potential \cite{Gerisch-Chaplain_2008,Armstrong-Painter_2006,Carillo_Huang_2018,Murakawa-Carillo-Sato_2019}. 
To enhance predictive power of such models, a careful calibration of the model parameters according to experimental lab data is required. This task can become increasingly difficult due to sparsity of available data or increasing complexity of the model (such as, for example, by the addition of an interaction term). 
The application of state-of-the-art methodologies for performing parameter estimation on PDE models is receiving lots of attention in the last years \cite{Baker_Simpson_Vittadello_2020, Falco_Baker_2023}, and, in this context, our goal is to investigate the feasibility of using the methodology introduced in this work for the task.

To achieve this goal, we propose a benchmark PDE model:
\begin{equation}\label{eq:pdemodel}
    \frac{\partial u}{\partial t} = \nabla \cdot [ u \nabla (\Pi'(u) + (W * u ) ], 
\end{equation}
which describes the time evolution of the density of cells $u(\mathbf{x}, t)>0$, $\mathbf{x} \in \mathbb{R}^d, t>0$ ($d = 1, 2, 3$ is the spatial dimension), as governed by a series of potentials. In \eqref{eq:pdemodel}, $\Pi(u)$ is a density of the internal energy and $W(\mathbf{x})$ is an interaction potential (also called an interaction kernel) describing cell adhesion. The convolution ($W * u$) gives rise to non-local interactions. Depending on the explicit form of the potentials, they can describe a wide range of biological phenomena such as the formation of ligands through long philopodia, modeled by nonlocal attractive potentials, and volume constraints, modeled by  strong repulsive nonlocal potentials or nonlinear diffusive terms. We refer to \cite{SIAMNews} for more information about the applications of these aggregation-diffusion models in the sciences. We consider the following explicit forms for $\Pi'(u)$ and $W(\mathbf{x})$:
\begin{equation}
\begin{split}\label{eq:pdemodel_potentials}
    &\Pi'(u) = \nu u^{m-1}\\
    &W(\mathbf{x}) =  -a e^{|\mathbf{x}|^2/2r_W}/\sqrt{2\pi r_W}, 
\end{split}
\end{equation}
where $\Pi'(u)$ gives rise to a diffusion term, governed by the diffusion exponent $m$. 
In the limit $m\to1$, one recovers linear diffusion, whereas if $m>1$ the term represents porous medium and leads to finite speed propagation fronts. The diffusion coefficient $\nu$ defines the cell mobility in the medium. On the other hand, the explicit form of $W(\mathbf{x})$ leads to symmetric attraction between cells due to ligands formation on the cell's membrane. The strength of the attraction is given by the parameter $a>0$, whereas $r_{W}>0$ controls the range of the interaction. 

\begin{table}[t]
\centering
\begin{tabular}{l l r r r r}
Model & Sampler 		& $N_{1.01}$ & $L$ & $\Delta t$ & $\varphi$ \\
\hline
%GHMC & 2250 & \multirow{2}{*}{$1$} & $\mathcal{U} (0.071, 0.103)$ & $\mathcal{U} (0.110, 0.659)$ \\
\multirow{3}{*}{3-parameter}	& \acs{AT-GHMC}	& 46300 & \multirow{2}{*}{1} & $\mathcal{U} (0.0024, 0.0035)$ & $\mathcal{U} (0.15, 0.88)$ \\
					& \acs{AT-HMC}	& 288900 & &  & - \\
					& RW    		& 1370500 & - & - & - \\
\hline
\multirow{3}{*}{4-parameter}	& \acs{AT-GHMC}	& 47000 & \multirow{2}{*}{1} & $\mathcal{U} (0.0025, 0.0035)$ & $\mathcal{U} (0.11, 0.66)$ \\
					& \acs{AT-HMC}	& 471400 & &  & - \\
					& RW    		& 718000 & - & - & - \\
\hline
\end{tabular}
\caption{Simulation parameters for the numerical experiments on the cell-cell adhesion model. The optimal settings for \acs{AT-GHMC} and \acs{AT-HMC} are found using the procedure described in Section \ref{sec:Algorithm}. 
For RW applied to the 4-parameter model, the convergence criterion used to determine $N_{1.01}$ is relaxed to $\mathrm{maxPSRF} < 1.1$.}
\label{tab:PDESimulationParams}
\end{table}

Our purpose is to perform Bayesian inference of the parameters of this model using three different MCMC methods, and to evaluate the efficiency of each tested sampler for this class of models. Parameter inference in a Bayesian framework requires sampling from the posterior distribution of parameters given by $\pi(\btheta|\by) \propto L(\by|\btheta)p(\btheta)$, where $\by$ are the experimental data and $\btheta$ is the vector of model parameters. For the proposed model, the data consists of a collection of measurements of the cell density at different spatial points and different times $\by=\{u_{\by}(\mathbf{x}_{i},t_{j})\}_{i,j}$. In cell modeling, it is usual to assume that the experimental data stems from the proposed model with additive Gaussian noise $\varepsilon\sim\mathcal{N}(0,\sigma)$, i.e. $u_y (\mathbf{x}_{i},t_{j}) = u(\mathbf{x}_{i},t_{j}) + \varepsilon$. Under this assumption, the likelihood takes the form:
\begin{equation}
    L(\by|\btheta) = \frac{1}{\sqrt{2\pi\sigma^{2}}}\exp\left(\sum_{j}\left(\frac{u_{\by}(\mathbf{x},t_{j})-u(\mathbf{x},t_{j})}{\sigma}\right)^{2}\right).
\end{equation}
In this work, we use synthetic data generated from model \eqref{eq:pdemodel} in a one dimensional case ($d = 1$). First, we numerically solve the model in the interval $t\in[0,5]$ by using the finite volume scheme proposed in \cite{Carrillo2015} with a spatial meshing of 50 cells and a time step of $\delta t=0.1$, resulting in a total of 51 time points. We denote the numerical solution at point $x_{i}$ and time $t_{j}$ as $\tilde u_{s}(x_{i},t_{j})$. We then generate the synthetic data by adding Gaussian white noise $\varepsilon_{ij}\sim\mathcal{N}(0,\sigma)$ to each cell and time point $\tilde u_{\by}(x_{i},t_{j}) = \tilde u_{s}(x_{i},t_{j}) + \varepsilon_{ij}$, where $\sigma$ is the noise parameter. 

We consider two different variations of the model for the numerical experiments. In the first one (3-parameter model) the parameters sampled are the diffusion coefficient $\nu$, the diffusion exponent $m$ from the internal energy potential in \eqref{eq:pdemodel_potentials}, and the model independent noise parameter $\sigma$. The remaining parameters in  \eqref{eq:pdemodel_potentials} are fixed to $a = 2$ and $r_{W} = 0.5$. 
In the second variation (4-parameter model), we extend the previous model to also consider $r_{W}$ as a sampled parameter, leaving $a = 2$ as the only fixed parameter in \eqref{eq:pdemodel_potentials}. 
%so the only fixed parameter in \eqref{eq:pdemodel_potentials} is $a = 2$. 
The chosen priors are $\nu\sim\text{Exp}(1)$, $m\sim\Gamma(10,3)$, $\sigma\sim\text{Exp}(0.1)$, and $r_{W}\sim\text{Exp}(1)$.

\begin{sloppypar}
%To perform inference, we consider three samplers: 
Three sampling methodologies are examined: 
Random Walk Metropolis-Hastings (RW) (the most commonly used method for Bayesian inference), \acs{AT-HMC} and \acs{AT-GHMC}. 
In this case, we adopt the cheaper $S$ fitting factor approach \eqref{eq:SfCases} (see \ref{app:NondimensionalizationAppendix} for details) to avoid the additional computational cost associated with Hessian evaluations in a PDE model.
Gradients are estimated using finite differences.
%, both combined with the optimal hyperparameter settings described in this Chapter. 
%as implemented in \textsf{HaiCS} (\cite{tijana_thesis}, and Section~4.1.4). 

We evaluate performance using the metrics introduced in Section~\ref{sec:PerformanceMetrics}. For this study, 
%of the samplers using ESS/T metrics, with ESS = \{minESS, meanESS, multiESS\}  (see Sec.~\ref{sec:PerformanceMetrics}) and T being a computational time. 
%T
the metrics are computed over the first $N_{1.01} + 5000$ iterations, where $N_{1.01}$ is the number of iterations needed for convergence. %(cf. Sec.~\ref{sec:PerformanceMetrics}). 
We note that the slower-converging RW requires a large number of iterations to meet the strict threshold $\text{maxPSRF} < 1.01$ ($N_{1.01} > 10^6$). Therefore, for this sampler, we relax a convergence threshold to 1.1, as suggested in \cite{gelman_rubin1992}. 
Since RW does not involve gradient evaluations (which constitute the bulk of the computational cost in HMC/GHMC), 
%Since RW does not perform gradient evaluations (the bulk of computational cost of HMC/GHMC), 
we normalize ESS metrics with respect to the simulation time T. 
Simulation settings are summarized in Table~\ref{tab:PDESimulationParams}.
\end{sloppypar}

\begin{figure}[t]
	    \centering
         \includegraphics[width=\textwidth]{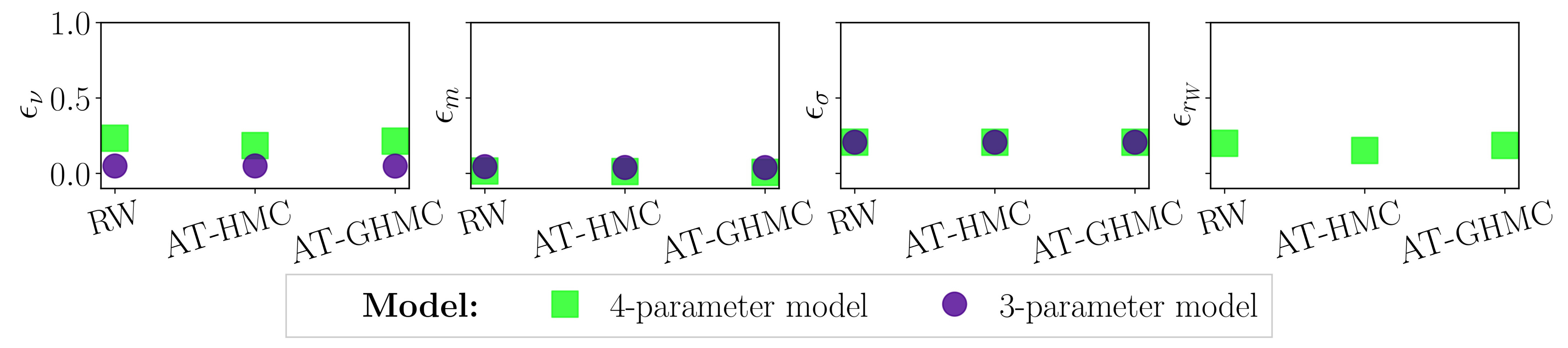}
	    \caption{Relative parameter of the diffusion coefficient (left), diffusion exponent (second to left), noise parameter (second to right) and range of interaction (right) for the 3-parameter model (purple circles) and 4-parameter model (green squares). All samplers are able to recover the true parameter value with low relative error, in particular AT-HMC and AT-GHMC show improved precision over RW.}
     \label{fig:pde_inference_parameter_errors}
	\end{figure}

Figure~\ref{fig:pde_inference_parameter_errors} shows the relative error for each parameter and both models, defined as $\epsilon_{\theta} = \frac{|\theta_\text{true} - \theta|}{\theta_ \text{true}}$, where $\theta_\text{true}$ is the true parameter value used for generating the synthetic data and $\theta$ is the mean parameter value obtained from the RW, \acs{AT-HMC} and \acs{AT-GHMC} simulations considering the first 5000 iterations after convergence. The three samplers are able to recover the true parameter values with high precision, as evidenced by the low relative errors, meaning all parameter can be confidently identified with small variance. In particular, \acs{AT-GHMC} and \acs{AT-HMC} exhibit higher precision than RW. Nevertheless, the number of iterations needed to reach the stationary distribution are vastly different across samplers: \acs{AT-GHMC} needs $\sim 10^{4}$ samples to converge for both models, whilst \acs{AT-HMC} requires $\sim10^{5}$ and RW $\sim10^{6}$ (even with the relaxed convergence threshold), confirming that \acs{AT-GHMC} has a much faster convergence rate.

%posterior distributions for the three sampled parameters observed in RW, \acs{AT-HMC} and \acs{AT-GHMC} simulations. The three samplers converge to a well-defined unimodal and narrow posterior distribution, meaning all parameters can be confidently identified with small variance. Nevertheless, the number of iterations needed to reach the stationary distribution are vastly different across samplers: \acs{AT-GHMC} needs $\approx60500$ samples to converge, whilst \acs{AT-HMC} requires $\approx344600$ and RW $\approx1.65\cdot10^{6}$ (even with the relaxed convergence threshold), confirming that \acs{AT-GHMC} has a much faster convergence rate.

\begin{figure}[t]
	    \centering
        \includegraphics[width=\textwidth]{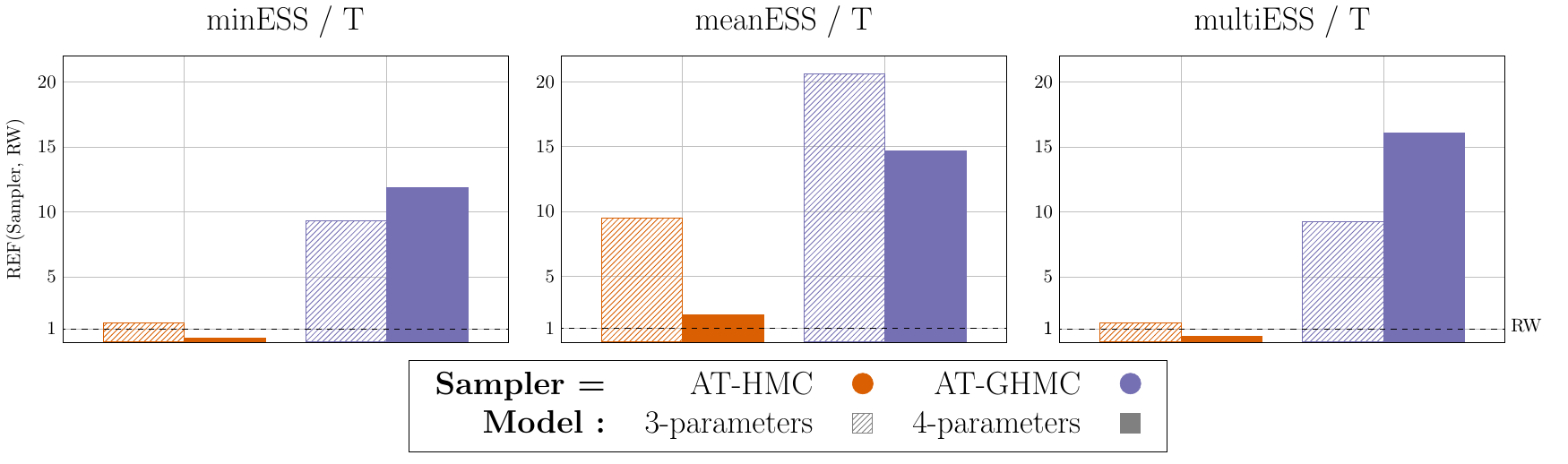}
	    \caption{Relative efficiency REF for minESS/T (left), meanESS/T (center), and multiESS/T (right) (T is a computational time) of \acs{AT-GHMC} and \acs{AT-HMC} with respect to RW. For the 3-parameter model, \acs{AT-GHMC} (purple) and \acs{AT-HMC} (orange) show improved performance over RW (dotted line) for all metrics. For the 4-parameter model, RW performs better than HMC for two out of three metrics. AT-GHMC demonstrates the best performance by a wide margin, outperforming RW by a factor of up to 20 in the 3-parameter model and up to 15 times in the 4-parameter model.}
     \label{fig:GHMC_HMC_pdemodel_performance}
	\end{figure}

In Figure~\ref{fig:GHMC_HMC_pdemodel_performance}, we present the comparison of the REF metric \eqref{eq:REF} for the three samplers (Sampler$_1$ = \{\acs{AT-GHMC}, \acs{AT-HMC}\}, Sampler$_2$ = RW) for both models, in terms of ESS/T, $\text{ESS} \in \{\text{minESS}, \text{meanESS}, \text{multiESS}\}$. \acs{AT-GHMC} exhibits high efficiency across all metrics and models, outperforming the popular RW and \acs{AT-HMC} by up to an order of magnitude in both models. Additionally, \acs{AT-GHMC} shows visibly higher relative performance in the 4-parameter model than in the 3-parameter model for two out of three considered metrics. This may indicate the potential of the method for high dimensional problems. %meaning this method has better scaling in higher dimensions. 
These results suggest that \acs{AT-GHMC} provides a suitable framework for PDE calibration. On the other hand, \acs{AT-HMC} displays worse performance than RW in two out of three metrics when increasing the dimension of the model.
%suggesting 
Most likely, an optimal choice of $L$ for GHMC \eqref{eq:LoptRandScheme} does not necessarily translate to optimal performance for HMC, especially for higher dimensions. 

%This is because, as as previously discussed in \ref{sec:BreastCancer}, the values of $L_{opt}$ used with \acs{AT-HMC} are the ones proposed for GHMC, but might not necessarily the optimal choice for HMC.
%
%The improved performance of AT-HMC and AT-GHMC over RW suggests that HMC-based samplers provide a suitable framework for PDE calibration. In particular, AT-GHMC exhibits high efficiency across all metrics, outperforming the popular RW and AT-HMC by a factor of up to 20 and 10, respectively. Also, the solid performance achieved by AT-HMC further confirms the applicability of the proposed optimal parameters for this sampling method as previously discussed in \ref{sec:BreastCancer}.
	
\subsection{Influenza A (H1N1) epidemics outbreak}
Mechanistic disease transmission models are mathematical frameworks used to describe how infectious diseases spread through populations. These models commonly employ compartmental structures 
that classify individuals into distinct states (such as susceptible, infected, recovered), and define transitions between these states over time using systems of ODEs. 
%to represent different states or groups of individuals (e.g., susceptible, infected, recovered) and employ systems of ordinary differential equations (ODEs) to define transitions between these states over time. 
Classical disease transmission models are often deterministic, 
meaning that the system's evolution is fully determined by its parameters and initial conditions, always yielding the same outcome for a given setup.
%meaning that the future state of the system is entirely determined by the model's parameters and initial conditions, always producing the same outcome. 
%While deterministic models offer valuable insights, they do not account for the inherent variability and stochasticity of real-world disease transmission. To address this limitation, mechanistic models can be enhanced with statistical models that use available data to estimate the underlying parameters governing them. This is accomplished by overlaying a probabilistic or generative model onto the deterministic structure, with Bayesian inference proving particularly effective for this purpose (\citet{Grinsztajn2020}, \citet{inouzhe2023}).
While deterministic models offer valuable insights into average system behavior, they fail to capture the inherent variability and stochasticity of real-world disease transmission. To address this limitation, mechanistic models can be enriched with statistical components that incorporate observational data to infer the underlying parameters. This is achieved by overlaying a probabilistic (or generative) model onto the deterministic structure. Bayesian inference, in particular, has proven highly effective for this purpose (refer to \cite{Grinsztajn2020, inouzhe2023} and references therein). 

As an illustrative example, we consider an outbreak of influenza A (H1N1) in 1978 at a British boarding school involving 763 students. The dataset, available in the \texttt{R} package \texttt{outbreaks} \cite{outbreaks20}, records the daily number of students confined to bed over a 14-day period. Following \cite{Grinsztajn2020}, we model the outbreak dynamics using a classical SIR (Susceptible-Infected-Recovered) model, coupled with a negative binomial observation model to account for overdispersion in the reported case counts.
%
%As illustration, we consider an outbreak of influenza A (H1N1) in 1978 at a British boarding school with 763 students. The dataset, available in the \textsf{R} package \texttt{outbreaks} (\citet{outbreaks20}), contains the daily number of students in bed over a 14-day period. Following \citet{Grinsztajn2020}, we model the transmission dynamics using a SIR (Susceptible-Infected-Recovered) model (\citet{SIR_paper} and Appendix~\ref{app:SIR}) coupled with a negative binomial distribution to represent the observed number of students in bed. 
The model includes the transmission rate, $\beta$, and the recovery rate, $\gamma$, which govern the transitions from susceptible to infected and from infected to recovered, respectively. For the prior distributions, we assume 
%The parameters of the model include the transmission rate, $\beta$, and the recovery rate, $\gamma$, which rule the transitions between the susceptible and infected, and the infected and recovered states, respectively. For prior distributions, we consider 
$\beta \sim \mathcal{N}^{0}(1, 1)$ and $\gamma \sim \mathcal{N}^{0}(0.4, 0.5)$, where $\mathcal{N}^{a}$ denotes a normal distribution truncated at $a$. 
%Unlike \citet{Grinsztajn2020}, we treat the initial number of infected students, $I_0$, as unknown and assign a prior of $\mathcal{N}^{0}(1,1)$. The prior specification is completed by assigning a distribution to the dispersion parameter, $\phi$, of the negative binomial distribution, and we consider $1/\phi \sim \text{Exp}(5)$.
The prior specification is completed by assigning an exponential prior to the inverse of the dispersion parameter, $\phi$, of the negative binomial distribution, i.e. $1/\phi \sim \text{Exp}(5)$. 
The initial conditions are set as $I_0 = 1$, $R_0 = 0$, and thus $S_0 = 762$. 
%by assigning a distribution to the dispersion parameter, $\phi$, of the negative binomial distribution, and we consider $1/\phi \sim \text{Exp}(5)$. Regarding the initial conditions of the system, we set $I_0 = 1$, $R_0 = 0$, and thus $S_0 = 762$. 
%Here, we follow the notations commonly used in epidemiological modeling. 

\begin{table}[t]
\centering
\begin{tabular}{l r r r r}
Sampler & $N_{1.01}$ & $L$ & $\Delta t$ & $\varphi$ \\
\hline
%GHMC & 2250 & \multirow{2}{*}{$1$} & $\mathcal{U} (0.071, 0.103)$ & $\mathcal{U} (0.110, 0.659)$ \\
\acs{AT-GHMC} & 2250 & $1$ & $\mathcal{U} (0.073, 0.105)$ & $\mathcal{U} (0.088, 0.527)$ \\
%HMC$_{L_\text{opt}}$ & 1500 & & $\mathcal{U} (0.078, 0.113)$ & - \\ %\cline{2-6} 
%NUTS & 140 & $\overline{L} = 5.45$ & $\overline{\Delta t} = 0.607$ & - \\
NUTS & 140 & $\overline{L} = 5$ & $\overline{\Delta t} = 0.607$ & - \\
 \hline
\end{tabular}
\caption{Simulation parameters for the numerical experiments on the Influenza A (H1N1) epidemics dataset with the SIR model. The optimal settings for \acs{AT-GHMC} %and HMC 
is found using the algorithm described in Section~\ref{sec:Algorithm}. 
\acs{AT-GHMC} 
%and HMC 
employs s-AIA3 integrator while NUTS uses standard Verlet. The mean values of $L$ and $\Delta t$ used by NUTS optimal routine are denoted as $\overline{L}$ and $\overline{\Delta t}$, respectively.}
\label{tab:SIRSimulationParams}
\end{table}

As in Section~\ref{sec:BreastCancer}, we compare the performance  
of two samplers: \acs{AT-GHMC} and 
NUTS \cite{hoffman_gelman2014}, as implemented in the \texttt{R} package \texttt{rstan} \cite{RStan}. 
We note that, in both cases, the (approximate) solution to the ODE system is obtained using the CVODES component of the SUNDIALS suite \cite{cvode_documentation}. Specifically, we use the ODE solver that relies on the backward differentiation formula (BDF) method. Similarly to the breast cancer study, both samplers are run with the same number of burn-in iterations, and their performance is assessed in terms of the grad/ESS metrics calculated over the first $N_{1.01} + 1000$ iterations after the burn-in (see Section~\ref{sec:PerformanceMetrics}).

\begin{figure}[t]
	    \centering
        \includegraphics[width=\textwidth]{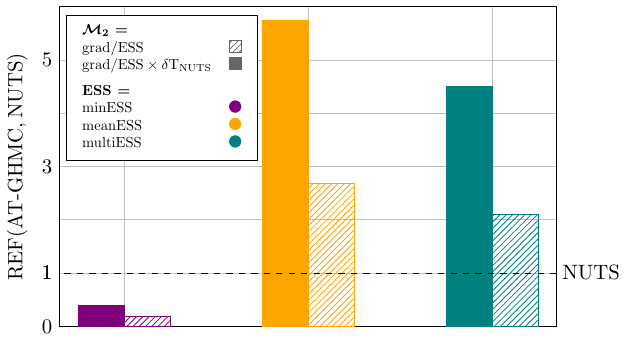}
	    \caption{Relative efficiency (REF) \eqref{eq:REF} in terms of grad/minESS (violet), grad/meanESS (yellow), and grad/multiESS (green) of \acs{AT-GHMC} with respect to Stan NUTS. 
	    Filled bars correspond to the performance metrics that take into account the overheads introduced by NUTS at the optimization stage (quantified by $\delta \text{T}_\text{NUTS}$). 
	    \acs{AT-GHMC} demonstrates superior performance with meanESS and multiESS, while NUTS outperforms \acs{AT-GHMC} for minESS. When accounting for computational time differences, \acs{AT-GHMC} shows around 5x improvement over NUTS for both meanESS and multiESS.}
     \label{fig:SIR_I0_plot}
	\end{figure}

\begin{sloppypar}
The simulation settings for the numerical experiments are summarized in Table~\ref{tab:SIRSimulationParams}. 
We emphasize that, as with the cell–cell adhesion model in Section~\ref{sec:CellCellModel}, we bypass Hessian computations and the resulting computational overhead by adopting the cheaper $S$ fitting factor approach \eqref{eq:SfCases} (see \ref{app:NondimensionalizationAppendix} for details).
Figure~\ref{fig:SIR_I0_plot} shows the relative efficiency REF \eqref{eq:REF} of \acs{AT-GHMC} over NUTS, evaluated in terms of grad/ESS, $\text{ESS} \in \{\text{minESS}, \text{meanESS}, \text{multiESS}\}$. 
In addition, to account for the extra computational cost incurred solely by NUTS due to its optimization process during the burn-in phase, we 
multiply the grad/ESS metrics for NUTS by a factor quantifying such overheads, $\delta \text{T}_\text{NUTS}$. This factor is carefully evaluated using the \texttt{rstan} package. 
\end{sloppypar}

\acs{AT-GHMC} with our proposed settings outperforms NUTS in terms of both meanESS and multiESS, with improvements increasing from a factor of 2x, when no NUTS computational overheads are considered, to 5x, otherwise. 
%up to 5x when the computational overheads of NUTS are considered. 
In contrast, NUTS achieves better sampling 
with the most restrictive metric, minESS, 
which is likely enabled by the use of a tuned mass matrix. 
This trend aligns with the findings from the breast cancer study (Figure~\ref{fig:GHMC_STAN_Comparison_breast_cancer}), where \acs{AT-GHMC} shows more significant improvements over NUTS for average ESS metrics than for minESS. 
This suggests that tuning a position-dependent mass matrix during the NUTS burn-in stage may have a pronounced effect on the 
efficiency of an HMC algorithm in sampling complex distributions. 
Nevertheless, the computational time analysis shows that, in this case study, NUTS requires approximately twice the computational time of a tuned GHMC to achieve on average comparable sampling performance, as highlighted in Figure~\ref{fig:SIR_I0_plot}.  

\section{Conclusion}\label{sec:Conclusion}

The challenge of tuning appropriate hyperparameters to enhance HMC sampling performance has been extensively addressed by the statistical community. In contrast, GHMC, despite its promising potential to improve sampling efficiency due to the inherent irreversibility of its Markov chains, has not garnered comparable attention. 

\begin{sloppypar}
Here
%we present a comprehensive procedure that, given a simulated sytem, provides a complete set of system-specific optimized HMC/GHMC hyperparameters, without introducing any additional computational overhead during the simulation process.
%
we present a computationally inexpensive adaptive tuning approach ATune
%have presented a comprehensive procedure 
which generates optimal parameter settings on the fly for HMC and GHMC simulations, bringing forth adaptively tuned \acs{AT-HMC} and \acs{AT-GHMC} methods. Given a simulated system, the new methodology delivers a set of system-specific optimized HMC/GHMC hyperparameters, without introducing any additional computational overhead during the production simulation stage. 
Owing to the flexibility of the HMC framework, AT-HMC and AT-GHMC are applicable to a broad spectrum of real-world Bayesian inference problems. Nevertheless, for non-differentiable posteriors or models with inaccessible or highly complex internal structure, gradient-free sampling approaches remain necessary, preventing such models from benefiting from the enhanced sampling efficiency offered by HMC and GHMC and their top-tier variants AT-HMC and AT-GHMC.
\end{sloppypar}

To determine an optimal integration step size, we build on the insights from the s-AIA adaptive integration approach introduced in \cite{sAIA_paper}. We propose a system-specific randomization interval that minimizes energy error near the center of an estimated stability interval -- the region typically associated with the most efficient performance \cite{bcss_paper, mazur}. 
Additionally, we confirm that the s-AIA3 integrator consistently outperforms fixed-parameter integrators when combined with both HMC and GHMC samplers. 
%We also confirm that the s-AIA3 integrator is the optimal choice, outperforming fixed-parameter integrators, combined with both HMC and GHMC samplers. 
We notice that even following the general recommendations for GHMC settings leads to its superior performance over manually tuned HMC. 
%This validation further underscores the overall superiority of GHMC over HMC in terms of sampling efficiency, particularly when reasonable choices of simulation step size and integration scheme are applied. 
These findings motivate a deeper exploration of the proper tuning of $\varphi$ and $L$ parameters for GHMC.

Leveraging the methodology proposed in \cite{akhmatskaya_etal_2017}, 
we develop an automated procedure for finding an optimal value for the random noise parameter in PMU \eqref{eq:PMUIntro}, based solely on the size of the simulated system (assuming the simulation step size and numerical integrator are chosen as discussed). 
When applied to high-dimensional systems, this approach leads to smaller modifications of momentum, which, combined with the high accuracy provided by s-AIA3 and an optimal choice of $\Delta t$, facilitates the use of shorter trajectories, thus enhancing sampling efficiency in GHMC \cite{skeel2014}. We note that, unlike the optimization procedures proposed for other hyperparameters, the selection of the optimal random noise $\varphi$ can be performed prior a burn-in stage, relying only on the known dimensionality of the system. 
%, relying only on the known dimensionality of the system without requiring data from a burn-in stage. 
As a result, we propose employing the optimal $\varphi$ from the very start of a GHMC simulation, rather than limiting it to the production stage only.

The introduced optimal hyperparameter settings exhibit superiority over other combinations of hyperparameters, leading to significant GHMC performance improvement across a set of popular benchmarks. 
%significant performance improvements over other combinations of hyperparameters across a set of popular benchmarks.
Additionally, the performance gains observed when comparing our results to state-of-the-art samplers in three distinct real-world case studies indicate the broad applicability and effectiveness of \acs{AT-GHMC} for diverse probabilistic models. It is worth noting that, although the optimal choice of $L$ is specifically tailored for GHMC, $L_\text{opt}$ also enhances sampling efficiency when applied to HMC in some cases, especially of small dimensions.

The proposed ATune algorithm can be easily implemented in any software package for statistical modeling using Bayesian inference with Hamiltonian Monte Carlo to offer the advantages of \acs{AT-GHMC} over conventional HMC.

\appendix

\section{Overestimation of a dimensional stability interval for 3-stage integrators}\label{app:3stageCorrection}
\subsection{Overestimation of the expected acceptance rate in the s-AIA analysis}
%\subsection{Lower bound of $\mathbb{E} [\Delta H]$}
We consider the high-dimensional asymptotic formula for expected acceptance rate $\mathbb{E} [\text{AR}]$
\cite{beskos_optimal_tuning} proven for Gaussian distributions in a general scenario \cite{calvo2021hmc}, i.e.
%From \cite{calvo_etal_2021} (Section 5.4), the Gupta expression for $\mathbb{E} [\text{AR}]$ reads as
\begin{equation}\label{eq:GuptaGeneral}
\mathbb{E} [\text{AR}] = 2 \, \Phi \left( -\sqrt{\mu / 2} \right),
\end{equation}
with $\mu = \mathbb{E}^D [\Delta H]$ and
\begin{equation}\label{eq:GuptaExpression}
2 \, \Phi \left( -\sqrt{\mu / 2} \right) = 1 - \frac{1}{2 \sqrt{\pi}} \sqrt{\mu} + \mathcal{O} (\mu^{\frac{3}{2}}), \qquad \mu \to 0, \, D \to \infty.
\end{equation}
Replacing \eqref{eq:GuptaGeneral} with its asymptotic expression 
% for $ \mathbb{E}^D [\Delta H] \to 0$,
\begin{equation}\label{eq:GuptaEDHto0}
\mathbb{E} [\alpha] = 1 - \frac{1}{2 \sqrt{\pi}} \sqrt{\mathbb{E}^D [\Delta H]}, \qquad \mathbb{E}^D [\Delta H] \to 0, \, D \to \infty,
\end{equation}
allows us to compute $\mathbb{E}^D [\Delta H]$ as proposed in \cite{sAIA_paper}: 
%we utilized the expression for $\mathbb{E}^D [\Delta H]$ from \eqref{e:GuptaEDHto0}, i.e.
\begin{equation}\label{eq:EdHGupta}
\mathbb{E}^D [\Delta H] \approx 4 \pi (1 - \mathbb{E} [\alpha])^2.
\end{equation}
Here, $\alpha$ is the acceptance probability of the Metropolis test. 
This gives rise to 
%to calculate 
the system-specific fitting factor $S_f$ (see \ref{app:NondimensionalizationAppendix}) and a system-specific $\Delta t_\text{CoLSI}$ (for more details see \cite{sAIA_paper}). In particular, we have
\begin{equation}
S_f \propto \mathbb{E}^D [\Delta H], \qquad \Delta t_\text{CoLSI} \propto \frac{1}{S_f},
\end{equation}
that brings to
\begin{equation}\label{eq:DeltaTproptoDeltaH}
\Delta t_\text{CoLSI} \propto \frac{1}{\mathbb{E}^D [\Delta H]}.
\end{equation}
The impact of the approximation error $\mathcal{O} (\mu^{\frac{3}{2}})$ in \eqref{eq:GuptaExpression} on the estimated $\Delta t_\text{CoLSI}$ can be evaluated by substituting the RHS of the equation \eqref{eq:GuptaExpression} into the inequality $ 0 \leqslant{\mathbb{E} [\alpha]} \leqslant{1}$ which holds by definition. Then, one obtains: 

\begin{equation}\label{eq:ARinequality}
\frac{1}{2 \sqrt{\pi}} \sqrt{\mu} -1 \leqslant{\mathcal{O} (\mu^{\frac{3}{2}})} \leqslant{\frac{1}{2 \sqrt{\pi}} \sqrt{\mu}}.
 \end{equation}

For $\mu \to 0$, as in our case, the inequality \eqref{eq:ARinequality} becomes

\begin{equation}%\label{eq:ARinequality}
-1 \leqslant{\mathcal{O} (\mu^{\frac{3}{2}})} \leqslant{0}.
 \end{equation}
 
Therefore, the expected acceptance rate in \eqref{eq:GuptaEDHto0} overestimates the real expected acceptance rate, leading to the underestimation of the energy error in \eqref{eq:EdHGupta}, which, in turn, results in the overestimation of $\Delta t_\text{CoLSI}$ in \eqref{eq:DeltaTproptoDeltaH}. We notice, however, that such an error is small due to the functional dependencies between the variables   involved in the estimation of $\Delta t_\text{CoLSI}$ and the condition $\mu \to 0$, though the order of such an approximation is out of the scope of this study. 

\subsection{Overestimation of $\Delta t_\text{CoLSI}$ for 3-stage integrators} 
%\subsection{Correction for 3-stage integrators} 
A closed-form expression for the expected energy error produced by 1-stage Velocity Verlet with $L$ = 1 in the case of a univariate Gaussian model, was derived in \cite{sAIA_paper}: 
\begin{equation}\label{eq:ExpectationDeltaH1sVV}
\mathbb{E}^1_{\text{VV}}[\Delta H] = \frac{h^6}{32},  \quad  h \in \left( 0, \, 2\right),
\end{equation}
where $h$ is the nondimensional counterpart of the integration step size $\Delta t$ (see \cite{%bcss_paper, 
sAIA_paper} and \ref{app:NondimensionalizationAppendix}).

Eq. \eqref{eq:ExpectationDeltaH1sVV} comes from the more general expression for the expected energy error for a univariate Gaussian model \cite{bcss_paper}
\begin{equation}\label{eq:ExpectationDeltaHforL1}
\mathbb{E}^1 [\Delta H] = \frac{(B_h + C_h)^2}{2}.
\end{equation}
Here, $B_h, C_h$ are coefficients depending on an integrator in use. %, and %$h_{\text{VV}}$ 
%$h$ is the nondimensional counterpart of a given dimensional simulation step size %$\Delta t_{\text{VV}}$
%$\Delta t$ (see \ref{app:NondimensionalizationAppendix}).

It is also possible to derive from \eqref{eq:ExpectationDeltaHforL1} similar expressions for the 2- and 3-stage cases (see \ref{app:VV2Coeffs} and \cite{sAIA_paper}, respectively).

In particular, 
\begin{itemize}
\item $k = 2$: the coefficients for the 2-stage Velocity Verlet are (see \ref{app:VV2Coeffs})
\begin{equation}
B_h = h - \frac{h^3}{8}, \qquad C_h = - h + \frac{3 h^3}{16} - \frac{h^5}{128},
\end{equation}
and, from \eqref{eq:ExpectationDeltaHforL1}, one gets
\begin{equation}\label{eq:ExpectationDeltaH2sVV}
\mathbb{E}^1_{\text{VV2}} [\Delta H] = \frac{h^6 (h^2 - 8)^2}{2^{15}}, \quad  h \in \left( 0, \, 4\right).
\end{equation}
\item $k = 3$: the coefficients for the 3-stage Velocity Verlet (see Appendix A in \cite{sAIA_paper}) 
\begin{equation}
B_h = \frac{h (h^2 - 9) (h^2 - 27)}{243}, \qquad C_h = \frac{h (h^2 - 9) (h^2 - 27) (h^2 - 36)}{8748},
\end{equation}
together with \eqref{eq:ExpectationDeltaHforL1} give
\begin{equation}\label{eq:ExpectationDeltaH3sVV}
\mathbb{E}^1_{\text{VV3}} [\Delta H] = \frac{h^6 (h^2 - 9)^2 (h^2 - 27)^2}{2^5 \, 3^{14}}, \quad  h \in \left( 0, \, 6\right).
\end{equation}
\end{itemize}
One can immediately notice that, contrary to the formula \eqref{eq:ExpectationDeltaH1sVV} for the standard 1-stage Velocity Verlet, both \eqref{eq:ExpectationDeltaH2sVV} and \eqref{eq:ExpectationDeltaH3sVV} do not allow for closed-form expressions for $h$, which are needed for estimating stability intervals with the procedure proposed in \cite{sAIA_paper}.

However, some analysis and comparison of $\mathbb{E}^1_{\text{VV}k} [\Delta H]$ in \eqref{eq:ExpectationDeltaH2sVV}--\eqref{eq:ExpectationDeltaH3sVV} with $\mathbb{E}^1_{\text{VV}} [\Delta H]$ in \eqref{eq:ExpectationDeltaH1sVV} is possible. For that, we first apply a change of variables $h=h'$, $h=2h'$ and $h=3h'$, where $h' \in \left( 0, \, 2\right)$ in \eqref{eq:ExpectationDeltaH1sVV}, \eqref{eq:ExpectationDeltaH2sVV} and \eqref{eq:ExpectationDeltaH3sVV}, respectively.  

Afterwards, we can equate both \eqref{eq:ExpectationDeltaH2sVV} and \eqref{eq:ExpectationDeltaH3sVV} to $\mathbb{E}^1_{\text{VV}} [\Delta H]$ in \eqref{eq:ExpectationDeltaH1sVV} and see which values of $h'$ solve the equation.
\begin{itemize}
\item $k=2$: by equating \eqref{eq:ExpectationDeltaH2sVV} with $h=2h'$ and \eqref{eq:ExpectationDeltaH1sVV} with $h=h'$, one obtains
\begin{equation}
\frac{2^6 h'^6 (4 h'^2 - 8)^2}{2^{15}} = \frac{h'^6}{32},
\end{equation}
which brings to
\begin{equation}\label{e:TwoStageRoots}
(h'^2 - 2)^2 = 1, 
\end{equation}
with two nonnegative roots, $h'_1 = 1$, $h'_2 = \sqrt{3}$, or, equivalently, $h_1 = 2$, $h_2 = 2 \sqrt{3}$ in \eqref{eq:ExpectationDeltaH2sVV}. Thus, Eq.~\eqref{eq:ExpectationDeltaH1sVV} and Eq.~\eqref{eq:ExpectationDeltaH2sVV} coincide at the CoLSI. Hence, we can conclude that the analysis made with \eqref{eq:ExpectationDeltaH1sVV} as proposed in \cite{sAIA_paper} is reliable for the estimation of a dimensional stability limit for a 2-stage integrator.
\item $k = 3$: by equating \eqref{eq:ExpectationDeltaH3sVV} with $h = 3h'$ and \eqref{eq:ExpectationDeltaH1sVV} with $h=h'$, one gets
\begin{equation}
\frac{3^6 h'^6 (9 h'^2 - 9)^2 (9 h'^2 - 27)^2}{2^5 \, 3^{14}} = \frac{h'^6}{32},
\end{equation}
leading to
\begin{equation}\label{e:ThreeStageRoots}
(h'^4 - 4 h'^2 + 3)^2 = 1.
\end{equation}
\eqref{e:ThreeStageRoots} has 3 nonnegative roots $h'_1 = \sqrt{2}$, $h'_2 = \sqrt{2 + \sqrt{2}}$ and $h'_3 = \sqrt{2 - \sqrt{2}}$. However, $h' = 1$, i.e. $h = 3$, is not a root of \eqref{e:ThreeStageRoots}. The closest to CoLSI $h = 3$ is  $3h'_3 \approx 2.296$. Clearly, the analysis carried out using \eqref{eq:ExpectationDeltaH1sVV} brings to an overestimation of a stability interval for 3-stage integrators and may not find an accurate estimation of a dimensional stability limit or CoLSI for a 3-stage integrator. 
We have to remark that the calculation of dimensional $\Delta t_\text{CoLSI}$ does not solely depend on $h_\text{CoLSI}$ but also on the simulated properties, such as AR, extracted at the HMC burn-in stage. Obviously, acceptance rates obtained with 1- and 3-stage integrators should differ, and thus it is not feasible to make an accurate prediction of the estimation error for $\Delta t_\text{CoLSI}$ within the proposed analysis. In any case, 
using the randomization interval suggested in 
\eqref{eq:h_BCSS3maxtoHSL} resolves this problem by probing the step sizes smaller or equal than $h=3$, including $h = 2.296$. 
\end{itemize}

\subsection{Calculation of 2-stage Velocity Verlet integration coefficients}\label{app:VV2Coeffs}
Consider the harmonic oscillator with Hamiltonian
\begin{equation}\label{HamiltonianHarmonicOscillator}
H = \frac{1}{2} (p^2 + \theta^2), \qquad \theta,p \in \mathbb{R},
\end{equation}
and equations of motions
\begin{equation}\label{EqsHarmonicOscillator}
\frac{d \theta}{dt} = p, \qquad \frac{d p}{dt} = - \theta.
\end{equation}
A 2-stage palindromic splitting integrator $\Psi_h$ ($h$ is the integration step size) acts on a configuration $(\theta_i, p_i)$ at the $i$-th iteration as
\begin{equation}\label{NumericalIntegratorEqsofmotion}
\Psi_h \left(
\begin{array}{c}
q_i \\ p_i
\end{array}
\right) =
\left( \begin{array}{c}
q_{i+1} \\ p_{i+1}
\end{array} \right)
= \left( \begin{matrix}
A_h & B_h \\
C_h & D_h
\end{matrix} \right)
\left( \begin{array}{c}
q_i \\
p_i
\end{array} \right),
\end{equation}
for suitable method-dependent coefficients $A_h$, $B_h$, $C_h$, $D_h$.

Following \cite{bcss_paper} and Appendix A in \cite{sAIA_paper}, we find $B_h$ and $C_h$ required for the calculation of $\mathbb{E}^1 [\Delta H]$ \eqref{eq:ExpectationDeltaHforL1}:
\begin{align}
B_h &= h - \frac{1}{2} (1/2 - b) h ^3,\\
C_h &= - h + b(1-b) h^3 - \frac{b^2}{2}(1/2 -b) h^5,
\end{align}
For the 2-stage Velocity Verlet ($b = 1/4$, see \cite{bcss_paper, sAIA_paper}), they are
\begin{equation}
B_h = h - \frac{h^3}{8}, \qquad C_h = - h + \frac{3 h^3}{16} - \frac{h^5}{128}.
\end{equation}

\section{Local maximum of $\rho_{\text{BCSS3}} (h)$ for $h < 3$}\label{app:rho_maxima}
The upper bound of the expected energy error for 3-stage integrators reads as (\cite{sAIA_paper}, Eq. (16))
\begin{equation}\label{eq:rho3stage}
\rho_3 (h, b) =
\scalebox{0.95}{$ \frac{h^4 \left( -3 b^4 + 8 b^3 -19/4 b^2 + b + b^2 h^2 \left( b^3 - 5/4 b^2 + b/2 - 1/16 \right) - 1/16 \right)^2}{2 \left( 3 b - b h^2 \left( b - 1/4 \right) - 1 \right) \left(1 - 3 b - b h^2 \left(b - 1/2 \right)^2 \right) \left( -9 b^2 + 6 b - h^2 \left( b^3 - 5/4 b^2 + b/2 - 1/16 \right) - 1 \right)}$}.
\end{equation}
We define $\rho_{\text{BCSS3}} (h) = \rho_3 (h, b_{\text{BCSS3}})$ at $b_\text{BCSS3} = 0.11888010966548$.
The derivative with respect to $h$ of $\rho_3 (h, b)$ in \eqref{eq:rho3stage} is
\begin{flalign}\label{eq:rhoBCSS3deriv}
\frac{\partial \rho_3 (h, b)}{\partial h} = \frac{f(h, b)}{g(h, b)},
\end{flalign}
where
\begin{flalign}
\begin{aligned}
f(h, b) = &\Bigg(4 h^3 \bigg(-3b^4 + 8b^3 - \frac{19}{4}b^2 + b + b^2h^2\left(b^3 - \frac{5}{4}b^2 + \frac{b}{2} - \frac{1}{16}\right) - \frac{1}{16}\bigg) \\
&\cdot \bigg(-3b^4 + 8b^3 - \frac{19}{4}b^2 + b + b^2h^2\left(b^3 - \frac{5}{4}b^2 + \frac{b}{2} - \frac{1}{16}\right) - \frac{1}{16}\bigg) \\
&\qquad + 2 h^4 \bigg(-3b^4 + 8b^3 - \frac{19}{4}b^2 + b + b^2h^2\left(b^3 - \frac{5}{4}b^2 + \frac{b}{2} - \frac{1}{16}\right) - \frac{1}{16}\bigg) \\
&\cdot 2 b^2 h \left(b^3 - \frac{5}{4}b^2 + \frac{b}{2} - \frac{1}{16}\right)\Bigg) \\
&\cdot \bigg(2 \bigg(3b - bh^2\left(b-\frac{1}{4}\right)-1\bigg) \\
&\cdot \bigg(1 - 3b - bh^2\left(b-\frac{1}{2}\right)^2\bigg) \\
&\cdot \bigg(-9b^2 + 6b - h^2\left(b^3 - \frac{5}{4}b^2 + \frac{b}{2} - \frac{1}{16}\right) - 1\bigg)\bigg) \\
&- \bigg(h^4 \bigg(-3b^4 + 8b^3 - \frac{19}{4}b^2 + b + b^2h^2\left(b^3 - \frac{5}{4}b^2 + \frac{b}{2} - \frac{1}{16}\right) - \frac{1}{16}\bigg) \\
&\cdot \bigg(-3b^4 + 8b^3 - \frac{19}{4}b^2 + b + b^2h^2\left(b^3 - \frac{5}{4}b^2 + \frac{b}{2} - \frac{1}{16}\right) - \frac{1}{16}\bigg)\bigg) \\
&\qquad \cdot \bigg(2 \bigg(- 2 b h \left(b - \frac{1}{4}\right) \bigg(1 - 3b - bh^2\left(b-\frac{1}{2}\right)^2\bigg) \\
&\cdot \bigg(-9b^2 + 6b - h^2\left(b^3 - \frac{5}{4}b^2 + \frac{b}{2} - \frac{1}{16}\right) - 1\bigg) \\
&+ \bigg(3b - bh^2\left(b-\frac{1}{4}\right)-1\bigg) \left( -2 b h \right) \left(b - \frac{1}{2}\right)^2 \\
&\cdot \bigg(-9b^2 + 6b - h^2\left(b^3 - \frac{5}{4}b^2 + \frac{b}{2} - \frac{1}{16}\right) - 1\bigg) \\
&+ \bigg(3b - bh^2\left(b-\frac{1}{4}\right)-1\bigg) \\
&\cdot \bigg(1 - 3b - bh^2\left(b-\frac{1}{2}\right)^2\bigg) \left(-2 h \right) \\
&\cdot \left(b^3 - \frac{5}{4}b^2 + \frac{b}{2} - \frac{1}{16}\right)\bigg)\bigg),
\end{aligned}
\end{flalign}
and
\begin{flalign}
g(h, b) = &\Bigg(2 \bigg(3b - bh^2\left(b-\frac{1}{4}\right)-1\bigg) \\
&\cdot \bigg(1 - 3b - bh^2\left(b-\frac{1}{2}\right)^2\bigg) \\
&\cdot \bigg(-9b^2 + 6b - h^2\left(b^3 - \frac{5}{4}b^2 + \frac{b}{2} - \frac{1}{16}\right) - 1\bigg)\Bigg) \\
&\cdot \bigg(2 \bigg(3b - bh^2\left(b-\frac{1}{4}\right)-1\bigg) \\
&\cdot \bigg(1 - 3b - bh^2\left(b-\frac{1}{2}\right)^2\bigg) \\
&\cdot \bigg(-9b^2 + 6b - h^2\left(b^3 - \frac{5}{4}b^2 + \frac{b}{2} - \frac{1}{16}\right) - 1\bigg)\bigg).
\end{flalign}

 By setting $\frac{\partial \rho_3 (h, b)}{\partial h} = 0$ and $b = b_{\text{BCSS3}}$ \cite{bcss_paper} in \eqref{eq:rhoBCSS3deriv}, one obtains the local maxima of $\rho_{\text{BCSS3}} (h)$ around 
$h_{\text{lower}}\equiv h_\text{BCSS3max}\approx 2.0772$.

\section{$h \to \Delta t$ map for a simulation step size}\label{app:NondimensionalizationAppendix}
Following the nondimensionalization approaches presented in \cite{sAIA_paper}, here we derive a map $h \to \Delta t$ which, given a dimensionless step size $h$, computes its dimensional counterpart $\Delta t$.

s-AIA \cite{sAIA_paper} uses a multivariate Gaussian model and simulation data obtained at the HMC burn-in stage to compute a system-specific \emph{fitting factor} $S_f$. Fitting factors quantify the step size limitations imposed by nonlinear stability in numerical schemes employed for the integration of the Hamiltonian dynamics. These factors (also called \emph{safety factors} in \cite{AIApaper2016}) are closely related to stability limits on products of frequencies and step sizes introduced for up to the 6th order resonances for applications in molecular simulations in \cite{schlick}. 
Those limitations are usually related to the frequency of the fastest oscillation $\tilde{\omega}$, and provide a nondimensionalization map for a simulation step size $\Delta t$, i.e.
\begin{equation}\label{eq:NondimensionalizationSafetyFactorAppendix}
h = \text{$S_f$} \, \tilde{\omega} \, \Delta t.
\end{equation}
In \cite{sAIA_paper}, Sec. 4, we introduce three optional nondimensionalization maps $\Delta t \to h$ with different levels of accuracy and computational effort.
First, we distinguish between the more demanding and accurate $S_f = S_\omega$, which requires the computation of the frequencies $\omega_i$ of the harmonic forces \cite{sAIA_paper}, and the cheaper and less precise $S_f = S$ (see Eq. \eqref{eq:SfCases}), where
\begin{equation}\label{eq:FittingFactorsBothAsMaxAppendix}
S_{\omega} = \max \left(1, \frac{2}{\Delta t_{\text{VV}}} \sqrt[6]{\frac{2 \pi (1 - \text{AR})^2}{\sum_{j=1}^D \omega_j^6}} \right), \, \, S = \max \left(1, \frac{2}{\tilde{\omega} \Delta t_{\text{VV}}} \sqrt[6]{\frac{2 \pi (1 - \text{AR})^2}{D}} \right).
\end{equation}
Here, $\Delta t_\text{VV}$ is the integration step size used during the HMC burn-in simulation, AR is the acceptance rate observed in such a simulation, and $D$ is the dimension of the simulated system.

On the other hand, if the system frequencies are available, the nondimensionalization approach adapts according to the standard deviation $\sigma$ of their distribution. More explicitly, if the distribution of the frequencies is disperse ($\sigma > 1$), we apply a correction to \eqref{eq:NondimensionalizationSafetyFactorAppendix}, and define a nondimensionalization factor
\begin{equation}\label{eq:CF}
\text{CF} =S_{\omega}(\tilde{\omega} - \sigma),
\end{equation}
to replace $\text{$S_f$} \, \tilde{\omega}$ in \eqref{eq:NondimensionalizationSafetyFactorAppendix}.
Otherwise, if $\sigma < 1$, the nondimensionalization factor reads
\begin{equation}\label{eq:CFnoSigmaCorrection}
\text{CF} = S_f \tilde{\omega},
\end{equation}
where $S_f$ is defined as in \eqref{eq:SfCases}.

In any case, given a nondimensional step size $h$, its dimensional counterpart for a specific simulated system is computed as
\begin{equation}
\Delta t = \frac{h}{\text{CF}}.
\end{equation}

%\newpage

\section{Sensitivity analysis: Optimal $h_\text{lower}$}\label{app:RandomizationSensitivityCheck}
For the three benchmarks considered in Section~\ref{sec:DeltaTRandomizationNumericalExperiments}, G1000, German and Musk, we perform a sensitivity analysis on $h_\text{lower}$, the lower bound of the optimal randomization interval for the simulation step size \eqref{eq:h_BCSS3maxtoHSL}. 

In particular, we introduce variations of $\pm 5 \%$ around $h_\text{lower}$, keep the same hyperparameter settings as in Section~\ref{sec:DeltaTRandomizationNumericalExperiments} for both HMC and GHMC, %. We consider the three benchmarks considered in Section~\ref{sec:DeltaTRandomizationNumericalExperiments} -- G1000, German, Musk -- 
and monitor the metrics proposed in Section~\ref{sec:PerformanceMetrics}. We use the adaptive s-AIA3 integrator for all numerical experiments. 

Table~\ref{tab:SensitivityCheckResults} reports the results. We observe that all three tested randomization intervals yield similar outcomes, with minor fluctuations attributable to stochastic variability.

\begin{table}[t]
\centering
\resizebox{\textwidth}{!}{\begin{tabular}{l l r r r r}
Benchmark & Sampler & $h_\text{lower}$ & grad/minESS & grad/meanESS & grad/multiESS \\
\hline
\multirow{6}{*}{G1000} 
 & \multirow{3}{*}{HMC} 
    & \eqref{eq:h_BCSS3maxtoHSL} 			  	& 5554.2 & 4198.1 & 4001.0 \\
 &  & \eqref{eq:h_BCSS3maxtoHSL} $- \, 5\%$ 	& 5034.1 & 4361.2 & 4001.0 \\
 &  & \eqref{eq:h_BCSS3maxtoHSL} $+ \, 5\%$ 	& 5107.4 & 4483.2 & 4001.0 \\\cline{2-6}
 & \multirow{3}{*}{GHMC} 
    & \eqref{eq:h_BCSS3maxtoHSL} 			  	& 2518.3 & 2081.1 & 2000.0 \\
 &  & \eqref{eq:h_BCSS3maxtoHSL} $- \, 5\%$ 	& 2818.4 & 2189.3 & 2000.0 \\
 &  & \eqref{eq:h_BCSS3maxtoHSL} $+ \, 5\%$ 	& 2772.3 & 2231.2 & 2000.0 \\
\hline
\multirow{6}{*}{German} 
 & \multirow{3}{*}{HMC} 
%    & \eqref{eq:h_BCSS3maxtoHSL} 			  	& 26.18 & 24.45 & 25.44 \\
    & \eqref{eq:h_BCSS3maxtoHSL} 			  	& 26.2 & 24.5 & 25.4 \\
% &  & \eqref{eq:h_BCSS3maxtoHSL} $- \, 5\%$ 	& 25.25 & 24.77 & 25.23 \\
 &  & \eqref{eq:h_BCSS3maxtoHSL} $- \, 5\%$ 	& 25.3 & 24.8 & 25.2 \\
% &  & \eqref{eq:h_BCSS3maxtoHSL} $+ \, 5\%$ 	& 25.39 & 25.00 & 25.30 \\\cline{2-6}
 &  & \eqref{eq:h_BCSS3maxtoHSL} $+ \, 5\%$ 	& 25.4 & 25.0 & 25.3 \\\cline{2-6}
 & \multirow{3}{*}{GHMC} 
%    & \eqref{eq:h_BCSS3maxtoHSL} 			  	& 12.38 & 10.05 & 10.65 \\
    & \eqref{eq:h_BCSS3maxtoHSL} 			  	& 12.4 & 10.1 & 10.7 \\
% &  & \eqref{eq:h_BCSS3maxtoHSL} $- \, 5\%$ 	& 11.39 & 10.18 & 10.56 \\
 &  & \eqref{eq:h_BCSS3maxtoHSL} $- \, 5\%$ 	& 11.4 & 10.2 & 10.6 \\
% &  & \eqref{eq:h_BCSS3maxtoHSL} $+ \, 5\%$ 	& 11.25 & 10.18 & 10.65 \\
 &  & \eqref{eq:h_BCSS3maxtoHSL} $+ \, 5\%$ 	& 11.3 & 10.2 & 10.7 \\
\hline
\multirow{6}{*}{Musk} 
 & \multirow{3}{*}{HMC} 
    & \eqref{eq:h_BCSS3maxtoHSL} 			  	& 290.3 & 176.3 & 159.5 \\
 &  & \eqref{eq:h_BCSS3maxtoHSL} $- \, 5\%$ 	& 288.6 & 175.8 & 162.2 \\
 &  & \eqref{eq:h_BCSS3maxtoHSL} $+ \, 5\%$ 	& 275.3 & 179.2 & 158.4 \\\cline{2-6}
 & \multirow{3}{*}{GHMC} 
%    & \eqref{eq:h_BCSS3maxtoHSL} 			  	& 179.3 & 73.48 & 39.37 \\
    & \eqref{eq:h_BCSS3maxtoHSL} 			  	& 179.3 & 73.5 & 39.4 \\
% &  & \eqref{eq:h_BCSS3maxtoHSL} $- \, 5\%$ 	& 171.0 & 70.74 & 34.56 \\
 &  & \eqref{eq:h_BCSS3maxtoHSL} $- \, 5\%$ 	& 171.0 & 70.7 & 34.6 \\
% &  & \eqref{eq:h_BCSS3maxtoHSL} $+ \, 5\%$ 	& 174.8 & 70.24 & 35.01 \\
 &  & \eqref{eq:h_BCSS3maxtoHSL} $+ \, 5\%$ 	& 174.8 & 70.2 & 35.0 \\
\hline
\end{tabular}}
\caption{Sensitivity analysis results for variations of $\pm 5 \%$ around $h_\text{lower}$. All three tested intervals produce similar outcomes.}
\label{tab:SensitivityCheckResults}
\end{table}

\section{$L_\text{opt}$ calculation}\label{app:LoptCalculation}
As proposed in Section \ref{sec:TuningL}, we consider
\begin{equation}
X = \frac{\mathbb{E}^{\text{$L$}}_{\text{$S_{f}>1$}} [\Delta H]}{\mathbb{E}^{\text{$L=1$}}_{\text{$S_{f}=1$}} [\Delta H]},
\end{equation}
where $\mathbb{E}^{\text{$L$}}_{\text{$S_{f}>1$}} [\Delta H]$ represents the expected energy error for 1-stage Velocity Verlet at an arbitrary $L \ge 1$ for anharmonic systems ($S_f > 1$), and $\mathbb{E}^{\text{$L=1$}}_{\text{$S_{f}=1$}} [\Delta H]$ \eqref{eq:Rho_VV_1} stands for the expected energy error for 1-stage Velocity Verlet derived at $L = 1$ for systems with harmonic behaviour ($S_f = 1$). $S_f$ is the fitting factor defined in \eqref{eq:FittingFactorsBothAsMaxAppendix}.

One can express the quantity $X$ as
\begin{equation}
X = \frac{\mathbb{E}^{\text{$L$}}_{\text{$S_{f}>1$}} [\Delta H]}{\mathbb{E}^{\text{$L=1$}}_{\text{$S_{f}>1$}} [\Delta H]} \cdot \frac{\mathbb{E}^{\text{$L=1$}}_{\text{$S_{f}>1$}} [\Delta H]}{\mathbb{E}^{\text{$L=1$}}_{\text{$S_{f}=1$}} [\Delta H]} = X_1 \cdot X_2,
\end{equation}
where
\begin{align}
X_1 &= \frac{\mathbb{E}^{\text{$L$}}_{\text{$S_{f}>1$}} [\Delta H]}{\mathbb{E}^{\text{$L=1$}}_{\text{$S_{f}>1$}} [\Delta H]} \overset{\text{\cite{bcss_paper} }}{=} \frac{\sin^2 \left(\eta_h L\right) \rho_\text{VV} (h, b)}{\sin^2 \left(\eta_h\right) \rho_\text{VV} (h, b)} = \frac{\sin^2 \left(\eta_h L\right)}{\sin^2 \left(\eta_h\right)}, \\
X_2 &= \frac{\mathbb{E}^{\text{$L=1$}}_{\text{$S_{f}>1$}} [\Delta H]}{\mathbb{E}^{\text{$L=1$}}_{\text{$S_{f}=1$}} [\Delta H]} \overset{\text{\eqref{eq:Rho_VV_1}}}{=} \frac{\left(\Delta t_\text{VV} \tilde{\omega} S_f\right)^6}{\left(\Delta t_\text{VV} \tilde{\omega}\right)^6} = S_f^6.
\end{align}
Therefore, the resulting expression for $X$ is
\begin{equation}\label{eq:XFactorDeveloped}
X = \frac{\sin^2 \left(\eta_h L\right)}{\sin^2 \left(\eta_h\right)} S_f^6.
\end{equation}
Here, $\eta_h$ is a coefficient that depends on the specific integrator in use (see \cite{bcss_paper}). In particular, it is determined by the number of stages in the integrator, the integration coefficients, and the step size.

We recall that $L$ is a free parameter. Then, to minimize the difference between $\mathbb{E}^{\text{$L$}}_{\text{$S_{f}>1$}} [\Delta H]$ and $\mathbb{E}^{\text{$L=1$}}_{\text{$S_{f}=1$}} [\Delta H]$ (as suggested in Section \ref{sec:TuningL}) one can find such a value of $L \ge 1$ that $X =1$. Thus, using Eq. \eqref{eq:XFactorDeveloped}, one obtains
\begin{equation}\label{eq:LoptFirst}
L_n = \frac{\arcsin \frac{\sin \eta_h}{S_f^3} \pm 2 \pi n}{\eta_h}, \qquad n = 0, 1, 2, ...
\end{equation}
In Section \ref{sec:TuningL}, we argue that an optimal value of $L$ for GHMC in the proposed settings should be small. Therefore, we take $n = 0$ in \eqref{eq:LoptFirst}, which simplifies to
\begin{equation}\label{eq:Loptn0}
L_0 = \frac{\arcsin \frac{\sin \eta_h}{S_f^3}}{\eta_h}.
\end{equation}
Combining $L \ge 1$ and equation \eqref{eq:Loptn0} yields $S_f \le 1$.  However, by definition, $S_f \ge 1$. Therefore, $S_f = 1$ and $L = 1$. This suggests a choice of $L_\text{opt} = 1$ for the systems with harmonic behaviour, i.e. with $S_f = 1$. 

On the other hand, as discussed in \ref{app:3stageCorrection}, $S_f$ is overestimated for sAIA3 by a factor of $\approx 1.3066$. Consequently, it is reasonable to consider $L_\text{opt} = 1$ for the systems with $S_f \lessapprox 1.3$. In practice, it might be fair to relax condition on $S_f$ and choose $L_\text{opt} = 1$ for systems with $S_f$ rounded up to 1, i.e. $S_f < 1.5$.

%Otherwise (i.e. if $S_f \ge 1.5$), we refer again to Eq. \eqref{eq:LoptFirst} and set $n=1$. This leads to
%\begin{equation*}
%L_\text{opt} = \frac{\arcsin \frac{\sin \eta_h}{S_f^3} + 2 \pi}{\eta_h}.
%\end{equation*}
%%
%In the settings of this study, the integration step size $h$ is drawn from $\mathcal{U} (h_\text{lower}, h_\text{CoLSI})$ (see Section \ref{sec:rho_maxima}). We therefore calculate $\eta_h$ (as detailed in \cite{bcss_paper} and Appendix A in \cite{sAIA_paper}) at the midpoint $h^* = \frac{h_\text{lower} + h_\text{CoLSI}}{2}$, which gives $\eta_{h^*} \approx 2.637354$. 
%%
%
Otherwise (i.e. if $S_f \ge 1.5$), we refer again to Eq. \eqref{eq:LoptFirst}. 
In the settings of this study, the integration step size $h$ is drawn from $\mathcal{U} (h_\text{lower}, h_\text{CoLSI})$ (see Section \ref{sec:rho_maxima}). We therefore calculate $\eta_h$ (as detailed in \cite{bcss_paper} and Appendix A in \cite{sAIA_paper}) at the midpoint $h^* = \frac{h_\text{lower} + h_\text{CoLSI}}{2}$, which gives $\eta_{h^*} \approx 2.637354$. 
We emphasize that, following \cite{skeel2014}, we focus on small values of $L$, i.e. of order 0, which implies $n = 1, 2, 3$. 
By evaluating \eqref{eq:LoptFirst} for $n = 1, 2, 3$, we obtain
\begin{equation}
L_1 \approx 2.4, \qquad L_2 \approx 4.8, \qquad L_3 \approx 7.2,
\end{equation}
that is:
%$L_2$ provides a value closer to an integer number, which results in $L_\text{opt} = 5$ for $S_f \ge 1.5$. 
%
%In conclusion:
%
%s the last possible choice leading to obviously small optimal L, L <= 5 (our objective) but it is slightly closer to the whole number than the optimal L obtained with n=1 (4.76 against 2.38). only the choices n=1 and n=2 sutisfy our condition for L to be of an order of 1, with n=2 providing a solution closer to the whole number.
%
%Alternatively, setting $n=2$ results in 
%\begin{equation*}
%L_\text{opt} = \frac{\arcsin \frac{\sin \eta_h}{S_f^3} + 4 \pi}{\eta_h} \approx 4.8
%\end{equation*}
%
%Thus, we obtain
\begin{equation}
L_\text{opt} =
\begin{cases}
1, \quad &\text{for} \quad 1.0\le S_f < 1.5,\\
2, 5, 7, \quad  &\text{otherwise}.
\end{cases}
\end{equation}

\section{Sampling performance of GHMC with different choices of $\varphi$ and $L$}\label{app:PhivsLNumericalExperiments}
\renewcommand{\thefigure}{F.\arabic{figure}}
\setcounter{figure}{0}
\begin{figure}[!]
    \centering
    \includegraphics[width=\textwidth]{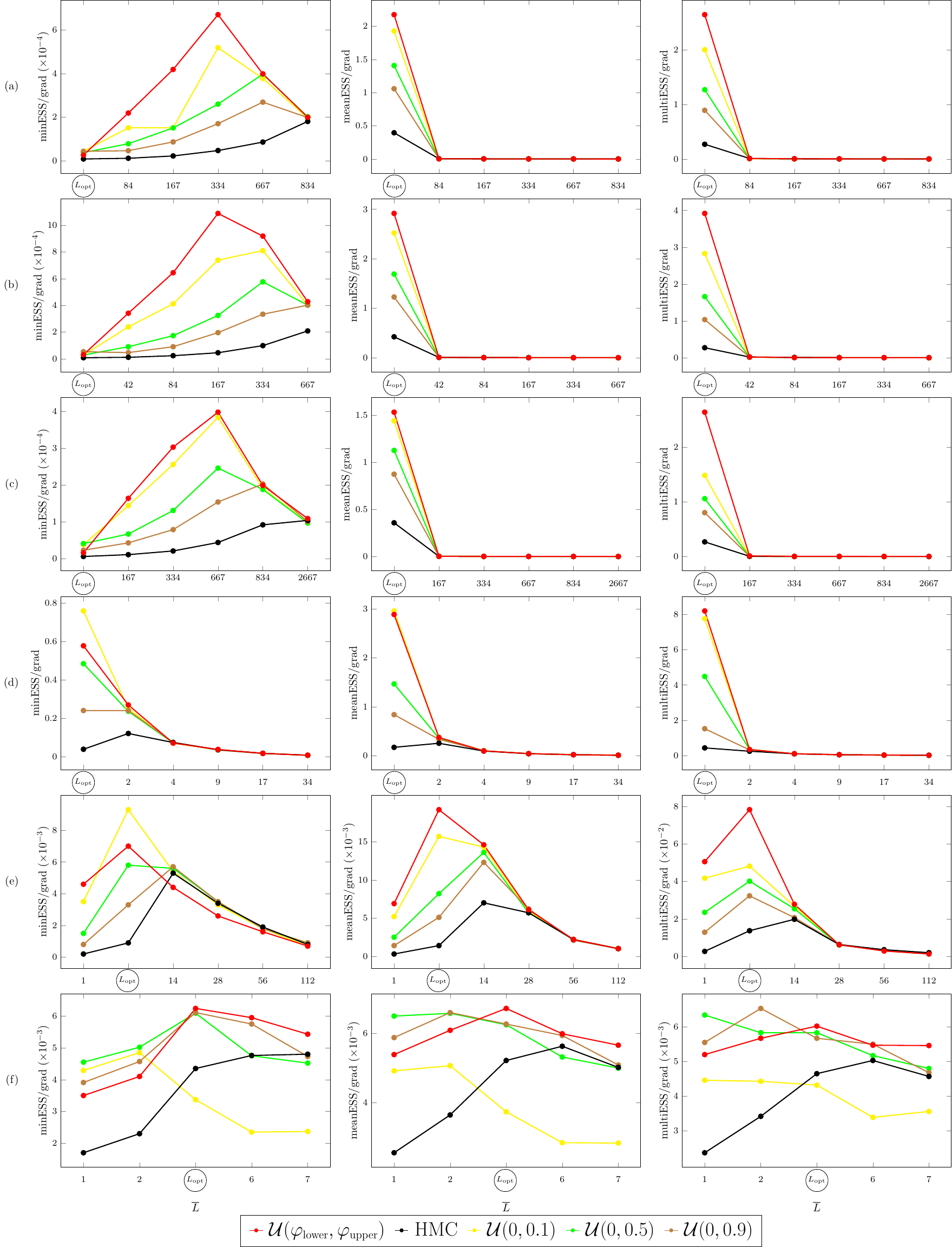}
        \caption{ESS/grad metric, ESS =\{minESS, meanESS, multiESS\} observed in G1000 (a), G500 (b), G2000 (c), German (d), Musk (e), Banana (f) with different choices of $\varphi$ and $\overline{L}$, the mean of a randomization interval for $L$. For each benchmark, the optimal randomization scheme for $L$ \eqref{eq:LoptRandScheme} is circled on the x-axis. The higher values of each metric indicate the better performance.}
    \label{fig:phivsL_all}
\end{figure}

Figure~\ref{fig:phivsL_all} presents the comparison of ESS/grad metric, ESS =\{minESS, meanESS, multiESS\}, obtained for the tested benchmarks using GHMC with different randomization schemes for $\varphi$ and $L$, and the optimal randomization scheme for the integration step size $\Delta t$ (Section~\ref{sec:DeltatOptimalRandomization}). The s-AIA3 integrator is used in all experiments.

Figures \ref{fig:phivsL_all}(a)--\ref{fig:phivsL_all}(c) illustrate that, for the Gaussian benchmarks, $\varphi \sim$ $\mathcal{U}(\varphi_\text{lower}, \varphi_\text{upper})$ ensures the best (or close to the best) results for minESS/grad metric among the ones obtained with all the tested randomization schemes for $\varphi$, regardless of the choice of $L$. 

For the average metrics meanESS and multiESS, a significant improvement with  $\varphi \sim$ $\mathcal{U}(\varphi_\text{lower}, \varphi_\text{upper})$ is observed only at the optimal $L_\text{opt} = 1$ (indicated by a circle). With longer trajectories, GHMC with any tested choice of $\varphi$, including $\varphi = 1$, i.e. HMC, exhibits reduced performance in terms of meanESS and multiESS.

The results for the BLR benchmarks, German and Musk, are displayed in Figures \ref{fig:phivsL_all}(d) and \ref{fig:phivsL_all}(e), respectively. 
For both, the $\mathcal{U}(0, 0.1)$ randomization scheme for $\varphi$ slightly outperforms our optimal setting in terms of the minESS/grad metric. For the German benchmark (Figure \ref{fig:phivsL_all}(d)), AT-GHMC performs comparably to $L_\text{opt}$ combined with $\varphi \sim \mathcal{U}(0, 0.1)$ on the average metrics.
%For the German benchmark (Figure \ref{fig:phivsL_all}(d)), the $\mathcal{U}(0, 0.1)$ randomization scheme for $\varphi$ slightly outperforms our optimal setting in terms of the minESS/grad metric and performs comparably to our approach on the average metrics. 
In contrast, for the Musk benchmark (Figure \ref{fig:phivsL_all}(e)), our method achieves the highest values across 
grad/meanESS and grad/multiESS.
%all ESS metrics, demonstrating superior performance.

Finally, for the Banana benchmark (Figure \ref{fig:phivsL_all}(f)), 
the proposed optimal settings for $\varphi$, $L$ and $\Delta t$ 
lead to the best performance with all tested metrics except grad/multiESS.
%lead either to the best or close to the best performance with all tested metrics.  

\section*{Acknowledgments}\label{sec:Acknowledgements}

We gratefully acknowledge Caetano Souto Maior for his valuable contributions at the early stages of this study, particularly to the implementation of the cell-cell adhesion PDE model in \textsf{HaiCS}.
We thank Maria dM Vivanco for generously sharing her knowledge on the resistance to endocrine therapy in breast cancer and for the fruitful discussions.

This research was supported by MICIU/AEI/10.13039/501100011033 and by ERDF A way for Europe under Grant PID2022-136585NB-C22; by the Basque Government through ELKARTEK Programme under Grants KK-2024/00062; through TT24 under 6/12/TT/2024/00003, and through the BERC 2022-2025 program.  We acknowledge the financial support by the Ministry of Science and Innovation through BCAM Severo Ochoa accreditation CEX2021-001142-S / MICIU/ AEI  / 10.13039/501100011033 and PRE2022-104791 (L. G. B.), and PLAN COMPLEMENTARIO MATERIALES AVANZADOS 2022-2025, PROYECTO 1101288.
The authors acknowledge the financial support received from the grant BCAM-IKUR HPC\&IA 2025-2026, HPCAI4, funded by the Basque Government by the IKUR Strategy and by the European Union NextGenerationEU/PRTR. 
JAC was supported by the Advanced Grant Non\-local-CPD (Nonlocal PDEs for Complex Particle Dynamics: Phase Transitions, Patterns and Synchronization) of the European Research Council Executive Agency (ERC) under the European Union’s Horizon 2020 research and innovation programme (grant agreement No. 883363). 
M.X. R-A was partially funded by the Spanish State Research Agency through the Ramón y Cajal grant RYC2019-027534-I. 
This work has been possible thanks to the support of the computing infrastructure of the BCAM in-house cluster Hipatia, i2BASQUE academic network, Barcelona Supercomputing Center (RES, QHS-2025-1-0027), DIPC Computer Center and the technical and human support provided by IZO-SGI SGIker of UPV/EHU. 

\section*{References}
\bibliographystyle{elsarticle-num}
\bibliography{ATune_paper_bibliography}

\end{document}